\newif\ifsinglecolumn
\singlecolumntrue
\ifsinglecolumn
\documentclass[review]{elsarticle}
\else
\documentclass[preprint,5p,10pt,authoryear]{elsarticle}
\fi

\usepackage[utf8]{inputenc}
\usepackage[T1]{fontenc}
\usepackage{amsmath}
\usepackage{amssymb}
\usepackage{mathrsfs}
\usepackage{color}
\usepackage[pdfencoding=auto]{hyperref}
\usepackage{algorithm,algpseudocode}
\usepackage{graphicx}
\hypersetup{colorlinks=true}
\usepackage[labelformat=simple]{subcaption}

\usepackage{multirow}
\usepackage{times}

\usepackage{placeins}
\usepackage{rotating}

\algnewcommand\algorithmicinput{\textbf{Input:}}
\algnewcommand\INPUT{\item[\algorithmicinput]}
\algnewcommand\algorithmicoutput{\textbf{Output:}}
\algnewcommand\OUTPUT{\item[\algorithmicoutput]}

\DeclareMathOperator*{\argmin}{arg\,min}

\def \pathsaa {images/Scenario_1/landsat8_landsat8_pan/}
\def \pathsab {images/Scenario_1/landsat8_landsat8_ms/}
\def \pathsac {images/Scenario_1/aviris_aviris_hs/}

\def \pathsba {images/Scenario_2/pan_ali_ms_sentinel2/}
\def \pathsbb {images/Scenario_2/pan_landsat8_hs_aviris/}

\def \pathsca {images/Scenario_3/ali_ms_sentinel2_ms/}
\def \pathscb {images/Scenario_3/landsat8_ms_sentinel2_ms/}

\def \pathsda {images/Scenario_4/landsat8_pan_landsat8_ms/}
\def \pathsdb {images/Scenario_4/ali_pan_landsat8_ms/}
\def \pathsdc {images/Scenario_4/landsat8_pan_ali_ms/}

\def \pathsea {images/Scenario_5/ali_ms_aviris_hs/}
\def \pathseb {images/Scenario_5/landsat8_ms_aviris_hs/}

\def \pathsfb {images/Scenario_6/ali_pan_landsat8_pan2/}

\def \pathsga {images/Scenario_7/sentinel2_ms_landsat8_pan/}

\def \pathsha {images/Scenario_8/ali_9_ms_landsat_8_ms/}

\def \pathsia {images/Scenario_9/sentinel_4_ms_landsat_5_ms/}

\def \pathsja {images/Scenario_10/sentinel_6_ms_ali_6_ms/}

\journal{one journal}

\begin{document}

\begin{frontmatter}

\title{Robust fusion algorithms for unsupervised change detection \\between multi-band optical images -- \\A comprehensive case study}

\author[n7]{Vinicius Ferraris\corref{cor1}}
\ead{vinicius.ferraris@enseeiht.fr}
\author[n7]{Nicolas Dobigeon}
\ead{nicolas.dobigeon@enseeiht.fr}
\author[n7]{Marie Chabert}
\ead{marie.chabert@enseeiht.fr}
\address[n7]{University of Toulouse, IRIT/INP-ENSEEIHT, 31071 Toulouse, France}

\begin{abstract}

Unsupervised change detection techniques are generally constrained to two multi-band optical images acquired at different times through sensors sharing the same spatial and spectral resolution. While the optical imagery modality is largely studied in the remote sensing community, this scenario is suitable for a straight comparison of homologous pixels such as pixel-wise differencing. However, in some specific cases such as emergency situations, punctual missions, defense and security, the only available images may be those acquired through different kinds of sensors with different resolutions. Recently some change detection techniques dealing with images with different spatial and spectral resolutions, have been proposed. Nevertheless, they are focused on a specific scenario where one image has a high spatial and low spectral resolution while the other has a low spatial and high spectral resolution. This paper addresses the problem of detecting changes between any two multi-band optical images disregarding their spatial and spectral resolution disparities. To overcome those limitations, state-of-the art methods consist in applying conventional change detection methods after preprocessing steps applied independently on the two images, e.g. resampling operations intended to reach the same spatial and spectral resolutions. Nevertheless, these preprocessing steps may waste relevant information since they do not take into account the strong interplay existing between the two images. Conversely, in this paper, we propose a method that more effectively uses the available information by modeling the two observed images as spatially and spectrally degraded versions of two (unobserved) latent images characterized by the same high spatial and high spectral resolutions. Covering the same scene, the latent images are expected to be globally similar except for possible changes in spatially sparse locations. Thus, the change detection task is envisioned through a robust fusion task which enforces the differences between the estimated latent images to be spatially sparse. We show that this robust fusion can be formulated as an inverse problem which is iteratively solved using an alternate minimization strategy. The proposed framework is implemented for an exhaustive list of applicative scenarios and applied to real multi-band optical images. A comparison with state-of-the-art change detection methods evidences the accuracy of the proposed robust fusion-based strategy.

\end{abstract}


\end{frontmatter}

%\linenumbers

%% main text
\section{Introduction}
Remote sensing consists in collecting measurements, without any physical contact, about an object or phenomenon. This paper focusses on applications to Earth observation and surface monitoring \citep{bellremote1995,elachiintroduction2006,richardsremote2006,campbellintroduction2011}. The type of acquired measurements, also referred to as modality, is intimately related to the sensor. Each modality provides a predefined amount and type of information about the scene. The technological growth and new data policies directly impact the availability of multi-temporal data (i.e., acquired at different time instants)  \citep{bovolotime2015}, which overcomes this limitation, yet, simultaneously introducing new challenges. Notably, multi-temporal data acquired over the same geographical location can be used to detect changes or variations. Thus, analyzing multi-temporal data has culminated in the development of an extremely important area for the remote sensing community, namely, change detection (CD).

CD refers to the techniques used to detect areas where potential changes have occurred between multiple multi-temporal and possibly multi-source (i.e., acquired by different sensors) images acquired over the same scene (geographical location) \citep{singhreview1989, DU2013}. CD is generally conducted within a supervised or unsupervised context \citep{bovolotime2015}. The former requires prior ground-truth knowledge in order to train algorithms maximizing the detection rate while minimizing the false alarm rate. Conversely the latter tries to infer changes after carefully designing a blind model-based distance operator. As ground-truth information is not easily available, significant efforts have been made so that unsupervised CD techniques reach the supervised CD performance. Nevertheless, almost all unsupervised CD methods only focus on a particular scenario, actually the most favorable one which considers two multi-band optical images with same spatial and spectral resolutions \citep{bovolotime2015}. There are two main reasons for considering such a scenario: i) multi-band optical images represent the most commonly used and largely studied remote sensing modality, according to \citep{UCSdatabase}, and ii) images with same spatial and spectral resolutions are pixel-wisely comparable, which makes easy the use of simple distance operators.

Multi-band optical sensors provide a particular representation of the observed scene according to some intrinsic characteristics, particularly, the ability of sensing the reflected electromagnetic spectrum of the incoming light. Well suited to map horizontal structures like land-cover type at large scales \citep{dalla_murachallenges2015}, its facility to be directly interpreted has contributed to popularize its use. Another important aspect of optical imaging is the widely admitted Gaussian modeling of the sensor noises, which has lead to a massive development of least-square like methods, specially for CD. Indeed, the properties of the noise model, for instance the symmetry of the Gaussian probability distribution function, allow CD techniques to be implemented through a simple operation of image differencing, as noticed in \cite{singhreview1989} and \cite{bovolotime2015}. Although CD differencing methods have been adapted to handle multi-band images by considering spectral change vectors \citep{bovolotheoretical2007,bovoloframework2012} and transform analysis \citep{nielsenmultivariate1998,nielsenregularized2007}, all of them rely on the crucial premise of a favorable scenario which assumes that the observed images share the same spatial and spectral resolutions.

%The possibility of detecting changes by exploiting both spatial and spectral information is one of the greatest advantages of this modality, but only some few works tries to exploit it \cite{ferrarisrobust2017,ferraris_Fusion_2017,FerrarisICASSP2017} entirely, always being limited to some specific scenario.
However, the need for flexible and reliable CD techniques that are able to handle more scenarios is real. In some situations, for instance consecutive to natural disasters or within punctual imagery missions, the limited availability of the sensors and the time constraints may preclude the use of the same sensor at two distinct time instants. Thus, in these cases, observed images are possibly from different modalities and do not share the same spatial and/or spectral resolutions. To make existing conventional CD methods usable in these cases, one strategy, hereafter referred to as the worst-case (WC) method, consists in individually and independently, spatially and/or spectrally, resampling the images to reach the same spatial and spectral resolutions. Although this WC technique allows off-the-shelf CD techniques to be used directly, it may remains suboptimal since i) resampling operations independently applied to each image do not take into account their joint characteristics and thus crucial information may be missed and ii) these spatial and spectral operations are generally from a higher to a lower resolution, which results in a significant loss of information. To overcome these limitations, \cite{ferrarisrobust2017} and \cite{ferraris_Fusion_2017} recently proposed two CD approaches specifically designed to deal with multi-band images with different spatial and spectral resolution. Both approaches rely on the inference of a latent (i.e., unobserved) image which results from the fusion of the two observed images. Fusing information contained in remote sensing images has motivated a lot of research works in the literature \citep{KOTWAL2013,KOTWAL2013_2,SONG2014,GHASSEMIAN2016,LI2017}. Within a CD context, the underlying assumption is  most of pixels of the fused image, which are supposed not to have been changed during the time interval, produce consistent information while the few remaining ones, locating in the change regions, produce aberrant information. More precisely, the method proposed in \cite{ferraris_Fusion_2017} is based on a $3$-step procedure (namely fusion, prediction and detection) which, instead of independently preprocess each observed image, recovers a latent high spatial and spectral resolution image containing changed and unchanged regions by fusing observed images. Then, it predicts pseudo-observed images by artificially degrading this estimated latent image using forward models underlying the actually observed images. The result is two pairs, each composed of a predicted image and an observed image with the same spatial and spectral resolutions. Then, any classical multi-band CD method can be finally applied to estimate two change images, that can be thresholded to build the change maps. Conversely, \cite{ferrarisrobust2017} propose a robust fusion-based CD technique which aims at recovering two high spatial and spectral resolution latent images related to the observed images via a double physically-inspired forward model. The robust fusion of multi-band images is formulated as an inverse problem. The difference between the two latent images is assumed to be spatially sparse, implicitly locating the changes at a high resolution scale. The resulting objective function is solved through the use of an alternating minimization algorithm, which iteratively optimizes with respect to (w.r.t.) one latent image and the change image. The CD map can be finally generated from the recovered change image. Both methods offer a way to conduct unsupervised CD between two multi-band optical images. Even if they have shown significant improvements in the detection rate when compared to WC method, they are also still limited to a single scenario: one high spatial low spectral resolution image and one low spatial high spectral resolution image.

In this paper, capitalizing on the method proposed in \cite{ferrarisrobust2017}, we show that the unsupervised CD task can be formulated in a general robust-fusion form for all multi-band optical image scenarios involving two observed images. Contrary to the previous technique, the proposed approach is not limited to one high spatial and low spectral resolution observed image and one low spatial high spectral resolution observed one. It generalizes the robust fusion model to handle all possible configurations of two multi-band optical images. Note that the scenario and the solution proposed in \cite{ferrarisrobust2017} is a specific instance of the framework developed in this paper. Therefore, the same assumptions regarding the two observed images is adopted. Namely they can be jointly approximated by a standard linear decomposition model complemented with an outlier term corresponding to the change image. The outlier term is still characterized by a spatial sparsity-inducing regularization. The resulting objective function, regardless the scenario proposed, is solved through the use of an alternate minimization algorithm. Remarkably, optimizing w.r.t. the latent image always relies on a closed-form solution, which ensures the convergence of the alternate minimization procedure. The CD map can be finally generated from the recovered change image.

The paper is organized as follows. Section \ref{sec:ps} formulates the change detection problem for multi-band optical image. Section \ref{sec:algorithm} presents the solution for the formulated problem based on robust fusion for each possible scenarios. Experimental CD examples are considered in Section \ref{sec:experiments} for each possible scenario described in Section \ref{sec:algorithm}. Section \ref{sec:conclusion} concludes the paper.

%\section{Problem Statement}
\section{Problem formulation}
\label{sec:ps}

\subsection{Generic forward model for multi-band optical images}
\label{subsec:forward}	
In digital image processing, the image formation process inherent to multi-band optical sensors can be generally modeled as a sequence of successive transformations and degradations. These transformations are applied over the original scene and resulting in an output image, commonly referred as the observed image and denoted $\mathbf{Y} \in \mathbb{R}^{m_{\lambda}\times m}$ where $m$ and $m_{\lambda}$ are the numbers of pixels and spectral bands in the observed image. It corresponds to the particular limited representation of the original scene according to the characteristics imposed by the image signal processor describing the sensor. The original scene cannot be exactly represented because of its continuous nature, but it can be conveniently approximated by an (unknown) latent image of higher spatial and spectral resolutions, $\mathbf{X} \in \mathbb{R}^{n_{\lambda}\times n}$, where $n \geq m$ and $n_{\lambda}\geq m_{\lambda}$ are the numbers of pixels and spectral bands, respectively. In what follows, as a well-admitted approximation, the observed and latent images are assumed to be related according to the generic forward model \citep{weibayesian2015-2,Yokoya2012,Simoes2014b}
 \begin{equation}
	\label{eq:model}
		\mathbf{Y} = \mathbf{L}\mathbf{X}\mathbf{R} + \mathbf{N} %\vspace{-0.3cm}
	\end{equation}
where
	\begin{itemize}
		\item $\mathbf{L} \in \mathbb{R}^{m_{\lambda} \times n_{\lambda}}$ is a spectral degradation matrix,
		\item $\mathbf{R} \in \mathbb{R}^{n \times m}$  is a spatial degradation matrix,
		\item $\mathbf{N}$ is an additive term comprising sensor noise and modeling errors.
	\end{itemize}
In \eqref{eq:model}, the left-multiplying matrix $\mathbf{L}$ and right-multiplying matrix $\mathbf{R}$ spectrally and spatially degrade the latent image, respectively, by combining some spectral bands of each pixel or by combining neighboring pixel measurements in each spectral band. More precisely, the spectral degradation $\mathbf{L}$ represents a spectral resolution reduction with respect to the latent image $\mathbf{X}$, as already considered in \cite{Yokoya2012}, \cite{Simoes2014b} and \cite{weifast2015-2}. In practice, this matrix is fully defined by spectral filters characterizing the optical sensors. When the specifications of the sensor are available, these filters are known. Otherwise, they can be learned by cross-calibration, e.g., following the strategies proposed in \cite{Simoes2014b} or \cite{yokoyacross-calibration2013}. On the other hand, the spatial degradation matrix $\mathbf{R}$ models the combination of different spatial transformations applied to the pixel measurements within each spectral band. These transformations are specific of the sensor architecture and include warp, blur, translation and decimation \citep{yokoyacross-calibration2013,weifast2015-2}. In this work, geometrical transformations such as warp and translation are assumed to have been previously corrected, e.g., using image spatial alignment techniques. Thus, similarly to the model considered in \cite{weifast2015-2}, the spatial degradation matrix $\mathbf{R}$ only stands for a spatially invariant blurring, followed by a decimation (i.e., downsampling) operation. Thus, in what follows, the spatial degradation matrix $\mathbf{R}$ will be assumed of the form
\begin{equation}
\label{eq:spatial_degradation_matrix}
  \mathbf{R} = \mathbf{B}\mathbf{S}.
\end{equation}
The sparse symmetric Toeplitz matrix $\mathbf{B} \in \mathbb{R}^{n\times n}$ in \eqref{eq:spatial_degradation_matrix} operates a cyclic convolution on each individual band to model a space-invariant blur associated with a symmetric convolution kernel. The decimation operation, denoted by the $n\times m$ matrix $\mathbf{S}$ in \eqref{eq:spatial_degradation_matrix}, corresponds to a uniform downsampling operator\footnote{The corresponding operator $\mathbf{S}^{T}$ represents an upsampling transformation by zero-interpolation from $m$ to $n$.} of factor $d = d_{r} \times d_{c}$ with $ m = n/d$ ones on the block diagonal and zeros elsewhere, such that $\mathbf{S}^{T}\mathbf{S} = \mathbf{I}_{m}$ \cite{weifast2015-2}.

The noise corrupting multi-band optical images is generally modeled as additive and Gaussian \citep{bovolotime2015,elachiintroduction2006,loncanhyperspectral2015,weifast2015-2}. Thus the noise matrix $\mathbf{N}$ in \eqref{eq:model} is assumed to be distributed according to the following matrix normal distribution (see Appendix \ref{app:matrix_normal_distribution})
	\begin{equation*}
        \label{eq:noise_stats}
			\mathbf{N} \sim \mathcal{M}\mathcal{N}_{m_{\lambda},m}(\mathbf{0}_{m_{\lambda}\times m},\mathbf{\Lambda},\mathbf{\Pi}).
	\end{equation*}
The row covariance matrix $\mathbf{\Lambda}$ carries information regarding the between-band spectral correlation. In what follows, similarly to the approach in \cite{weifast2015-2}, this covariance matrix $\mathbf{\Lambda}$ will be assumed to be diagonal, which implies that the noise is spectrally independent and characterized by a specific variance in each band. Conversely, the column covariance matrix $\mathbf{\Pi}$ models the noise correlation w.r.t. to the pixel locations. Following a hypothesis widely admitted in the literature, this matrix is assumed to be identity, $\mathbf{\Pi}=\mathbf{I}_{m}$, to reflect the fact the noise is spatially independent. In real applications, both matrices $\mathbf{\Lambda}$ and $\mathbf{\Pi}$ can be estimated by  calibration \citep{yokoyacross-calibration2013}.

\subsection{Problem statement}
\label{subsec:ps}

%As in definition, CD considers observed images (here two) acquired in different times by different (or not) sensors. Let us denote $\mathbf{P}_{1}$ and $\mathbf{P}_{2}$ the acquisition sensors with respective acquisition times $t_1$ and $t_2$ of two co-registered multi-band optical images. It is not assumed any specific information about time ordering of acquisition, either $t_2<t_1$ or $t_2>t_1$ are possible cases. Hence, without loss of generality, one can denote $\mathbf{Y}_{1} \in \mathbb{R}^{m_{\lambda_{1}} \times m_{1}}$ the image acquired by the sensor $\mathbf{P}_{1}$ at time $t_1$ while $\mathbf{Y}_{2} \in \mathbb{R}^{m_{\lambda_{2}} \times m_{2}}$ the image acquired by the sensor $\mathbf{P}_{2}$ at time $t_2$.

Let us consider two co-registered multi-band optical images $\mathbf{Y}_{1} \in \mathbb{R}^{m_{\lambda_{1}} \times m_{1}}$ and $\mathbf{Y}_{2} \in \mathbb{R}^{m_{\lambda_{2}} \times m_{2}}$ acquired by two sensors $\textsf{S}_1$ and $\textsf{S}_2$ at times $t_1$ and $t_2$, respectively. It is not assumed any specific information about time ordering of acquisitions, either $t_2<t_1$ or $t_2>t_1$ are possible cases. The problem addressed in this paper consists in detecting significant changes between these two multi-band optical images. This is a challenging task mainly due to the possible spatial and/or spectral resolution dissimilarity (i.e., $m_{\lambda_{1}}\neq m_{\lambda_{2}}$ and/or $ m_{1} \neq  m_{2}$), which prevents any use of simple yet efficient differencing operation \citep{singhreview1989,bovolotime2015}. To alleviate this issue, this work proposes to generalize the CD framework introduced in \cite{ferrarisrobust2017} to handle all possible combinations (scenarios) of two multi-band optical images. More precisely, following the widely admitted forward model described in Section \ref{subsec:forward} and adopting consistent notations, the observed images $\mathbf{Y}_{1}$ and $\mathbf{Y}_{2}$ can be related to two latent images $\mathbf{X}_{1} \in \mathbb{R}^{n_{\lambda}\times n}$ and $\mathbf{X}_{2} \in \mathbb{R}^{n_{\lambda}\times n}$ with the same spatial and spectral resolutions
\begin{subequations}
\label{eq:jointobsmodel}
		\begin{align}
			&\mathbf{Y}_{1} = \mathbf{L}_{1}\mathbf{X}_{1}\mathbf{R}_{1} + \mathbf{N}_{1}  \label{eq:jointobsmodelAlpha}\\
			&\mathbf{Y}_{2} = \mathbf{L}_{2}\mathbf{X}_{2}\mathbf{R}_{2} + \mathbf{N}_{2} \label{eq:jointobsmodelBeta}.
		\end{align}
\end{subequations}
%$\mathbf{Y}_{1}^{t_1}$ $\mathbf{Y}_{2}^{t_2}$
where $\mathbf{L}_{1}$ and $\mathbf{L}_{2}$ denote two spectral degradation operators and $\mathbf{R}_{1}$ and $\mathbf{R}_{2}$ are two spatial degradation operators that can be decomposed according to \eqref{eq:spatial_degradation_matrix}. Note that \eqref{eq:jointobsmodelAlpha} and \eqref{eq:jointobsmodelBeta} are a specific double instance of the model \eqref{eq:model}. In particular, the two multi-band latent images $\mathbf{X}_{1}$ and $\mathbf{X}_{2}$ share the same spectral and spatial resolutions, generally higher than those of the observed images:
\begin{equation}
\label{eq:resolutions_assumption}
  n_{\lambda} \geq \max\left\{m_{\lambda_1},m_{\lambda_2}\right\}\ \text{and/or} \ n \geq \max\left\{m_{1},m_{2}\right\}.
\end{equation}
Thereby, after inferring the latent images, any classical differencing technique can be subsequently implemented to compute a change image $\Delta\mathbf{X}=\left[\Delta\mathbf{x}_1,\ldots,\Delta\mathbf{x}_n\right] \in \mathbb{R}^{n_{\lambda},n}$ defined by
\begin{equation}
\label{eq:assumption}
 \Delta\mathbf{X} = \mathbf{X}_{2} - \mathbf{X}_{1}
\end{equation}
where $\Delta\mathbf{x}_p\in\mathbb{R}^{n_{\lambda}}$ denotes the spectral change vector in the $p$th pixel ($p=1,\ldots,n$). It is worth noting that, under the assumptions \eqref{eq:resolutions_assumption}, these changes can be identified at a high spatial and spectral resolutions. Finally this change image can be further exploited by conducting a pixel-wise change vector analysis (CVA) which exhibits the polar coordinates (i.e., magnitude and direction) of the spectral change vectors \citep{johnsonchange1998}. Then, to spatially locate the changes, a natural approach consists in monitoring the information contained in the magnitude part of this representation, summarized by the change energy image \citep{singhreview1989, bovolotheoretical2007, bovoloframework2012}
\begin{equation*}
\label{eq:spectral_change_energy_image}
  \mathbf{e} =\left[e_1,\ldots,e_n\right]\in \mathbb{R}^{n}
\end{equation*}
with
\begin{equation*}
  e_p = \left\|\Delta\mathbf{x}_p\right\|_2, \quad p=1,\ldots,n.
\end{equation*}
When the CD problem in the $p$th pixel is formulated as the binary hypothesis testing
\begin{equation*}
\label{eq:test}
 \left\{
		\begin{array}{rcl}
			\mathcal{H}_{0,p} &:& \text{no change occurs in the $p$th pixel}  \\
			\mathcal{H}_{1,p} &:& \text{a change occurs in the $p$th pixel}
		\end{array}
        \right.
\end{equation*}
a pixel-wise statistical test can be written by thresholding the change energy image pixels
\begin{equation*}
    \label{eq:decision_rule}
  e_p \overset{\mathcal{H}_{1,p}}{\underset{\mathcal{H}_{0,p}}{\gtrless}} \tau.
\end{equation*}
The final binary CD map denoted ${\mathbf{d}} = \left[d_1,\ldots,d_n\right] \in \{0,1\}^n$ can be derived as
	\begin{equation*}
	\label{eq:CVArule}
 {d}_p = \left\{\begin{array}{lll}
             1 & \mbox{if } e_p \geq \tau & (\mathcal{H}_{1,p})\\
			 0 & \mbox{otherwise}          & (\mathcal{H}_{0,p}).
				\end{array}\right.
\end{equation*}
%When complementary information needs to be extracted from the change image $\Delta\mathbf{X}$, e.g., to identify different types of changes, the whole polar representation (i.e., both magnitude and direction) can be fully exploited \cite{bovolotheoretical2007, bovoloframework2012}.
As a consequence, to solve the multi-band image CD problem, the key issue lies in the joint estimation of the pair of HR latent images $\left\{\mathbf{X}_{1},\mathbf{X}_{2}\right\}$ from the joint forward model \eqref{eq:jointobsmodel} or, equivalently, the joint estimation of one latent image and the difference image, i.e., $\left\{\mathbf{X}_{1},\Delta\mathbf{X}\right\}$. Finally, the next paragraph introduces the CD-driven optimization problem to be solved.

\subsection{Optimization problem}
Following a Bayesian approach, the joint maximum a posteriori (MAP) estimator $\left\{\hat{\mathbf{X}}_{1,\mathrm{MAP}}, \mathbf{\Delta}\hat{\mathbf{X}}_{\mathrm{MAP}}\right\}$ of the latent and change images can be derived by maximizing the posterior distribution
	\begin{multline*}
\label{eq:posterior}
		 p(\mathbf{X}_{1},\Delta\mathbf{X}|\mathbf{Y}_{2},\mathbf{Y}_{1}) \propto p(\mathbf{Y}_{2},\mathbf{Y}_{1}|\mathbf{X}_{1},\Delta\mathbf{X}) p(\mathbf{X}_{1}) p(\Delta\mathbf{X})
	\end{multline*}
where $p(\mathbf{Y}_{2},\mathbf{Y}_{1}|\mathbf{X}_{1},\Delta\mathbf{X})$ is the joint likelihood function and $p(\mathbf{X}_{1})$ and $p(\Delta\mathbf{X})$ correspond to the prior distributions associated with the latent and change images, respectively, assumed to be a priori independent. Because of the additive nature and statistical properties of the noise discussed in Section \ref{subsec:forward}, this boils down to solve the following minimization problem
	\begin{equation}
		\label{eq:MAP}
		\left\{\hat{\mathbf{X}}_{1,\mathrm{MAP}}, \Delta\hat{\mathbf{X}}_{\mathrm{MAP}}\right\} \in \operatornamewithlimits{argmin}_{\mathbf{X}_{1},\Delta\mathbf{X}}   \mathcal{J}\left(\mathbf{X}_{1},\Delta\mathbf{X}\right) %\vspace{-0.25cm}
	\end{equation}
with
\begin{equation}
\label{eq:objective}
\begin{aligned}
  \mathcal{J}\left(\mathbf{X}_{1},\Delta\mathbf{X}\right)&=\frac{1}{2}\left\|\mathbf{\Lambda}_{2}^{-\frac{1}{2}} \left(\mathbf{Y}_{2} - \mathbf{L}_{2}\left(\mathbf{X}_{1}+\Delta\mathbf{X}\right)\mathbf{R}_{2}  \right) \right\|_{\mathrm{F}}^{2}  \\
		&+ \frac{1}{2}\left\|\mathbf{\Lambda}_{1}^{-\frac{1}{2}} \left(\mathbf{Y}_{1} - \mathbf{L}_{1}\mathbf{X}_{1}\mathbf{R}_{1} \right) \right\|_{\mathrm{F}}^{2} \\
        &+ \lambda \phi_1\left(\mathbf{X}_{1}\right) + \gamma \phi_2 \left(\Delta\mathbf{X}\right).
\end{aligned}
\end{equation}
where $\left\|\cdot\right\|_{\mathrm{F}}$ denotes the Frobenius norm.
The regularizing functions $\phi_{1}(\cdot)$ and $\phi_{2}(\cdot)$ can be related to the negative log-prior distributions of the latent and change images, respectively, and the parameters $\lambda$ and $\gamma$ tune the amount of corresponding penalizations in the overall objective function $\mathcal{J}(\mathbf{X}_{1},\Delta\mathbf{X})$. These functions should be carefully designed to exploit any prior knowledge regarding the parameters of interest. As discussed in Section \ref{subsec:robust_fusion}, numerous regularizations can be advocated for the latent image $\mathbf{X}_{1}$. In this work, to maintain computational efficiency while providing accurate results \citep{loncanhyperspectral2015}, a Tikhonov regularization proposed in \cite{weibayesian2015-2} has been adopted
\begin{equation*}
\label{eq:phi_1}
  \phi_1\left(\mathbf{X}_{1}\right) = \left\|\mathbf{X}_{1} - \bar{\mathbf{X}}_1\right\|_{\mathrm{F}}^2
\end{equation*}
where $\bar{\mathbf{X}}_1$ refers to a crude estimate of $\mathbf{X}_{1}$.%, e.g., resulting from a naive spatial interpolation of the observed LR-HS image $\mathbf{Y}_{1}$. This choice has been proven to maintain computational efficiency while providing accurate results \cite{loncanhyperspectral2015}. Additionally, a subspace-based representation can also be adopted to enforce $\mathbf{X}_{1}$ to live in a previously identified subspace, as advocated in \cite{weibayesian2015} and \cite{Simoes2014b}.

Regarding the regularizing function $\phi_2(\cdot)$, as already mentioned in the previous section, it should reflect the fact that most of the pixels are expected to remain unchanged i.e., most of the columns of the change image $\Delta \mathbf{X}$ are expected to be null vectors. Thus, the regularizing function $\phi_2(\cdot)$ is chosen as in \cite{Fevotte2015} as the $\ell_{2,1}$-norm of the change image
\begin{equation}
\label{eq:phi_2}
\phi_{2}\left(\Delta\mathbf{X}\right) = \left\|\Delta\mathbf{X}\right\|_{2,1} = \sum_{p=1}^{n} \left\|\Delta \mathbf{x}_p\right\|_2.
\end{equation}
The next section describes the general iterative algorithm scheme which solves the minimization problem \eqref{eq:MAP}.

\section{Robust multi-band image fusion algorithm: generic formulation}
\label{sec:algorithm}

%Computing the joint MAP estimator of the latent image $\mathbf{X}_{1}$ at time $t_1$ and of the change image $\Delta\mathbf{X}$ can be achieved by solving the minimization problem in \eqref{eq:MAP}. However, no closed-form solution can be derived for this problem for all the scenarios of interest. Thus this section introduce a minimization algorithm which iteratively converges to this solution. It consists in sequentially solving the two problems associate with each individual variables $\mathbf{X}_{1}$ and $\Delta\mathbf{X}$. This alternating minimization (AM) algorithm is summarized in Algo. \ref{AM} whose main steps (referred to as \emph{fusion} and \emph{correction}) are detailed in what follows according to each specific scenario. It is worth noting that the difficulty of conducting these two steps is highly related to the spatial and/or spectral degradations operated on the two latent images, according to the applicative scenarios detailed in paragraph \ref{subsec:tax}. Interestingly, we will show that these steps generally reduce to ubiquitous (multi-band) image processing tasks, namely denoising, spectral deblurring or spatial super-resolution from a single or multiple images, for which efficient and reliable strategies have been already proposed in the literature.

Computing the joint MAP estimator of the latent image $\mathbf{X}_{1}$ at time $t_1$ and of the change image $\Delta\mathbf{X}$ can be achieved by solving the minimization problem in \eqref{eq:MAP}. However, no closed-form solution can be derived for this problem for all the scenarios of interest. Thus this section introduces a minimization algorithm which iteratively converges to this solution. This alternating minimization (AM) algorithm, summarized in Algo. \ref{AM}, consists in iteratively minimizing the objective function \eqref{eq:objective} w.r.t. $\mathbf{X}_{1}$ and $\Delta\mathbf{X}$, within so-called \emph{fusion} and \emph{correction} discussed below.
\begin{algorithm}[h!]
	\caption{Algorithm for robust multi-band image fusion}
	\label{alg:AM}
	\begin{algorithmic}[1]
		\INPUT $\mathbf{Y}_{1}$, $\mathbf{Y}_{2}$, $\mathbf{L}_1$, $\mathbf{L}_2$, $\mathbf{R}_1$, $\mathbf{R}_2$, $\mathbf{\Lambda}_{1}$, $\mathbf{\Lambda}_{2}$.
        \State Set $\Delta\mathbf{X}_{1}$.
		\For{$k = 1,\ldots,K$}
        \State \% \emph{Fusion step}
		\State $ \mathbf{X}_{1}^{(k+1)} = \argmin_{\mathbf{X}_{1}}\mathcal{J}(\mathbf{X}_{1},\Delta\mathbf{X}^{(k)})$
        \State \% \emph{Correction step}
		\State $ \Delta\mathbf{X}^{(k+1)} =  \argmin_{\Delta\mathbf{X}}\mathcal{J}(\mathbf{X}_{1}^{(k+1)},\Delta\mathbf{X})$
		\EndFor
		\OUTPUT $\hat{\mathbf{X}}_{1,\mathrm{MAP}}\triangleq\mathbf{X}_{1}^{(K+1)}$ and $\Delta\hat{\mathbf{X}}_{\mathrm{MAP}}\triangleq \Delta\hat{\mathbf{X}}^{(K+1)}$
	\end{algorithmic}
	\label{AM}
\end{algorithm}
\subsection{Fusion step}
\label{subsec:robust_fusion}
As mentioned above, the forward model \eqref{eq:jointobsmodel} relying on the pair $\left\{\mathbf{X}_{1},\mathbf{X}_{2}\right\}$ of latent images can be rewritten as a function of $\left\{\mathbf{X}_{1},\Delta\mathbf{X}\right\}$, i.e.,
\begin{subequations}
\label{eq:jointobsmodel_bis}
		\begin{align}
			&\mathbf{Y}_{1} = \mathbf{L}_{1}\mathbf{X}_{1}\mathbf{R}_{1} + \mathbf{N}_{1} \label{eq:jointobsmodel_bisHR}\\
			&\mathbf{Y}_{2} = \mathbf{L}_{2}\left(\mathbf{X}_{1}+\Delta\mathbf{X}\right)\mathbf{R}_{2} + \mathbf{N}_{2} \label{eq:jointobsmodel_bisLR}.
		\end{align}
\end{subequations}
Generalizing the strategy proposed in \cite{ferrarisrobust2017}, given the change image $\Delta\mathbf{X}$ and the image $\mathbf{Y}_{1}$ observed at time $t_1$, a \emph{corrected} image denoted $\tilde{\mathbf{Y}}_{2}$ that would be acquired by the sensor $\textsf{S}_{2}$ at time $t_1$ can be defined as
\begin{equation}
	\label{eq:pseudoObs}
		\tilde{\mathbf{Y}}_{2} = \mathbf{Y}_{2} - \mathbf{L}_{2}\Delta\mathbf{X}\mathbf{R}_{2}.
	\end{equation}
With this notation, the forward model \eqref{eq:jointobsmodel_bis} can be easily rewritten, leading to
\begin{subequations}
\label{eq:jointobsmodel_ter}
\begin{align}
			&{\mathbf{Y}}_{1} = \mathbf{L}_{1}\mathbf{X}_{1}\mathbf{R}_{1} + \mathbf{N}_{1}\\
            &\tilde{\mathbf{Y}}_{2} = \mathbf{L}_{2}\mathbf{X}_{1}\mathbf{R}_{2} + \mathbf{N}_{2}.
\end{align}
\end{subequations}
Thus, the fusion step, at iteration $k$, consists in minimizing \eqref{eq:objective} w.r.t. $\mathbf{X}_{1}$, i.e.,
			\begin{equation*}
        \hat{\mathbf{X}}_{1}^{(k+1)} = \operatornamewithlimits{argmin}_{\mathbf{X}_{1}} \mathcal{J}_1\left(\mathbf{X}_{1}\right)\triangleq \mathcal{J}\left(\mathbf{X}_{1},\Delta\mathbf{X}^{(k)}\right)
      \end{equation*}
      with
      \begin{equation}
\label{eq:objective_x}
\begin{aligned}
  \mathcal{J}_1\left(\mathbf{X}_{1}\right)&=\frac{1}{2}\left\|\mathbf{\Lambda}_{2}^{-\frac{1}{2}} \left(\tilde{\mathbf{Y}}_{2}^{(k)} - \mathbf{L}_{2}\mathbf{X}_{1}\mathbf{R}_{2}  \right) \right\|_{\mathrm{F}}^{2}  \\
		&+ \frac{1}{2}\left\|\mathbf{\Lambda}_{1}^{-\frac{1}{2}} \left(\mathbf{Y}_{1} - \mathbf{L}_{1}\mathbf{X}_{1}\mathbf{R}_{1} \right) \right\|_{\mathrm{F}}^{2} \\
        &+ \lambda \left\|\mathbf{X}_{1} - \bar{\mathbf{X}}_1\right\|_{\mathrm{F}}^2.
\end{aligned}
\end{equation}
The double forward model \eqref{eq:jointobsmodel_ter}, as well as the optimization problem \eqref{eq:objective_x}, underly the estimation of an image $\mathbf{X}_{1}$ from an observed image $\mathbf{Y}_{1}$ and a pseudo-observed image $\tilde{\mathbf{Y}}_{2}$. Various instances of this pixel-level fusion problem have been widely considered in the literature \citep{KOTWAL2013,SONG2014,GHASSEMIAN2016,LI2017}. For instance, \cite{JYW2010} and \cite{zhaofast2016} have addressed the problem of single mono-band image superresolution from a single observed image $\mathbf{Y}_{1}$, i.e., with $\mathbf{L}_{1} = \mathbf{I}_{m_{\lambda_1}}$ and $m_{\lambda_1}=n_{\lambda}=1$. The problem of fusing several degraded mono-band images to recover a common high resolution latent image has been considered in \cite{Elad1997}. Similarly, the model \eqref{eq:jointobsmodel_ter} generalizes the conventional observational model widely adopted by the remote sensing community to conduct multi-band image fusion \citep{Hardie2004,Eismann2005,Zhang2009,Yokoya2012,KOTWAL2013_2,Simoes2014b,weihyperspectral2015,weibayesian2015-2,weifast2015-2}. Within this specific scenario, a high spatial and high spectral resolution latent image $\mathbf{X}_{1}$ is estimated from two observed images, one of low spatial and high spectral resolutions (i.e., $\mathbf{L}_{1} = \mathbf{I}_{m_{\lambda_1}}$) and the other of high spatial and low spectral resolutions (i.e., $\mathbf{R}_{2} = \mathbf{I}_{n_{2}}$).

In this context, the CD task considered in this paper can be cast as a so-called \emph{robust} fusion problem since the multi-band image fusion model \eqref{eq:jointobsmodel_ter} implicitly depends on the (unknown) change image $\Delta\mathbf{X}$. More precisely, since the two latent images $\mathbf{X}_{1}$  and $\mathbf{X}_{2}$ are related to the same scene observed at two time instants, they are expected to share a high level of similarity, i.e., the change image $\Delta\mathbf{X}$ is expected to be spatially sparse. Thus, this additional unknown change image $\Delta\mathbf{X}$ to be inferred can be considered as an outlier term, akin to those encountered in several robust factorizing models such as robust principal component analysis (RPCA) \citep{candesrobust2011} and robust nonnegative factorization \citep{Fevotte2015}. A particular instance of this strategy has been successfully adopted in \cite{ferrarisrobust2017} to detect changes between two complementary multi-band images, i.e., in the particular case of $\mathbf{L}_{1} = \mathbf{I}_{m_{\lambda_1}}$ and $\mathbf{R}_{2} = \mathbf{I}_{n_{2}}$. In this work, we propose to follow a similar route while generalizing the approach to the much more generic model \eqref{eq:jointobsmodel} to handle all practical scenarios of CD. These different scenarios are discussed in the next paragraph.

\newcommand{\notspecblur}{$-$}
\newcommand{\notspatblur}{$-$}
\setlength{\tabcolsep}{3pt}
%\renewcommand{\arraystretch}{}
%\begin{table*}
\begin{sidewaystable}
%\begin{footnotesize}
    \centering
    \begin{tabular}{c|c|c|c|c|c|}
    \cline{2-6}
 & \multicolumn{2}{c|}{{\textbf{Forward model $\sharp 1$}}} & \multicolumn{2}{c|}{{\textbf{Forward model $\sharp 2$}}} & \multirow{3}{*}{\textbf{Comments}}\\
\cline{2-5}
    \multicolumn{1}{c|}{} & \multirow{1}{*}{\textbf{Spectral}} & \multirow{1}{*}{\textbf{Spatial}} & \multirow{1}{*}{\textbf{Spectral}} & \multirow{1}{*}{\textbf{Spatial}} &  \\
    \multicolumn{1}{c|}{} & {\textbf{degradation }} & \textbf{degradation} & \textbf{degradation}& \textbf{degradation} & \\
    \hline
	\hline
		\multicolumn{1}{|c|}{\multirow{2}{*}{\rotatebox{00}{$\mathcal{S}_1$}}} & \multirow{2}{*}{\notspecblur} & \multirow{2}{*}{\notspatblur} &   \multirow{2}{*}{\notspecblur} & \multirow{2}{*}{\notspatblur} & Conventional CD framework --\\
    \multicolumn{1}{|c|}{} & & & &  &  $\mathbf{Y}_1$ and $\mathbf{Y}_2$ of same spatial and spectral resolutions\\
        \hline
		\multicolumn{1}{|c|}{\multirow{2}{*}{\rotatebox{00}{$\mathcal{S}_2$}}} & \multirow{2}{*}{$\mathbf{L}_{1}$} & \multirow{2}{*}{\notspatblur} &   \multirow{2}{*}{\notspecblur} & \multirow{2}{*}{\notspatblur} & {$\mathbf{Y}_1$ of lower spectral resolution}\\
       \multicolumn{1}{|c|}{} & & & &  & $\mathbf{Y}_1$ and $\mathbf{Y}_2$ of same spatial resolutions \\
        \hline
		\multicolumn{1}{|c|}{\multirow{2}{*}{\rotatebox{00}{$\mathcal{S}_3$}}} & \multirow{2}{*}{\notspecblur} & \multirow{2}{*}{$\mathbf{R}_{1}$} &   \multirow{2}{*}{\notspecblur} & \multirow{2}{*}{\notspatblur} &{$\mathbf{Y}_1$ of lower spatial resolution}\\
       \multicolumn{1}{|c|}{} & & & &  &  $\mathbf{Y}_1$ and $\mathbf{Y}_2$ of same spectral resolutions\\
        \hline
		\multicolumn{1}{|c|}{\multirow{2}{*}{\rotatebox{00}{$\mathcal{S}_4$}}} & \multirow{2}{*}{\notspecblur} & \multirow{2}{*}{$\mathbf{R}_{1}$} &   \multirow{2}{*}{$\mathbf{L}_{2}$} & \multirow{2}{*}{\notspatblur} & \multirow{2}{*}{$\mathbf{Y}_1$ and $\mathbf{Y}_2$ of complementary resolutions}\\
     \multicolumn{1}{|c|}{} & & & &  & \\
        \hline
		\multicolumn{1}{|c|}{\multirow{2}{*}{\rotatebox{00}{$\mathcal{S}_5$}}} & \multirow{2}{*}{$\mathbf{L}_{1}$} & \multirow{2}{*}{$\mathbf{R}_{1}$} &   \multirow{2}{*}{\notspecblur} & \multirow{2}{*}{\notspatblur} & \multirow{2}{*}{$\mathbf{Y}_1$ of low spatial and spectral resolutions}\\
      \multicolumn{1}{|c|}{}  & & & &  &  \\
        \hline

       \multicolumn{1}{|c|}{\multirow{2}{*}{\rotatebox{00}{$\mathcal{S}_6$}}} & \multirow{2}{*}{\notspecblur} &  \multirow{2}{*}{$\mathbf{R}_{1}$} & \multirow{2}{*}{\notspecblur} & \multirow{2}{*}{$\mathbf{R}_{2}$} &  Generalization of $\mathcal{S}_3$ with non-integer\\
      \multicolumn{1}{|c|}{}  & & & &  & relative spatial downsampling factor \\
        \hline
         \multicolumn{1}{|c|}{\multirow{2}{*}{\rotatebox{00}{$\mathcal{S}_7$}}} & \multirow{2}{*}{$\mathbf{L}_{1}$}  &  \multirow{2}{*}{$\mathbf{R}_{1}$} & \multirow{2}{*}{\notspecblur} & \multirow{2}{*}{$\mathbf{R}_{2}$} &   Generalization of $\mathcal{S}_4$ with non-integer\\
     \multicolumn{1}{|c|}{}   & & & &  &  relative spatial downsampling factor\\
        \hline
        \multicolumn{1}{|c|}{\multirow{2}{*}{\rotatebox{00}{$\mathcal{S}_8$}}}
        & \multirow{2}{*}{$\mathbf{L}_{1}$} &  \multirow{2}{*}{\notspatblur} & \multirow{2}{*}{$\mathbf{L}_{2}$} & \multirow{2}{*}{\notspatblur} &  Generalization of $\mathcal{S}_2$ with some \\
      \multicolumn{1}{|c|}{}  & & & &  & non-overlapping spectral bands\\
        \hline
        \multicolumn{1}{|c|}{\multirow{2}{*}{\rotatebox{00}{$\mathcal{S}_9$}}}
        & \multirow{2}{*}{$\mathbf{L}_{1}$}&  \multirow{2}{*}{$\mathbf{R}_{1}$} & \multirow{2}{*}{$\mathbf{L}_{2}$} & \multirow{2}{*}{\notspatblur} &  Generalization of $\mathcal{S}_4$ with some \\
     \multicolumn{1}{|c|}{}   & & & &  & non-overlapping spectral bands\\
        \hline
				\multicolumn{1}{|c|}{\multirow{3}{*}{\rotatebox{00}{$\mathcal{S}_{10}$}}}
        & \multirow{3}{*}{$\mathbf{L}_{1}$} &  \multirow{3}{*}{$\mathbf{R}_{1}$} & \multirow{3}{*}{$\mathbf{L}_{2}$}& \multirow{3}{*}{$\mathbf{R}_{2}$} &  Generalization of $\mathcal{S}_4$ with some non-overlapping\\
       \multicolumn{1}{|c|}{} & & & &  &  spectral bands and non-integer relative\\
     \multicolumn{1}{|c|}{} & & & &  &  spatial downsampling factor\\
         \hline
    \hline
    \end{tabular}
        \caption{Overviews of the spectral and spatial degradations w.r.t. to experimental scenarios. The symbol $-$ stands for ``no degradation''.}
  \label{table:SEN_mat}
%  \end{footnotesize}
%\end{table*}
\end{sidewaystable}

\subsection{Correction step}

Given the current state $\mathbf{X}_{1}$ of the latent image, the predicted image that would be observed by the sensor $\textsf{S}_{2}$ at time $t_1$ can be defined as
\begin{equation}
	\label{eq:pseudoObs_dX}
  \check{\mathbf{Y}}_{2}^{(k)} = \mathbf{L}_{2}\mathbf{X}_{1}^{(k)}\mathbf{R}_{2}
\end{equation}
leading to the predicted change image
\begin{equation}	
	\label{eq:pseudoObs_dX_2}
	\Delta{\check{\mathbf{Y}}}_{2} = \mathbf{Y}_{2} - \check{\mathbf{Y}}_{2}.
\end{equation}
Then, the correction step in Algo. \ref{alg:AM} consists in solving
\begin{equation}
        \Delta\hat{\mathbf{X}}^{(k+1)} = \operatornamewithlimits{argmin}_{\Delta\mathbf{X}} \mathcal{J}_2\left(\Delta\mathbf{X}\right)	\triangleq \mathcal{J}\left(\mathbf{X}_{1}^{(k)},\Delta\mathbf{X}\right)
\end{equation}
			with
\begin{equation}
	\label{eq:objective_dX}
	\begin{aligned}
		\mathcal{J}_2\left(\Delta\mathbf{X}\right)= \left\|\mathbf{\Lambda}_{2}^{-\frac{1}{2}} \left(\Delta{\check{\mathbf{Y}}}_{2}^{(k)} - 	 \mathbf{L}_{2}\Delta\mathbf{X}\mathbf{R}_{2}\right)\right\|_{\mathrm{F}}^{2} +  \gamma \left\|\Delta\mathbf{X}\right\|_{2,1}.
	\end{aligned}
\end{equation}
This correction can be interpreted as a joint spatial and spectral deblurring of the predicted change image $\Delta{\check{\mathbf{Y}}}_{2}^{(k)}$. Note that this ill-posed inverse problem is regularized through an $\ell_{2,1}$-norm penalization, which promotes the spatial sparsity of the change image $\Delta\mathbf{X}$.

It is worth noting that the difficulty of conducting the two steps of the AM algorithm detailed above is highly related to the spatial and/or spectral degradations operated on the two latent images, according to applicative scenarios which are detailed in the next section. Interestingly, the following section will also show that these steps generally reduce to ubiquitous (multi-band) image processing tasks, namely denoising, spectral deblurring or spatial super-resolution from a single or multiple images, for which efficient and reliable strategies have been already proposed in the literature.

\section{Algorithmic implementations for applicative scenarios}
\label{implementation}			

The general model presented in \eqref{eq:jointobsmodel} and the AM algorithm proposed in Section \ref{alg:AM} can be implemented to handle all scenarios derived from two multi-band optical images. These scenarios differ by the corresponding spatial and spectral degradations relating the pair of observed images $\left\{\mathbf{Y}_1,\mathbf{Y}_2\right\}$ and the pair of latent images $\left\{\mathbf{X}_1,\mathbf{X}_2\right\}$. Table \ref{table:SEN_mat} summarizes the 10 distinct scenarios (denoted $\mathcal{S}_1$ to $\mathcal{S}_{10}$) according to the degradations operated on the two latent images $\mathbf{X}_1$ and $\mathbf{X}_2$. The specificities of these scenarios are also discussed in what follows.

\begin{description}
\item[$\mathcal{S}_1$] is devoted to a pair of observed images sharing the same spatial and spectral resolutions. In this case, CD can be conducted by pixel-wise comparisons, as classically addressed in the literature, e.g., in \cite{singhreview1989} and \cite{bovolotime2015}.
\item[$\mathcal{S}_2$] consists in conducting CD between two images with the same spatial resolution but different spectral resolution, considered in \cite{nielsenmultivariate1998} and \cite{nielsenregularized2007}.
\item[$\mathcal{S}_3$] consists in conducting CD between two images with the same spectral resolution but different spatial resolution.
\item[$\mathcal{S}_4$] relies on two complementary images: the first image with high spectral and low spatial resolutions, the second image with low spectral and high spatial resolutions. This is the CD scenario considered in \cite{FerrarisICASSP2017,ferraris_Fusion_2017,ferrarisrobust2017}. When the two observed images have been acquired at the same time instants ($t_i=t_j$), this scenario corresponds to the multi-band image fusion task considered in numerous works, e.g., in \cite{weibayesian2015-2}, \cite{Yokoya2012} and \cite{Simoes2014b}.
\item[$\mathcal{S}_5$] represents an even less favorable instance of $\mathcal{S}_2$ and $\mathcal{S}_3$ where one image is of high spatial and spectral resolutions while the other is of low spatial and spectral resolutions.
\item[$\mathcal{S}_6$] generalizes $\mathcal{S}_3$. As for $\mathcal{S}_3$, both observed images have the same spectral resolutions and different spatial resolutions. However, contrary to $\mathcal{S}_3$, the relative downsampling factor between images is non-integer, which precludes the use of a unique spatial degradation matrix $\mathbf{R} = \mathbf{B}\mathbf{S}$. As a consequence, the latent images $\mathbf{X}_1$ and $\mathbf{X}_2$ are characterized by a common spatial resolution which is higher than those of both observed images. The choice of this virtual downsampling factor is based on the greatest common divisor between spatial resolutions.
\item[$\mathcal{S}_7$] generalizes $\mathcal{S}_4$ with a non-integer relative downsampling factor (as for $\mathcal{S}_6$).
\item[$\mathcal{S}_8$] generalizes $\mathcal{S}_2$ where the two observed image share the same spatial resolution but have distinct spectral resolutions. However, contrary to $\mathcal{S}_2$, this difference in spectral resolutions cannot be expressed using a unique spectral degradation matrix. This may happen when the two spectral ranges of observed images contain non-overlapping bands.
\item[$\mathcal{S}_9$] generalizes $\mathcal{S}_4$, but the difference in spectral resolutions cannot be expressed using a single degradation matrix (as for $\mathcal{S}_8$).
\item[$\mathcal{S}_{10}$] generalizes $\mathcal{S}_4$, but  the difference in spatial resolutions cannot be expressed using a unique spatial degradation matrix (as for $\mathcal{S}_6$) and the difference in spectral resolutions cannot be expressed using a single spectral degradation matrix (as for $\mathcal{S}_8$).
%\item[$\mathcal{S}_8$ to $\mathcal{S}_{10}$] can be represented by the general proposed model. Nevertheless, they do not correspond to any real application. Indeed, they can be cast in one of the previous scenarios.
\end{description}

Finally, the following paragraphs instantiate the AM algorithm for each scenario. These specific instantiations will relate the fusion and correction steps with ubiquitous image processing tasks that can be performed efficiently thanks to recent contributions proposed in the image processing literature. Table \ref{table:SEN_alg} summarizes these implementations w.r.t. the discussed scenarios.

\newcommand{\proxOp}{$\ell_{2,1}$-prox. mapping}
\newcommand{\spatsuper}{Spatial super-resolution}
\newcommand{\leastsquares}{Least squares}
\newcommand{\forwardbackward}{Forward-backward}
\newcommand{\spectraldeblurring}{Spectral deblurring}
\newcommand{\imagefusion}{Multi-band image fusion}

\renewcommand{\arraystretch}{0.45}
\setlength{\tabcolsep}{10pt}

%\begin{table*}
\begin{sidewaystable}
    \centering
    \begin{tabular}{c|cc|c|cc|c|}
\cline{2-7}
 & \multicolumn{3}{c|}{\multirow{2}{*}{\textbf{Fusion Step}}} & \multicolumn{3}{c|}{\multirow{2}{*}{\textbf{Correction Step}}} \\ & \multicolumn{3}{c|}{} & \multicolumn{3}{c|}{} \\
\cline{2-7}
 & \multicolumn{2}{c|}{\multirow{2}{*}{\textbf{Algorithm}}} & \multirow{2}{*}{\textbf{Operation}}  & \multicolumn{2}{c|}{\multirow{2}{*}{\textbf{Algorithm}}} & \multirow{2}{*}{\textbf{Operation}}\\&&&&&&\\
\cline{2-7}
\hline
\hline
		\multicolumn{1}{|c|}{\multirow{2}{*}{\rotatebox{00}{$\mathcal{S}_1$}}} & \multicolumn{2}{c|}{\multirow{2}{*}{\leastsquares}} & \multirow{2}{*}{Denoising} & \multicolumn{2}{c|}{\multirow{2}{*}{\proxOp}}  & \multirow{2}{*}{Denoising} \\ \multicolumn{1}{|c|}{} & & & & & &  \\
\hline

\multicolumn{1}{|c|}{\multirow{2}{*}{\rotatebox{00}{$\mathcal{S}_2$}}} & \multicolumn{2}{c|}{\multirow{2}{*}{\leastsquares}} & \multirow{2}{*}{\spectraldeblurring} & \multicolumn{2}{c|}{\multirow{2}{*}{\proxOp}} & \multirow{2}{*}{Denoising} \\ \multicolumn{1}{|c|}{} & & & & & & \\
\hline

\multicolumn{1}{|c|}{\multirow{2}{*}{\rotatebox{00}{$\mathcal{S}_3$}}} & \multicolumn{2}{c|}{\multirow{2}{*}{\cite{zhaofast2016}}} & \multirow{2}{*}{\spatsuper} & \multicolumn{2}{c|}{\multirow{2}{*}{\proxOp}} & \multirow{2}{*}{Denoising} \\ \multicolumn{1}{|c|}{} & & & & & &  \\
\hline

\multicolumn{1}{|c|}{\multirow{2}{*}{\rotatebox{00}{$\mathcal{S}_4$}}} & \multicolumn{2}{c|}{\multirow{2}{*}{\cite{weifast2015-2}}} & \multirow{2}{*}{\imagefusion} & \multicolumn{2}{c|}{\multirow{2}{*}{\forwardbackward}} & \multirow{2}{*}{\spectraldeblurring} \\ \multicolumn{1}{|c|}{} & & & & & & \\
\hline

\multicolumn{1}{|c|}{\multirow{8}{*}{\rotatebox{00}{$\mathcal{S}_5$}}}  &\multicolumn{1}{|c}{\multirow{8}{*}{\rotatebox{90}{{ADMM}}}} &  \multirow{6}{*}{\leastsquares} &  \multirow{6}{*}{\spectraldeblurring} & \multicolumn{2}{c|}{\multirow{8}{*}{\proxOp}} & \multirow{8}{*}{Denoising} \\ \multicolumn{1}{|c|}{} &  &  \multirow{8}{*}{\cite{zhaofast2016}} & \multirow{8}{*}{\spatsuper} & & & \\
 \multicolumn{1}{|c|}{} & & & & & &  \\
 \multicolumn{1}{|c|}{} & & & & & & \\
 \multicolumn{1}{|c|}{} & & & & & & \\
 \multicolumn{1}{|c|}{} & & & & & & \\
 \multicolumn{1}{|c|}{} & & & & & & \\
 \multicolumn{1}{|c|}{} & & & & & & \\
\hline

\multicolumn{1}{|c|}{\multirow{8}{*}{\rotatebox{00}{$\mathcal{S}_6$}}}  &\multicolumn{1}{|c}{\multirow{8}{*}{\rotatebox{90}{{ADMM}}}} &  \multirow{6}{*}{\cite{zhaofast2016}} &  \multirow{6}{*}{\spatsuper}  & \multicolumn{1}{|c}{\multirow{8}{*}{\rotatebox{90}{{ADMM}}}} & \multirow{6}{*}{\cite{zhaofast2016}} & \multirow{6}{*}{\spatsuper} \\ \multicolumn{1}{|c|}{} &  &  \multirow{8}{*}{\cite{zhaofast2016}} & \multirow{8}{*}{\spatsuper} & & \multirow{8}{*}{\proxOp} & \multirow{8}{*}{Denoising}\\
 \multicolumn{1}{|c|}{} & & & & & &  \\
 \multicolumn{1}{|c|}{} & & & & & & \\
 \multicolumn{1}{|c|}{} & & & & & & \\
 \multicolumn{1}{|c|}{} & & & & & & \\
 \multicolumn{1}{|c|}{} & & & & & & \\
 \multicolumn{1}{|c|}{} & & & & & & \\
\hline

\multicolumn{1}{|c|}{\multirow{8}{*}{\rotatebox{00}{$\mathcal{S}_7$}}} &\multicolumn{1}{|c}{\multirow{8}{*}{\rotatebox{90}{{ADMM}}}} &  \multirow{6}{*}{\cite{weifast2015-2}} &  \multirow{6}{*}{\imagefusion} & \multicolumn{1}{|c}{\multirow{8}{*}{\rotatebox{90}{{ADMM}}}} & \multirow{6}{*}{\cite{zhaofast2016}} & \multirow{6}{*}{\spatsuper} \\ \multicolumn{1}{|c|}{} &  &  \multirow{8}{*}{\cite{zhaofast2016}} & \multirow{8}{*}{\spatsuper} & & \multirow{8}{*}{\proxOp} & \multirow{8}{*}{Denoising}\\
 \multicolumn{1}{|c|}{} & & & & & &  \\
 \multicolumn{1}{|c|}{} & & & & & & \\
 \multicolumn{1}{|c|}{} & & & & & & \\
 \multicolumn{1}{|c|}{} & & & & & & \\
 \multicolumn{1}{|c|}{} & & & & & & \\
 \multicolumn{1}{|c|}{} & & & & & & \\
\hline

\multicolumn{1}{|c|}{\multirow{2}{*}{\rotatebox{00}{$\mathcal{S}_8$}}}  & \multicolumn{2}{c|}{\multirow{2}{*}{\leastsquares}} & \multirow{2}{*}{\spectraldeblurring}  & \multicolumn{2}{c|}{\multirow{2}{*}{\forwardbackward}} & \multirow{2}{*}{\spectraldeblurring}\\ \multicolumn{1}{|c|}{} & & & & & & \\
\hline

\multicolumn{1}{|c|}{\multirow{8}{*}{\rotatebox{00}{$\mathcal{S}_9$}}}  &\multicolumn{1}{|c}{\multirow{8}{*}{\rotatebox{90}{{ADMM}}}} &  \multirow{6}{*}{\leastsquares} &  \multirow{6}{*}{\spectraldeblurring} & \multicolumn{2}{c|}{\multirow{8}{*}{\forwardbackward}}  & \multirow{8}{*}{\spectraldeblurring}\\ \multicolumn{1}{|c|}{}  &  &  \multirow{8}{*}{\cite{zhaofast2016}} & \multirow{8}{*}{\imagefusion} & & & \\
 \multicolumn{1}{|c|}{} & & & & & &  \\
 \multicolumn{1}{|c|}{} & & & & & & \\
 \multicolumn{1}{|c|}{} & & & & & & \\
 \multicolumn{1}{|c|}{} & & & & & & \\
 \multicolumn{1}{|c|}{} & & & & & & \\
 \multicolumn{1}{|c|}{} & & & & & & \\
\hline

\multicolumn{1}{|c|}{\multirow{8}{*}{\rotatebox{00}{$\mathcal{S}_{10}$}}}  &\multicolumn{1}{|c}{\multirow{8}{*}{\rotatebox{90}{{ADMM}}}} &  \multirow{4}{*}{\cite{weifast2015-2}} &  \multirow{4}{*}{\imagefusion}  & \multicolumn{1}{|c}{\multirow{8}{*}{\rotatebox{90}{{ADMM}}}} & \multirow{4}{*}{\proxOp} & \multirow{4}{*}{Denoising}\\ \multicolumn{1}{|c|}{}  &  &  \multirow{6}{*}{\cite{zhaofast2016}} & \multirow{6}{*}{\spatsuper}  & & \multirow{6}{*}{\cite{zhaofast2016}} & \multirow{6}{*}{\spatsuper}\\ \multicolumn{1}{|c|}{}  &  &  \multirow{8}{*}{\leastsquares} & \multirow{8}{*}{\spectraldeblurring}  & & \multirow{8}{*}{\leastsquares} & \multirow{8}{*}{\spectraldeblurring}\\
 \multicolumn{1}{|c|}{} & & & & & & \\
 \multicolumn{1}{|c|}{} & & & & & & \\
 \multicolumn{1}{|c|}{} & & & & & & \\
 \multicolumn{1}{|c|}{} & & & & & & \\
 \multicolumn{1}{|c|}{} & & & & & & \\
\hline
     \hline
    \end{tabular}
        \caption{Overview of the steps of the AM algorithm w.r.t. applicative scenarios.}
  \label{table:SEN_alg}
%\end{table*}
\end{sidewaystable}

	\subsection{Scenario \texorpdfstring{$\mathcal{S}_1$}{S1}}
	\label{subsec:sen1}

	Considering the degradation matrices specified in Table \ref{table:SEN_mat} for this scenario, the forward model \eqref{eq:jointobsmodel_bis} can be rewritten as
\begin{subequations}
	\label{eq:jointobsmodel_Sen1}
		\begin{align}
			&\mathbf{Y}_{1} = \mathbf{X}_{1} + \mathbf{N}_{1} \label{eq:jointobsmodel_Sen1_alpha}\\
			&\mathbf{Y}_{2} = \left(\mathbf{X}_{1}+\Delta\mathbf{X}\right) + \mathbf{N}_{2}            \label{eq:jointobsmodel_Sen1_beta}
		\end{align}
\end{subequations}
	As expected, for this scenario, the observed, latent and change images share the same spatial and spectral resolutions. The resulting objective function, initially in \eqref{eq:objective}, is simplified as
\begin{equation}
	\label{eq:objective_Sen1}
	\begin{aligned}
		 \mathcal{J}_{\mathcal{S}_1}\left(\mathbf{X}_{1},\Delta\mathbf{X}\right)&=\frac{1}{2}\left\|\mathbf{\Lambda}_{2}^{-\frac{1}{2}} \left(\mathbf{Y}_{2} - \left(\mathbf{X}_{1}+\Delta\mathbf{X}\right)\right) \right\|_{\mathrm{F}}^{2}  \\
		&+ \frac{1}{2}\left\|\mathbf{\Lambda}_{1}^{-\frac{1}{2}} \left(\mathbf{Y}_{1} - \mathbf{X}_{1} \right) \right\|_{\mathrm{F}}^{2} \\
        &+ \lambda \left\|\mathbf{X}_{1} - \bar{\mathbf{X}}_1\right\|_{\mathrm{F}}^2 + \gamma \left\|\Delta\mathbf{X}\right\|_{2,1}.
\end{aligned}
\end{equation}
	The two steps of the AM algorithm are detailed below.
	
	\subsubsection{Fusion: optimization w.r.t. \texorpdfstring{$\mathbf{X}_{1}$}{X1}}
	\label{subsubsec:est_sen1}

At the $k$th iteration of the AM algorithm, let assume that the current value of the change image is denoted by $\Delta\mathbf{X}^{(k)}$. As suggested in Section \ref{subsec:robust_fusion}, a corrected image $\tilde{\mathbf{Y}}_{2}^{(k)}$ that would be observed at time $t_1$ by the sensor $\textsf{S}_{2}$ given the image $\mathbf{Y}_{2}$ observed at time $t_2$ and the change image $\Delta\mathbf{X}^{(k)}$ can be introduced as
	\begin{equation}
		\label{eq:pseudoObs_Sen1_1}
		\tilde{\mathbf{Y}}_{2}^{(k)} = \mathbf{Y}_{2} - \Delta\mathbf{X}^{(k)}.
	\end{equation}
Updating the latent image $\mathbf{X}_{1}$ consists in minimizing w.r.t. $\mathbf{X}_{1}$ the partial function
	\begin{equation*}
		\label{eq:object_Sen1_1}
    \begin{aligned}
			\mathcal{J}_{\mathcal{S}_1,1}\left(\mathbf{X}_{1}\right) &\triangleq \mathcal{J}_{\mathcal{S}_1}\left(\mathbf{X}_{1},\Delta\mathbf{X}^{(k)}\right) \\
&=  \left\|\mathbf{\Lambda}_{1}^{-\frac{1}{2}} \left(\mathbf{Y}_{1} - \mathbf{X}_{1}\right)\right\|_{\mathrm{F}}^{2} \\
&+ \left\|\mathbf{\Lambda}_{2}^{-\frac{1}{2}} \left(\tilde{\mathbf{Y}}_{2}^{(k)} - \mathbf{X}_{1}\right)\right\|_{\mathrm{F}}^{2}
+ \lambda\left\|\mathbf{X}_{1} - \bar{\mathbf{X}}_1\right\|_{\mathrm{F}}^2.
    \end{aligned}
	\end{equation*}
This formulation shows that recovering $\mathbf{X}_{1}$ in Scenario $\mathcal{S}_1$ reduces to a denoising problem from an observed image $\mathbf{Y}_{1}$ and a pseudo-observed image $\tilde{\mathbf{Y}}_{2}^{(k)}$. A closed-form solution of this $\ell_2$-penalized least-square problem can be easily and efficiently computed.
%
%As noticed earlier, this sub-problem boils down to a close-related problem of multi-band image fusion with the particularity that both observed images share the same spectral and spatial resolution. Since both latent images have also the same resolution as both observed images, this problem can be cast as a denoising problem. The high dimension of the optimization problem \eqref{eq:object_Sen1_1}, consecutive to the size of all images, may produce some difficulties in order to compute its solution. Using the particular choice \eqref{eq:phi_1} of the regularization function $\phi_1(\cdot)$ adopted in this paper, a closed-form solution can still be derived and efficiently implemented either adopting a standard least square solution or using a modified version of the FuSE method proposed in \cite{weifast2015-2} which provides a closed-form solution for the more general problem of multi-band optical image fusion.

\subsubsection{Correction: optimization w.r.t. \texorpdfstring{$\Delta\mathbf{X}$}{DX}}
\label{subsubsec:corr_sen1}

Following the same strategy proposed in \cite{ferrarisrobust2017}, let $\check{\mathbf{Y}}_{2}^{(k)}$ denote the \emph{predicted} image that would be observed by the sensor $\textsf{S}_{2}$ at time $t_1$  given the current state of the latent image $\mathbf{X}_{1}^{(k)}$. Since the two sensors share the same spatial and spectral characteristics, one has
\begin{equation}
	\label{eq:pseudoObs_Sen1_2}
  \check{\mathbf{Y}}_{2}^{(k)} = \mathbf{X}_{1}^{(k)}.
\end{equation}
Similarly to \eqref{eq:assumption}, the \emph{predicted} change image can thus be defined as
\begin{equation}
	\label{eq:pseudoObs_Sen1_3}
		\Delta{\check{\mathbf{Y}}}_{2}^{(k)} = \mathbf{Y}_{2} - \check{\mathbf{Y}}_{2}^{(k)}.
\end{equation}
The objective function \eqref{eq:objective_Sen1} w.r.t. $\Delta\mathbf{X}$ is then rewritten by combining \eqref{eq:pseudoObs_Sen1_2} and \eqref{eq:pseudoObs_Sen1_3} with \eqref{eq:objective_Sen1}, leading to
\begin{equation}
	\label{eq:object_Sen1_2}
  \begin{aligned}
			\mathcal{J}_{\mathcal{S}_1,2}(\Delta\mathbf{X})&\triangleq \mathcal{J}_{\mathcal{S}_1}(\mathbf{X}_{1}^{(k)},\Delta\mathbf{X})   \\ &=\left\|\mathbf{\Lambda}_{2}^{-\frac{1}{2}} \left(\Delta{\check{\mathbf{Y}}}_{2}^{(k)} - \Delta\mathbf{X}\right)\right\|_{\mathrm{F}}^{2} +  \gamma \left\|\Delta\mathbf{X}\right\|_{2,1}.
  \end{aligned}
\end{equation}
Again, since the observed, latent and change images share the same spatial and spectral resolutions, this correction step reduces to a denoising task of the predicted change image $\Delta{\check{\mathbf{Y}}}_{2}^{(k)}$. With the particular CD-driven choice of $\phi_{2}\left(\cdot\right)$ in \eqref{eq:phi_2}, minimizing $\mathcal{J}_{\mathcal{S}_1,2}(\Delta\mathbf{X})$ is an $\ell_{2,1}$-penalized least square problem. Minimizing \eqref{eq:object_Sen1_2} also defines the proximal operator associated with the $\ell_{2,1}$-norm and can be directly achieved by applying a group-soft thresholding on the predicted change image $\Delta{\check{\mathbf{Y}}}_{2}^{(k)}$.
%Otherwise, in order to solve such convex optimization problem various algorithms have been proposed including forward-backward splitting \cite{Combettes2005,Combettes2011}, Douglas-Rachford splitting \cite{Combettes2007,Combettes2011} and alternating direction method of multipliers \cite{Boyd2010,Parikh2013}.
%
%Since the proximal operator related to both terms can be efficiently computed, in this work, we propose to resort to an iterative forward-backward algorithm which has shown to provide the fastest yet reliable results.

\subsection{Scenario \texorpdfstring{$\mathcal{S}_2$}{S2}}
	\label{subsec:sen2}
	
	In this scenario, the two observed images are of same spatial resolution (as for scenario $\mathcal{S}_1$) but with different optical spectral information, which preclude a simple comparison between pixels. For this scenario, the joint forward observation model derived from \eqref{eq:jointobsmodel_bis} can be written as
\begin{subequations}
	\label{eq:jointobsmodel_Sen2}
	\begin{align}
			&\mathbf{Y}_{1} =  \mathbf{L}_{1}\mathbf{X}_{1} + \mathbf{N}_{1} \label{eq:jointobsmodel_Sen2_alpha},\\
			&\mathbf{Y}_{2} = \left(\mathbf{X}_{1}+\Delta\mathbf{X}\right) + \mathbf{N}_{2} \label{eq:jointobsmodel_Sen2_beta},
	\end{align}
\end{subequations}
which results in the objective function
\begin{equation*}
	\label{eq:objective_Sen2}
	\begin{aligned}
  \mathcal{J}_{\mathcal{S}_2}\left(\mathbf{X}_{1},\Delta\mathbf{X}\right)&=\frac{1}{2}\left\|\mathbf{\Lambda}_{2}^{-\frac{1}{2}} \left(\mathbf{Y}_{2} - \left(\mathbf{X}_{1}+\Delta\mathbf{X}\right)\right) \right\|_{\mathrm{F}}^{2}  \\
		&+ \frac{1}{2}\left\|\mathbf{\Lambda}_{1}^{-\frac{1}{2}} \left(\mathbf{Y}_{1} - \mathbf{L}_{1}\mathbf{X}_{1} \right) \right\|_{\mathrm{F}}^{2} \\
        &+ \lambda \left\|\mathbf{X}_{1} - \bar{\mathbf{X}}_1\right\|_{\mathrm{F}}^2 + \gamma \left\|\Delta\mathbf{X}\right\|_{2,1}.
	\end{aligned}
\end{equation*}
%
%Note that the estimated images always keep, at least, the higher spatial and spectral resolution of the pair of observed images. Therefore, the choice of writing model \eqref{eq:jointobsmodel_Sen2} considering the spectral degradation in image acquired by sensor $\mathbf{P}_{1}$ the term, which does not take into account the change image \eqref{eq:jointobsmodel_Sen2_beta}, is encouraged so that the estimation of $\Delta\mathbf{X}$ becomes more easy and accurate avoiding additional debluring steps as in \cite{ferrarisrobust2017}.
%
Within an AM algorithmic schemes, the two sub-problems of interest are detailed below.

\subsubsection{Fusion: optimization w.r.t. \texorpdfstring{$\mathbf{X}_{1}$}{X1}}
		\label{subsubsec:est_sen2}

The same strategy as for scenario $\mathcal{S}_1$ in paragraph \ref{subsubsec:est_sen1} is adopted. As model \eqref{eq:jointobsmodel_Sen2_beta} is the the same as model \eqref{eq:jointobsmodel_Sen1_beta}, the corrected image $\tilde{\mathbf{Y}}_{2}^{(k)}$ is defined following \eqref{eq:pseudoObs_Sen1_1}. Then, updating the latent image $\mathbf{X}_{1}$ consists in minimizing the partial objective function
	\begin{equation}
		\label{eq:object_Sen2_1}
    \begin{aligned}
			\mathcal{J}_{\mathcal{S}_2,1}\left(\mathbf{X}_{1}\right) &\triangleq \mathcal{J}_{\mathcal{S}_2}\left(\mathbf{X}_{1},\Delta\mathbf{X}^{(k)}\right) \\
&=  \left\|\mathbf{\Lambda}_{1}^{-\frac{1}{2}} \left(\mathbf{Y}_{1} - \mathbf{L}_{1}\mathbf{X}_{1}\right)\right\|_{\mathrm{F}}^{2} \\
&+ \left\|\mathbf{\Lambda}_{2}^{-\frac{1}{2}} \left(\tilde{\mathbf{Y}}_{2}^{(k)} - \mathbf{X}_{1}\right)\right\|_{\mathrm{F}}^{2}
+ \lambda\left\|\mathbf{X}_{1} - \bar{\mathbf{X}}_1\right\|_{\mathrm{F}}^2.
    \end{aligned}
	\end{equation}
This problem can be interpreted as a spectral deblurring of the observed image $\mathbf{Y}_{1}$ where the corrected image $\tilde{\mathbf{Y}}_{2}^{(k)}$ plays the role of prior information. Minimizing \eqref{eq:object_Sen2_1} can be easily conducted by computing the standard least square solution.
\subsubsection{Correction: optimization w.r.t. \texorpdfstring{$\Delta\mathbf{X}$}{DX}}
		\label{subsubsec:corr_sen2}
As both models \eqref{eq:jointobsmodel_Sen2_beta} and \eqref{eq:jointobsmodel_Sen1_beta} are the same, optimizing w.r.t $\Delta\mathbf{X}$ can be conducted following the procedure detailed in paragraph \ref{subsubsec:corr_sen1} (i.e., denoising of the predicted change image).
\subsection{Scenario \texorpdfstring{$\mathcal{S}_3$}{S3}}
	\label{subsec:sen3}
In this scenario, the two observed images share the same spectral resolution but differ by their spatial resolutions. These spatial resolutions are related by an integer relative downsampling factor, which allows a unique spatial degradation matrix $\mathbf{R}_{1}$ to be used\footnote{The case of observed images with non-integer relative spatial downsampling factor is discussed in scenario $\mathcal{S}_6$.}. The joint forward observation model derived from \eqref{eq:jointobsmodel_bis} using the specific degradation matrices presented in Table \ref{table:SEN_mat} can be written as
\begin{subequations}
	\label{eq:jointobsmodel_Sen3}
	\begin{align}
			&\mathbf{Y}_{1} = \mathbf{X}_{1}\mathbf{R}_{1} + \mathbf{N}_{1} \label{eq:jointobsmodel_Sen3_alpha}.\\
			&\mathbf{Y}_{2} = \left(\mathbf{X}_{1}+\Delta\mathbf{X}\right) + \mathbf{N}_{2} \label{eq:jointobsmodel_Sen3_beta}.
	\end{align}
\end{subequations}
with the objective function
\begin{equation*}
	\label{eq:objective_Sen3}
	\begin{aligned}
  \mathcal{J}_{\mathcal{S}_3}\left(\mathbf{X}_{1},\Delta\mathbf{X}\right)&=\frac{1}{2}\left\|\mathbf{\Lambda}_{2}^{-\frac{1}{2}} \left(\mathbf{Y}_{2} - \left(\mathbf{X}_{1}+\Delta\mathbf{X}\right)\right) \right\|_{\mathrm{F}}^{2}  \\
		&+ \frac{1}{2}\left\|\mathbf{\Lambda}_{1}^{-\frac{1}{2}} \left(\mathbf{Y}_{1} - \mathbf{X}_{1}\mathbf{R}_{1} \right) \right\|_{\mathrm{F}}^{2} \\
        &+ \lambda \left\|\mathbf{X}_{1} - \bar{\mathbf{X}}_1\right\|_{\mathrm{F}}^2 + \gamma \left\|\Delta\mathbf{X}\right\|_{2,1}.
	\end{aligned}
\end{equation*}
%
%	Similarly as for \emph{Scenario} 2, the choice to considers the spatial degradation over the observed image acquired by sensor $\mathbf{P}_{1}$ makes easier the problem estimating $\Delta\mathbf{X}$. Also, it would provide more accurate results avoiding additional spatial debluring steps as in \cite{ferrarisrobust2017}. The solution for the problem in \eqref{eq:objective_Sen3} according to AM strategies are present in following.
	
\subsubsection{Fusion: optimization w.r.t. \texorpdfstring{$\mathbf{X}_{1}$}{X1}}
	\label{subsubsec:est_sen3}

The same strategy as for previous scenarios is adopted here. As model \eqref{eq:jointobsmodel_Sen3_beta} is the same as model \eqref{eq:jointobsmodel_Sen1_beta}, the corrected image $\tilde{\mathbf{Y}}_{2}^{(k)}$ is defined following \eqref{eq:pseudoObs_Sen1_1}. Then, updating the latent image consists in minimizing w.r.t. $\mathbf{X}_{1}$ the partial function
	\begin{equation*}
		\label{eq:object_Sen3_1}
    \begin{aligned}
			\mathcal{J}_{\mathcal{S}_3,1}\left(\mathbf{X}_{1}\right) &\triangleq \mathcal{J}_{\mathcal{S}_3}\left(\mathbf{X}_{1},\Delta\mathbf{X}^{(k)}\right) \\
&=  \left\|\mathbf{\Lambda}_{1}^{-\frac{1}{2}} \left(\mathbf{Y}_{1} - \mathbf{X}_{1}\mathbf{R}_{1}\right)\right\|_{\mathrm{F}}^{2} \\
&+ \left\|\mathbf{\Lambda}_{2}^{-\frac{1}{2}} \left(\tilde{\mathbf{Y}}_{2}^{(k)} - \mathbf{X}_{1}\right)\right\|_{\mathrm{F}}^{2}
+ \lambda\left\|\mathbf{X}_{1} - \bar{\mathbf{X}}_1\right\|_{\mathrm{F}}^2.
    \end{aligned}
	\end{equation*}
This fusion task can be interpreted as a set of $n_{\lambda}$ super-resolution problems associated with each band of the observed image $\mathbf{Y}_{1}$, where the corrected image $\tilde{\mathbf{Y}}_{2}^{(k)}$ acts here as a prior information. Closed-form expressions of these $n_{\lambda}$ solutions are given by  \cite{zhaofast2016}.

\subsubsection{Correction: optimization w.r.t. \texorpdfstring{$\Delta\mathbf{X}$}{DX}}
	\label{subsubsec:corr_sen3}
As the model \eqref{eq:jointobsmodel_Sen3_beta} is the same as model \eqref{eq:jointobsmodel_Sen1_beta} of scenarios $\mathcal{S}_1$ and $\mathcal{S}_2$, optimizing w.r.t. $\Delta\mathbf{X}$ can be conducted following the procedure detailed in paragraph \ref{subsubsec:corr_sen1} (i.e., denoising of the predicted change image).
\subsection{Scenario \texorpdfstring{$\mathcal{S}_4$}{S4}}
	\label{subsec:sen4}
Scenario $\mathcal{S}_4$ is specifically addressed in \cite{ferrarisrobust2017} with the joint forward model
\begin{equation*}
	\label{eq:jointobsmodel_Sen4}
	\begin{aligned}
			&\mathbf{Y}_{1} = \mathbf{X}_{1}\mathbf{R}_{1} + \mathbf{N}_{1},\\
			&\mathbf{Y}_{2} = \mathbf{L}_{2}\left(\mathbf{X}_{1}+\Delta\mathbf{X}\right) + \mathbf{N}_{2}.
	\end{aligned}
\end{equation*}
The two observed images have complementary information since $\mathbf{Y}_1$ and $\mathbf{Y}_2$ are of high spectral and spatial resolutions, respectively.  The resulting objective function writes
\begin{equation}
	\label{eq:objective_Sen4}
	\begin{aligned}
		 \mathcal{J}_{\mathcal{S}_4}\left(\mathbf{X}_{1},\Delta\mathbf{X}\right)&=\frac{1}{2}\left\|\mathbf{\Lambda}_{2}^{-\frac{1}{2}} \left(\mathbf{Y}_{2} - \mathbf{L}_{2} \left(\mathbf{X}_{1}+\Delta\mathbf{X}\right)\right) \right\|_{\mathrm{F}}^{2}  \\
		&+ \frac{1}{2}\left\|\mathbf{\Lambda}_{1}^{-\frac{1}{2}} \left(\mathbf{Y}_{1} - \mathbf{X}_{1}\mathbf{R}_{1} \right) \right\|_{\mathrm{F}}^{2} \\
        &+ \lambda \left\|\mathbf{X}_{1} - \bar{\mathbf{X}}_1\right\|_{\mathrm{F}}^2 + \gamma \left\|\Delta\mathbf{X}\right\|_{2,1}.
	\end{aligned}
\end{equation}
When these images have been acquired at the same time instant, the change image is $\Delta\mathbf{X}=\boldsymbol{0}$ and this configuration boils down to a multiband image fusion problem addressed in \cite{weifast2015-2}. Thus, minimizing \eqref{eq:objective_Sen4} can be conducted following the AM strategy by combining a multiband image fusion step \citep{weifast2015-2} and a spectral deblurring step of the predicted change image. The interested reader is invited to consult the work in \cite{ferrarisrobust2017} for a comprehensive description of the resolution.
\subsection{Scenario \texorpdfstring{$\mathcal{S}_5$}{S5}}
	\label{subsec:sen5}
Under this scenario, the observed image $\mathbf{Y}_2$ is of higher spatial and spectral resolutions than the observed image $\mathbf{Y}_1$. Within a conventional fusion context, one would probably discard $\mathbf{Y}_1$ since it would not bring additional information to the one provided by $\mathbf{Y}_2$. Conversely, within a CD context, both observed images are of interest and can be exploited. More precisely, here, the joint forward observation model derived from \eqref{eq:jointobsmodel_bis} is specifically written
\begin{subequations}
	\label{eq:jointobsmodel_Sen5}
		\begin{align}
			&\mathbf{Y}_{1} = \mathbf{L}_{1} \mathbf{X}_{1}\mathbf{R}_{1} + \mathbf{N}_{1} \label{eq:jointobsmodel_Sen5_alpha},\\
			&\mathbf{Y}_{2} = \left(\mathbf{X}_{1}+\Delta\mathbf{X}\right) + \mathbf{N}_{2} \label{eq:jointobsmodel_Sen5_beta},
		\end{align}
\end{subequations}
with the resulting objective function
\begin{equation*}
	\label{eq:objective_Sen5}
	\begin{aligned}
  \mathcal{J}_{\mathcal{S}_5}\left(\mathbf{X}_{1},\Delta\mathbf{X}\right)&=\frac{1}{2}\left\|\mathbf{\Lambda}_{2}^{-\frac{1}{2}} \left(\mathbf{Y}_{2} - \left(\mathbf{X}_{1}+\Delta\mathbf{X}\right)\right) \right\|_{\mathrm{F}}^{2}  \\
		&+ \frac{1}{2}\left\|\mathbf{\Lambda}_{1}^{-\frac{1}{2}} \left(\mathbf{Y}_{1} - \mathbf{L}_{1}\mathbf{X}_{1}\mathbf{R}_{1} \right) \right\|_{\mathrm{F}}^{2} \\
        &+ \lambda \left\|\mathbf{X}_{1} - \bar{\mathbf{X}}_1\right\|_{\mathrm{F}}^2 + \gamma \left\|\Delta\mathbf{X}\right\|_{2,1}.
	\end{aligned}
\end{equation*}
Its minimization relies on the two steps detailed below.
\subsubsection{Fusion: optimization w.r.t. \texorpdfstring{$\mathbf{X}_{1}$}{X1}}
	\label{subsubsec:est_sen5}
The same strategy as for previous scenarios is adopted here. After defining the corrected image $\tilde{\mathbf{Y}}_{2}^{(k)}$ by \eqref{eq:pseudoObs_Sen1_1}, updating the the latent image $\mathbf{X}_{1}$ consists in minimizing
\begin{equation}
	\label{eq:object_Sen5_1}
  \begin{aligned}
			\mathcal{J}_{\mathcal{S}_5,1}\left(\mathbf{X}_{1}\right) &\triangleq \mathcal{J}_{\mathcal{S}_5}\left(\mathbf{X}_{1},\Delta\mathbf{X}^{(k)}\right) \\
&=  \left\|\mathbf{\Lambda}_{1}^{-\frac{1}{2}} \left(\mathbf{Y}_{1} - \mathbf{L}_{1}\mathbf{X}_{1}\mathbf{R}_{1}\right)\right\|_{\mathrm{F}}^{2} \\
&+ \left\|\mathbf{\Lambda}_{2}^{-\frac{1}{2}} \left(\tilde{\mathbf{Y}}_{2}^{(k)} - \mathbf{X}_{1}\right)\right\|_{\mathrm{F}}^{2}
+ \lambda\left\|\mathbf{X}_{1} - \bar{\mathbf{X}}_1\right\|_{\mathrm{F}}^2.
  \end{aligned}
\end{equation}
Minimizing \eqref{eq:object_Sen5_1} can be interpreted as a simultaneous spatial super-resolution and spectral deblurring of the multiband image $\mathbf{Y}_{1}$, with prior information brought by $\tilde{\mathbf{Y}}_{2}^{(k)}$. This minimization is a much more challenging task than the fusion steps encountered for scenarios $\mathcal{S}_1$--$\mathcal{S}_4$. Indeed, the simultaneous spatial and spectral degradations applied to $\mathbf{X}_{1}$ prevents a closed-form solution to be efficiently computed. Thus, one proposes to resort to an iterative algorithm, namely the alternating direction method of multipliers (ADMM). It consists in introducing the splitting variable $\mathbf{U} \in \mathbb{R}^{m_{\lambda_{1}}\times n} = \mathbf{L}_{1}\mathbf{X}_{1}$. The resulting scaled augmented Lagrangian for the problem is expressed as
%
%\begin{equation*}
	%\mathbf{U} = \mathbf{L}_{1}\mathbf{X}_{1}.
	%\label{eq:ADMM_Sen5_split_variable}
%\end{equation*}
%
%
\begin{equation}
	\label{eq:ADMM_Sen5}
	\begin{aligned}
		&\mathcal{L}_{\mu}(\mathbf{X}_{1},\mathbf{U},\mathbf{V}) = \left\|\mathbf{\Lambda}_{1}^{-\frac{1}{2}} \left(\mathbf{Y}_{1} - \mathbf{U}\mathbf{R}_{1}\right)\right\|_{\mathrm{F}}^{2} + \\& \left\|\mathbf{\Lambda}_{2}^{-\frac{1}{2}} \left(\tilde{\mathbf{Y}}_{2}^{(k)} - \mathbf{X}_{1}\right)\right\|_{\mathrm{F}}^{2}  + \\& \lambda\left\|\mathbf{X}_{1} - \bar{\mathbf{X}}_1\right\|_{\mathrm{F}}^2 + \frac{\mu}{2}\left\|\mathbf{L}_{1}\mathbf{X}_{1} -  \mathbf{U} + \mathbf{V} \right\|_{\mathrm{F}}^{2}.
	\end{aligned}
\end{equation}
The ADMM iteratively minimizes $\mathcal{L}_{\mu}$ w.r.t. $\mathbf{U}$ and $\mathbf{X}_{1}$ and updates the dual variable $\mathbf{V}$. By comparing the partial objective function \eqref{eq:object_Sen5_1} and its augmented counterpart \eqref{eq:ADMM_Sen5}, it clearly appears that the splitting strategy allows the spectral and spatial degradations to be decoupled. Thus, each of these steps can be easily conducted. More precisely, optimizing w.r.t.  $\mathbf{U}$ consists in conducting a super-resolution step achieved as for scenario $\mathcal{S}_3$ by resorting to the algorithm proposed in \cite{zhaofast2016}. Conversely, optimizing w.r.t. $\mathbf{X}_{1}$ consists in solving a least-square problem whose closed-form solution can be computed (akin to scenario $\mathcal{S}_2$).
\subsubsection{Correction: optimization w.r.t. \texorpdfstring{$\Delta\mathbf{X}$}{DX}}
		\label{subsubsec:corr_sen5}
Again, as the forward model \eqref{eq:jointobsmodel_Sen5_beta} is the same as \eqref{eq:jointobsmodel_Sen1_beta} of Scenario $\mathcal{S}_1$, optimizing w.r.t. $\Delta\mathbf{X}$ can be conducted following the procedure detailed in paragraph \ref{subsubsec:corr_sen1} (i.e., denoising of the predicted change image).
\subsection{Scenario \texorpdfstring{$\mathcal{S}_6$}{S6}}
\label{subsec:sen6}
As for scenario $\mathcal{S}_3$, scenario $\mathcal{S}_6$ considers two observed images of same spectral resolutions but with distinct spatial resolutions. However, contrary to scenario $\mathcal{S}_3$, this difference in spatial resolutions cannot be expressed thanks to a unique spatial degradation matrix $\mathbf{R}_1$ due to a non-integer relative downsampling factor. Thus the forward model is written
\begin{subequations}
	\label{eq:jointobsmodel_Sen6}
	\begin{align}
			&\mathbf{Y}_{1} = \mathbf{X}_{1}\mathbf{R}_{1} + \mathbf{N}_{1} \label{eq:jointobsmodel_Sen6_alpha}.\\
			&\mathbf{Y}_{2} = \left(\mathbf{X}_{1}+\Delta\mathbf{X}\right)\mathbf{R}_{2} + \mathbf{N}_{2} \label{eq:jointobsmodel_Sen6_beta}.
	\end{align}
\end{subequations}
with the following objective function
\begin{equation}
	\label{eq:objective_Sen6}
	\begin{aligned}
  \mathcal{J}_{\mathcal{S}_6}\left(\mathbf{X}_{1},\Delta\mathbf{X}\right)&=\frac{1}{2}\left\|\mathbf{\Lambda}_{2}^{-\frac{1}{2}} \left(\mathbf{Y}_{2} - \left(\mathbf{X}_{1}+\Delta\mathbf{X}\right)\mathbf{R}_{2}\right) \right\|_{\mathrm{F}}^{2}  \\
		&+ \frac{1}{2}\left\|\mathbf{\Lambda}_{1}^{-\frac{1}{2}} \left(\mathbf{Y}_{1} - \mathbf{X}_{1}\mathbf{R}_{1} \right) \right\|_{\mathrm{F}}^{2} \\
        &+ \lambda \left\|\mathbf{X}_{1} - \bar{\mathbf{X}}_1\right\|_{\mathrm{F}}^2 + \gamma \left\|\Delta\mathbf{X}\right\|_{2,1}.
	\end{aligned}
\end{equation}
In \eqref{eq:jointobsmodel_Sen6}, both latent images are supposed to suffer from spatial degradations. Thus, choosing which spatial degradation affects the change image $\Delta\mathbf{X}$ results in a particular spatial resolution for this change map. To derive a CD map at a high spatial resolution, the spatial degradation applied to $\Delta\mathbf{X}$ should be chosen as the one with the lowest virtual downsampling factor. The minimization of \eqref{eq:objective_Sen6} according to the AM strategy is addressed in the following paragraphs.
\subsubsection{Fusion: optimization w.r.t. \texorpdfstring{$\mathbf{X}_{1}$}{X1}}
	\label{subsubsec:est_sen6}
For this scenario, the corrected image in \eqref{eq:pseudoObs} is defined as
\begin{equation*}
	\label{eq:pseudoObs_Sen6_1}
		\tilde{\mathbf{Y}}_{2}^{(k)} = \mathbf{Y}_{2} - \Delta\mathbf{X}^{(k)}\mathbf{R}_{2}.
\end{equation*}
Then, updating the latent image $\mathbf{X}_{1}$ consists in minimizing w.r.t. $\mathbf{X}_{1}$ the partial function
\begin{equation}
	\label{eq:object_Sen6_1}
    \begin{aligned}
			\mathcal{J}_{\mathcal{S}_6,1}\left(\mathbf{X}_{1}\right) &\triangleq \mathcal{J}_{\mathcal{S}_6}\left(\mathbf{X}_{1},\Delta\mathbf{X}^{(k)}\right) \\
&=  \left\|\mathbf{\Lambda}_{1}^{-\frac{1}{2}} \left(\mathbf{Y}_{1} - \mathbf{X}_{1}\mathbf{R}_{1}\right)\right\|_{\mathrm{F}}^{2} \\
&+ \left\|\mathbf{\Lambda}_{2}^{-\frac{1}{2}} \left(\tilde{\mathbf{Y}}_{2}^{(k)} - \mathbf{X}_{1}\mathbf{R}_{2}\right)\right\|_{\mathrm{F}}^{2}
+ \lambda\left\|\mathbf{X}_{1} - \bar{\mathbf{X}}_1\right\|_{\mathrm{F}}^2.
    \end{aligned}
\end{equation}
As for scenario $\mathcal{S}_3$, minimizing \eqref{eq:object_Sen6_1} can be interpreted as recovering a spatially super-resolved image $\mathbf{X}_{1}$ from the observed image $\mathbf{Y}_{1}$ and the corrected image $\tilde{\mathbf{Y}}_{2}^{(k)}$. However, contrary to scenario $\mathcal{S}_3$, here, $\tilde{\mathbf{Y}}_{2}^{(k)}$ rather defines an additional data-fitting term instead of a prior information \citep{Elad1997}. Moreover, this sub-problem cannot be solved directly since no closed-form solution can be efficiently derived, mainly due to the simultaneous presence of the two spatial degradation operators. Thus, as for scenario $\mathcal{S}_5$, one resorts to the ADMM scheme by introducing the splitting variable $\mathbf{U} \in \mathbb{R}^{n_{\lambda}\times n} = \mathbf{X}_{1}$. The resulting scaled augmented Lagrangian can be written as
%\begin{equation*}
	%\mathbf{U} = \mathbf{X}_{1}.
	%\label{eq:ADMM_Sen6_split_variable}
%\end{equation*}
%
\begin{equation}
	\label{eq:ADMM_Sen6}
	\begin{split}
		\mathcal{L}_{\mu}(\mathbf{X}_{1},\mathbf{U},\mathbf{V}) &= \left\|\mathbf{\Lambda}_{1}^{-\frac{1}{2}} \left(\mathbf{Y}_{1} - \mathbf{U}\mathbf{R}_{1}\right)\right\|_{\mathrm{F}}^{2} \\
&+\left\|\mathbf{\Lambda}_{2}^{-\frac{1}{2}} \left(\tilde{\mathbf{Y}}_{2}^{(k)} - \mathbf{X}_{1}\mathbf{R}_{2}\right)\right\|_{\mathrm{F}}^{2} + \lambda\left\|\mathbf{X}_{1} - \bar{\mathbf{X}}_1\right\|_{\mathrm{F}}^2\\
& + \frac{\mu}{2}\left\|\mathbf{X}_{1} -  \mathbf{U} + \mathbf{V} \right\|_{\mathrm{F}}^{2}.
	\end{split}
\end{equation}
Both minimizations of \eqref{eq:ADMM_Sen6} w.r.t. $\mathbf{U}$ and $\mathbf{X}_{1}$ can be conducted band-by-band following the strategy proposed in \cite{zhaofast2016}, which provides closed-form solutions of the underlying single-image super-resolution problems and also ensures the convergence of the AM algorithm.

\subsubsection{Correction: optimization w.r.t. \texorpdfstring{$\Delta\mathbf{X}$}{DX}}
	\label{subsubsec:corr_sen6}

For this scenario, a predicted image that would be observed by the sensor $\textsf{S}_{2}$ at time $t_1$ can be defined as
\begin{equation}
	\label{eq:pseudoObs_Sen6_2}
  \check{\mathbf{Y}}_{2}^{(k)} = \mathbf{X}_{1}^{(k)}\mathbf{R}_{2}
\end{equation}
with the resulting predicted change image
\begin{equation}	
	\label{eq:pseudoObs_Sen6_3}
	\Delta{\check{\mathbf{Y}}}_{2}^{(k)} = \mathbf{Y}_{2} - \check{\mathbf{Y}}_{2}^{(k)}.
\end{equation}
The objective function \eqref{eq:objective} w.r.t. $\Delta\mathbf{X}$ is then rewritten by combining \eqref{eq:pseudoObs_Sen6_2} and \eqref{eq:pseudoObs_Sen6_3} with \eqref{eq:objective}, leading to
\begin{equation}
	\label{eq:object_Sen6_2}
  \begin{aligned}
  	\mathcal{J}_{\mathcal{S}_6,2}(\Delta\mathbf{X})&\triangleq \mathcal{J}_{\mathcal{S}_6}(\mathbf{X}_{1}^{(k)},\Delta\mathbf{X}) \\
  &=  \left\|\mathbf{\Lambda}_{2}^{-\frac{1}{2}} \left(\Delta{\check{\mathbf{Y}}}_{2}^{(k)} - \Delta\mathbf{X}\mathbf{R}_{2}\right)\right\|_{\mathrm{F}}^{2} +  \gamma \left\|\Delta\mathbf{X}\right\|_{2,1}.
   \end{aligned}
\end{equation}
The minimization of \eqref{eq:object_Sen6_2} can be interpreted as a super-resolution problem. Even if a forward-backward algorithm could be used to iteratively minimize this objective function, the size of the spatial degradation matrix $\mathbf{R}_{2}$ suggests to resort to an ADMM. By introducing the splitting variable $\mathbf{W} \in \mathbb{R}^{n_{\lambda} \times m_{2}} = \Delta\mathbf{X}\mathbf{R}_{2}$, the resulting scaled augmented Lagrangian for the problem is expressed as
%\begin{equation*}
	%\mathbf{W} = \Delta\mathbf{X}\mathbf{R}_{2}.\\
	%\label{eq:ADMM_Sen6_split_variable_cor}
%\end{equation*}
%
\begin{equation}
		\label{eq:ADMM_Sen6_cor}
		\begin{aligned}
		\mathcal{L}_{\mu}(\Delta\mathbf{X},\mathbf{W},\mathbf{V}) = &\left\|\mathbf{\Lambda}_{2}^{-\frac{1}{2}} \left(\Delta{\check{\mathbf{Y}}}_{2}^{(k)} - \mathbf{W}\right)\right\|_{\mathrm{F}}^{2} + \lambda\left\|\Delta\mathbf{X}\right\|_{\mathrm{2,1}} \\& + \frac{\mu}{2}\left\|\Delta\mathbf{X}\mathbf{R}_{1} -  \mathbf{W} + \mathbf{V}\right\|_{\mathrm{F}}^{2}.
			\end{aligned}
\end{equation}
Closed-form expressions of the minimizers of \eqref{eq:ADMM_Sen6_cor} w.r.t. $\Delta\mathbf{X}$ and $\mathbf{W}$ can be derived, following a group soft-thresholding operation and the technique proposed in \cite{zhaofast2016}, respectively.

\subsection{Scenario \texorpdfstring{$\mathcal{S}_7$}{S7}}
	\label{subsec:sen7}

Scenario $\mathcal{S}_7$ generalizes scenario $\mathcal{S}_4$ with the specific case of a non-integer relative spatial downsampling factor, which precludes the use of a unique spatial degradation matrix. The resulting joint observation model is
\begin{subequations}
	\label{eq:jointobsmodel_Sen7}
	\begin{align}
			&\mathbf{Y}_{1} = \mathbf{L}_{1}\mathbf{X}_{1}\mathbf{R}_{1} + \mathbf{N}_{1} \label{eq:jointobsmodel_Sen7_alpha}.\\
			&\mathbf{Y}_{2} = \left(\mathbf{X}_{1}+\Delta\mathbf{X}\right)\mathbf{R}_{2} + \mathbf{N}_{2} \label{eq:jointobsmodel_Sen7_beta}
	\end{align}
\end{subequations}
which leads to the following objective function
\begin{equation*}
	\label{eq:objective_Sen7}
	\begin{aligned}
  \mathcal{J}_{\mathcal{S}_7}\left(\mathbf{X}_{1},\Delta\mathbf{X}\right)&=\frac{1}{2}\left\|\mathbf{\Lambda}_{2}^{-\frac{1}{2}} \left(\mathbf{Y}_{2} - \left(\mathbf{X}_{1}+\Delta\mathbf{X}\right)\mathbf{R}_{2} \right)\right\|_{\mathrm{F}}^{2}  \\
		&+ \frac{1}{2}\left\|\mathbf{\Lambda}_{1}^{-\frac{1}{2}}\left(\mathbf{Y}_{1} -  \mathbf{L}_{1}\mathbf{X}_{1}\mathbf{R}_{1} \right) \right\|_{\mathrm{F}}^{2} \\
        &+ \lambda \left\|\mathbf{X}_{1} - \bar{\mathbf{X}}_1\right\|_{\mathrm{F}}^2 + \gamma \left\|\Delta\mathbf{X}\right\|_{2,1}.
	\end{aligned}
\end{equation*}
The choice of assuming that the image acquired by the sensor $\textsf{S}_{2}$ does not suffers from spectral degradation is motivated by an easier and more accurate  estimation of the change image $\Delta\mathbf{X}$ by avoiding additional spectral deblurring steps. The two sub-problems underlying the AM algorithm are detailed below.

\subsubsection{Fusion: optimization w.r.t. \texorpdfstring{$\mathbf{X}_{1}$}{X1}}
	\label{subsubsec:est_sen7}

By defining the corrected image as for Scenario $\mathcal{S}_6$, i.e.,
\begin{equation*}
		\tilde{\mathbf{Y}}_{2}^{(k)} = \mathbf{Y}_{2} - \Delta\mathbf{X}^{(k)}\mathbf{R}_{2},
\end{equation*}
updating the latent image $\mathbf{X}_{1}$  consists in minimizing  the partial function
\begin{equation}
	\label{eq:object_Sen7_1}
   \begin{aligned}
			\mathcal{J}_{\mathcal{S}_{7},1}\left(\mathbf{X}_{1}\right) &\triangleq \mathcal{J}_{\mathcal{S}_7}\left(\mathbf{X}_{1},\Delta\mathbf{X}^{(k)}\right) \\
&=  \left\|\mathbf{\Lambda}_{1}^{-\frac{1}{2}} \left(\mathbf{Y}_{1} - \mathbf{L}_{1}\mathbf{X}_{1}\mathbf{R}_{1}\right)\right\|_{\mathrm{F}}^{2} \\
&+ \left\|\mathbf{\Lambda}_{2}^{-\frac{1}{2}} \left(\tilde{\mathbf{Y}}_{2}^{(k)} - \mathbf{X}_{1}\mathbf{R}_{2}\right)\right\|_{\mathrm{F}}^{2}
+ \lambda\left\|\mathbf{X}_{1} - \bar{\mathbf{X}}_1\right\|_{\mathrm{F}}^2.
    \end{aligned}
\end{equation}
Unfortunately, it is not possible to derive a closed-form solution of the minimizer \eqref{eq:object_Sen7_1}. As for Scenarios $\mathcal{S}_5$ and $\mathcal{S}_6$, capitalizing on the convexity of the objective function, an ADMM strategy is followed. By defining the splitting variable  $\mathbf{U} \in \mathbb{R}^{m_{\lambda_{1}}\times n} = \mathbf{L}_{1}\mathbf{X}_{1}$. The scaled augmented Lagrangian can be written
%
%\begin{equation}
 %\mathbf{U}= \mathbf{L}_{1}\mathbf{X}_{1}
%\label{eq:ADMM_Sen7_split_variable}
%\end{equation}
%
\begin{equation}
	\label{eq:ADMM_Sen7}
	\begin{split}
		\mathcal{L}_{\mu}(\mathbf{X}_{1},\mathbf{U},\mathbf{V}) &= \left\|\mathbf{\Lambda}_{1}^{-\frac{1}{2}} \left(\mathbf{Y}_{1} - \mathbf{U}\mathbf{R}_{1}\right)\right\|_{\mathrm{F}}^{2} \\
&+\left\|\mathbf{\Lambda}_{2}^{-\frac{1}{2}} \left(\tilde{\mathbf{Y}}_{2}^{(k)} - \mathbf{X}_{1}\mathbf{R}_{2}\right)\right\|_{\mathrm{F}}^{2} + \lambda\left\|\mathbf{X}_{1} - \bar{\mathbf{X}}_1\right\|_{\mathrm{F}}^2 \\
&+ \frac{\mu}{2}\left\|\mathbf{L}_{1}\mathbf{X}_{1} -   \mathbf{U} + \mathbf{V} \right\|_{\mathrm{F}}^{2}.
	\end{split}
\end{equation}
Iterative minimizations of \eqref{eq:ADMM_Sen7} w.r.t. both $\mathbf{U}$ and $\mathbf{X}_{1}$ can be conducted efficiently. More precisely, optimizing w.r.t. $\mathbf{U}$ consists in solving a set of super-resolution problems whose closed-form solutions are given band-by-band in \cite{zhaofast2016}. Regarding the minimization w.r.t. $\mathbf{X}_{1}$, it consists in solving a $\ell_2$-penalized super-resolution problem, whose closed-form solution is given in \cite{weifast2015-2}.

\subsubsection{Correction: optimization w.r.t. \texorpdfstring{$\Delta\mathbf{X}$}{DX}}
	\label{subsubsec:corr_sen7}

Since the observation model \eqref{eq:jointobsmodel_Sen7_beta} related to $\Delta\mathbf{X}$ is the same as the one of Scenario $\mathcal{S}_6$ (see \eqref{eq:jointobsmodel_Sen6_beta}), optimizing w.r.t. $\Delta\mathbf{X}$ can be achieved thanks to ADMM, as described in paragraph \ref{subsubsec:corr_sen6} (spatial super-resolution of the predicted change image).
	
\subsection{Scenario \texorpdfstring{$\mathcal{S}_8$}{S8}}
		\label{subsec:sen8}

This scenario is similar to the Scenario $\mathcal{S}_2$ described in paragraph \ref{subsec:sen2}. It relies on two images of same spatial resolution but of distinct spectral resolution. However, contrary to Scenario $\mathcal{S}_2$, this difference in spectral resolutions cannot be expressed with a unique spectral degradation matrix, e.g., due to respective spectral ranges with non-overlapping bands. In this case the joint forward observation model is
\begin{subequations}
	\label{eq:jointobsmodel_Sen8}
	\begin{align}
			&\mathbf{Y}_{1} =  \mathbf{L}_{1}\mathbf{X}_{1} + \mathbf{N}_{1} \label{eq:jointobsmodel_Sen8_alpha}.\\
			&\mathbf{Y}_{2} = \mathbf{L}_{2}\left(\mathbf{X}_{1}+\Delta\mathbf{X}\right) + \mathbf{N}_{2} \label{eq:jointobsmodel_Sen8_beta}.
		\end{align}
\end{subequations}
with the resulting objective function
\begin{equation*}
	\label{eq:objective_Sen8}
	\begin{aligned}
  \mathcal{J}_{\mathcal{S}_8}\left(\mathbf{X}_{1},\Delta\mathbf{X}\right)&=\frac{1}{2}\left\|\mathbf{\Lambda}_{2}^{-\frac{1}{2}} \left(\mathbf{Y}_{2} - \mathbf{L}_{2}\left(\mathbf{X}_{1}+\Delta\mathbf{X}\right)\right) \right\|_{\mathrm{F}}^{2}  \\
		&+ \frac{1}{2}\left\|\mathbf{\Lambda}_{1}^{-\frac{1}{2}} \left(\mathbf{Y}_{1} - \mathbf{L}_{1}\mathbf{X}_{1} \right) \right\|_{\mathrm{F}}^{2} \\
        &+ \lambda \left\|\mathbf{X}_{1} - \bar{\mathbf{X}}_1\right\|_{\mathrm{F}}^2 + \gamma \left\|\Delta\mathbf{X}\right\|_{2,1}.
	\end{aligned}
\end{equation*}
The choice of which degradation matrices applies to the change image $\Delta\mathbf{X}$ is driven by considering the matrix with larger number of bands, which results in a change image of higher spectral resolution. The associated sub-problems are described in what follows.

\subsubsection{Fusion: optimization w.r.t. \texorpdfstring{$\mathbf{X}_{1}$}{X1}}
		\label{subsubsec:est_sen8}

Similarly to Scenario $\mathcal{S}_4$, by defining the corrected image as $\tilde{\mathbf{Y}}_{2}^{(k)}= \mathbf{Y}_{2} - \mathbf{L}_{2}\Delta\mathbf{X}^{(t)}$, updating the latent image $\mathbf{X}_{1}$ consists in minimizing
	\begin{equation}
		\label{eq:object_Sen8_1}
    \begin{aligned}
			\mathcal{J}_{\mathcal{S}_8,1}\left(\mathbf{X}_{1}\right) &\triangleq \mathcal{J}_{\mathcal{S}_8}\left(\mathbf{X}_{1},\Delta\mathbf{X}^{(k)}\right) \\
&=  \left\|\mathbf{\Lambda}_{1}^{-\frac{1}{2}} \left(\mathbf{Y}_{1} - \mathbf{L}_{1}\mathbf{X}_{1}\right)\right\|_{\mathrm{F}}^{2} \\
&+ \left\|\mathbf{\Lambda}_{2}^{-\frac{1}{2}} \left(\tilde{\mathbf{Y}}_{2}^{(k)} - \mathbf{L}_{2}\mathbf{X}_{1}\right)\right\|_{\mathrm{F}}^{2}
+ \lambda\left\|\mathbf{X}_{1} - \bar{\mathbf{X}}_1\right\|_{\mathrm{F}}^2.
    \end{aligned}
	\end{equation}
Minimizing \eqref{eq:object_Sen8_1} formulates a joint spectral deblurring problem from an observed image $\mathbf{Y}_{1}$ and a pseudo-observed image $\tilde{\mathbf{Y}}_{2}^{(k)}$. Thanks to its quadratic form, this least-square problem can be easily solved.

\subsubsection{Correction: optimization w.r.t. \texorpdfstring{$\Delta\mathbf{X}$}{DX}}
	\label{subsubsec:corr_sen8}

The predicted image that would be observed by sensor $\textsf{S}_2$ at time $t_1$ can be defined as
\begin{equation*}
	\label{eq:pseudoObs_Sen8_2}
  \check{\mathbf{Y}}_{2}^{(k)} = \mathbf{L}_{2}\mathbf{X}_{1}^{(k)}
\end{equation*}
with the resulting predicted change image
\begin{equation*}	
	\label{eq:pseudoObs_Sen8_3}
	\Delta{\check{\mathbf{Y}}}_{2}^{(k)} = \mathbf{Y}_{2} - \check{\mathbf{Y}}_{2}^{(k)}.
\end{equation*}
The objective function \eqref{eq:objective} w.r.t. $\Delta\mathbf{X}$ is then rewritten by combining \eqref{eq:pseudoObs_Sen6_2} and \eqref{eq:pseudoObs_Sen6_3} with \eqref{eq:objective}, leading to
\begin{equation}
	\label{eq:object_Sen8_2}
  \begin{aligned}
  	\mathcal{J}_{\mathcal{S}_8,2}(\Delta\mathbf{X})&\triangleq \mathcal{J}_{\mathcal{S}_8}(\mathbf{X}_{1}^{(k)},\Delta\mathbf{X}) \\
  &=  \left\|\mathbf{\Lambda}_{2}^{-\frac{1}{2}} \left(\Delta{\check{\mathbf{Y}}}_{2}^{(k)} - \mathbf{L}_{2}\Delta\mathbf{X}\right)\right\|_{\mathrm{F}}^{2} +  \gamma \left\|\Delta\mathbf{X}\right\|_{2,1}.
   \end{aligned}
\end{equation}
As for scenario $\mathcal{S}_4$, minimizing \eqref{eq:object_Sen8_2} is a spectral deblurring of the  predicted change image $\Delta{\check{\mathbf{Y}}}_{2}^{(k)}$, which can be achieved using a forward-backward algorithm as proposed in \cite{ferrarisrobust2017}.

\subsection{Scenario \texorpdfstring{$\mathcal{S}_9$}{S9}}
	\label{subsec:sen9}

This scenario generalizes scenario $\mathcal{S}_4$, but with relative spectral responses involving non-overlapping bands. The joint forward observation model is then
\begin{subequations}
\label{eq:jointobsmodel_Sen9}
		\begin{align}
			&\mathbf{Y}_{1} =  \mathbf{L}_{1}\mathbf{X}_{1}\mathbf{R}_{1} + \mathbf{N}_{1} \label{eq:jointobsmodel_Sen9_alpha}.\\
			&\mathbf{Y}_{2} = \mathbf{L}_{2}\left(\mathbf{X}_{1}+\Delta\mathbf{X}\right) + \mathbf{N}_{2} \label{eq:jointobsmodel_Sen9_beta}.
		\end{align}
\end{subequations}
which yields the objective function
\begin{equation*}
	\label{eq:objective_Sen9}
	\begin{aligned}
  \mathcal{J}_{\mathcal{S}_9}\left(\mathbf{X}_{1},\Delta\mathbf{X}\right)&=\frac{1}{2}\left\|\mathbf{\Lambda}_{2}^{-\frac{1}{2}} \left(\mathbf{Y}_{2} - \mathbf{L}_{2}\left(\mathbf{X}_{1}+\Delta\mathbf{X}\right)\right) \right\|_{\mathrm{F}}^{2}  \\
		&+ \frac{1}{2}\left\|\mathbf{\Lambda}_{1}^{-\frac{1}{2}} \left(\mathbf{Y}_{1} - \mathbf{L}_{1}\mathbf{X}_{1}\mathbf{R}_{1} \right) \right\|_{\mathrm{F}}^{2} \\
        &+ \lambda \left\|\mathbf{X}_{1} - \bar{\mathbf{X}}_1\right\|_{\mathrm{F}}^2 + \gamma \left\|\Delta\mathbf{X}\right\|_{2,1}.
	\end{aligned}
\end{equation*}
Note that the estimated latent and change images are defined at the highest spatial resolution while benefiting from the spectral resolutions of both observed images. The choice of assuming that the image acquired by sensor $\textsf{S}_{2}$ does not suffer from spatial degradation has been motivated by an easier and accurate estimation of the change image $\Delta\mathbf{X}$ by avoiding additional spatial super-resolution steps. The resulting sub-problems involved in the  AM algorithm are detailed below.

\subsubsection{Fusion: optimization w.r.t. \texorpdfstring{$\mathbf{X}_{1}$}{X1}}
	\label{subsubsec:est_sen9}

As for scenarios $\mathcal{S}_4$ and $\mathcal{S}_8$, the corrected image $\tilde{\mathbf{Y}}_{2}^{(k)}$ can be defined as $\tilde{\mathbf{Y}}_{2}^{(k)}= \mathbf{Y}_{2} - \mathbf{L}_{2}\Delta\mathbf{X}^{(k)}$. Thus, updating the latent image $\mathbf{X}_{1}$ consists in minimizing
	\begin{equation}
		\label{eq:object_Sen9_1}
    \begin{aligned}
			\mathcal{J}_{\mathcal{S}_9,1}\left(\mathbf{X}_{1}\right) &\triangleq \mathcal{J}_{\mathcal{S}_9}\left(\mathbf{X}_{1},\Delta\mathbf{X}^{(k)}\right) \\
&=  \left\|\mathbf{\Lambda}_{1}^{-\frac{1}{2}} \left(\mathbf{Y}_{1} - \mathbf{L}_{1}\mathbf{X}_{1}\mathbf{R}_{1}\right)\right\|_{\mathrm{F}}^{2} \\
&+ \left\|\mathbf{\Lambda}_{2}^{-\frac{1}{2}} \left(\tilde{\mathbf{Y}}_{2}^{(k)} - \mathbf{L}_{2}\mathbf{X}_{1}\right)\right\|_{\mathrm{F}}^{2}
+ \lambda\left\|\mathbf{X}_{1} - \bar{\mathbf{X}}_1\right\|_{\mathrm{F}}^2.
    \end{aligned}
	\end{equation}
Minimizing \eqref{eq:object_Sen9_1} is challenging mainly due to the simultaneous presence of spatial and spectral degradation matrices $\mathbf{R}_{1}$ and $\mathbf{L}_{2}$ with an additional spatial degradation $\mathbf{L}_{1}$. Therefore, there is no closed-form solution for this problem, which can be eventually solved thanks to ADMM. By introducing the splitting variable $\mathbf{U} \in \mathbb{R}^{m_{\lambda}\times m_{1}} = \mathbf{X}_{1}\mathbf{R}_{1}$. The resulting scaled augmented Lagrangian is
%\begin{equation}
%\mathbf{U} = \mathbf{X}_{1}\mathbf{R}_{1}
%\label{eq:ADMM_Sen9_split_variable}
%\end{equation}
%
\begin{equation}
	\label{eq:ADMM_Sen9}
	\begin{split}
		\mathcal{L}_{\mu}(\mathbf{X}_{1},\mathbf{U},\mathbf{V}) &= \left\|\mathbf{\Lambda}_{1}^{-\frac{1}{2}} \left(\mathbf{Y}_{1} - \mathbf{L}_{1}\mathbf{U}\right)\right\|_{\mathrm{F}}^{2} \\
&+\left\|\mathbf{\Lambda}_{2}^{-\frac{1}{2}} \left(\tilde{\mathbf{Y}}_{2}^{(k)} - \mathbf{L}_{2}\mathbf{X}_{1}\right)\right\|_{\mathrm{F}}^{2} + \lambda\left\|\mathbf{X}_{1} - \bar{\mathbf{X}}_1\right\|_{\mathrm{F}}^2
\\ & + \frac{\mu}{2}\left\|\mathbf{X}_{1}\mathbf{R}_{1} -  \mathbf{U}  + \mathbf{V} \right\|_{\mathrm{F}}^{2}.
	\end{split}
\end{equation}
Closed-form expression of the minimizers of \eqref{eq:ADMM_Sen9} w.r.t. $\mathbf{X}_{1}$ and $\mathbf{U}$ can be derived, following \citep{weifast2015-2} and a least-square formulation, respectively.

\subsubsection{Correction: optimization w.r.t. \texorpdfstring{$\Delta\mathbf{X}$}{DX}}
	\label{subsubsec:corr_sen9}

As both models \eqref{eq:jointobsmodel_Sen9_beta} and \eqref{eq:jointobsmodel_Sen8_beta} are the same, optimizing w.r.t. $\Delta\mathbf{X}$ can be achieved following the strategy detailed in paragraph \ref{subsubsec:corr_sen8}, i.e., by spectrally deblurring a predicted change image $\Delta{\check{\mathbf{Y}}}_{2}^{(k)}$ thanks to the forward-backward algorithm proposed in \cite{ferrarisrobust2017}.

\subsection{Scenario \texorpdfstring{$\mathcal{S}_{10}$}{S10}}
		\label{subsec:sen10}

This scenario generalizes all the previous scenario with the particular difficulties of non-overlapping
bands in the spectral responses and non-integer relative spatial downsampling factor of the respective spatial degradations. The joint forward observation model is given by \eqref{eq:jointobsmodel_bis}, which results in the objective function $\mathcal{J}_{\mathcal{S}_{10}}$ in \eqref{eq:objective}. Again, as for scenarios $\mathcal{S}_7$ and $\mathcal{S}_9$, the choice of the spatial and spectral degradations applied to the change image $\Delta\mathbf{X}$ should be motivated by reaching the highest spatial and spectral resolutions of this change image. The optimization sub-problems are finally discussed below.
%
%\begin{equation*}
	%\label{eq:objective_Sen10}
	%\begin{aligned}
  %\mathcal{J}_{\mathcal{S}_{10}}\left(\mathbf{X}_{1},\Delta\mathbf{X}\right)&=\frac{1}{2}\left\|\mathbf{\Lambda}_{2}^{-\frac{1}{2}} \left(\mathbf{Y}_{2} - \mathbf{L}_{2}\left(\mathbf{X}_{1}+\Delta\mathbf{X}\right)\mathbf{R}_{2}\right) \right\|_{\mathrm{F}}^{2}  \\
		%&+ \frac{1}{2}\left\|\mathbf{\Lambda}_{1}^{-\frac{1}{2}} \left(\mathbf{Y}_{1} - \mathbf{L}_{1}\mathbf{X}_{1}\mathbf{R}_{1} \right) \right\|_{\mathrm{F}}^{2} \\
        %&+ \lambda \left\|\mathbf{X}_{1}\right\|_{\mathrm{F}}^2 + \gamma \left\|\Delta\mathbf{X}\right\|_{2,1}.
	%\end{aligned}
%\end{equation*}
%
%
\subsubsection{Fusion: optimization w.r.t. \texorpdfstring{$\mathbf{X}_{1}$}{X1}}
	\label{subsubsec:est_sen10}

For this scenario, the corrected image $\tilde{\mathbf{Y}}_{2}^{(k)}$ is given by \eqref{eq:pseudoObs}, leading to an updating rule of the $\mathbf{X}_{1}$ consists in minimizing \eqref{eq:objective_x}. This minimization cannot be conducted in a straightforward manner, since it requires to conduct a spectral deblurring and a spatial super-resolution simultaneously. However, the optimal solution can be reached by resorting to a ADMM with two splitting variables $\mathbf{U}_{1} = \mathbf{L}_{1}\mathbf{X}_{1}\in \mathbb{R}^{m_{\lambda_{1}} \times n} $ and $\mathbf{U}_{2} = \mathbf{X}_{1}\mathbf{R}_{2} \in \mathbb{R}^{n_{\lambda}\times m_{2}} $. The resulting scaled augmented Lagrangian for the problem is expressed as
	%\begin{equation*}
		%\label{eq:object_Sen10_1}
    %\begin{aligned}
			%\mathcal{J}_{\mathcal{S}_{10},1}\left(\mathbf{X}_{1}\right) &\triangleq \mathcal{J}_{\mathcal{S}_{10}}\left(\mathbf{X}_{1},\Delta\mathbf{X}_{k}\right) \\
%&=  \left\|\mathbf{\Lambda}_{1}^{-\frac{1}{2}} \left(\mathbf{Y}_{1} - \mathbf{L}_{1}\mathbf{X}_{1}\mathbf{R}_{1}\right)\right\|_{\mathrm{F}}^{2} \\
%&+ \left\|\mathbf{\Lambda}_{2}^{-\frac{1}{2}} \left(\tilde{\mathbf{Y}}_{2}^{(k)} - \mathbf{L}_{2}\mathbf{X}_{1}\mathbf{R}_{2}\right)\right\|_{\mathrm{F}}^{2}
%+ \lambda\left\|\mathbf{X}_{1}\right\|_{\mathrm{F}}^2.
    %\end{aligned}
	%\end{equation*}
%
%\begin{equation}
%\mathbf{U}_{1} = \mathbf{L}_{1}\mathbf{X}_{1} , \mathbf{U}_{2} = \mathbf{X}_{1}\mathbf{R}_{2}.\\
%\label{eq:ADMM_Sen10_split_variable}
%\end{equation}
%
\begin{equation}
	\label{eq:ADMM_Sen10}
	\begin{aligned}
		 &\mathcal{L}_{\mu}(\mathbf{X}_{1},\mathbf{U}_{1},\mathbf{U}_{2},\mathbf{V}_{1},\mathbf{V}_{2}) = \left\|\mathbf{\Lambda}_{1}^{-\frac{1}{2}} \left(\mathbf{Y}_{1} - \mathbf{U}_{1}\mathbf{R}_{1}\right)\right\|_{\mathrm{F}}^{2} \\
		& + \left\|\mathbf{\Lambda}_{2}^{-\frac{1}{2}} \left(\tilde{\mathbf{Y}}_{2}^{(k)} - \mathbf{L}_{2}\mathbf{U}_{2}\right)\right\|_{\mathrm{F}}^{2} + \lambda\left\|\mathbf{X}_{1} - \bar{\mathbf{X}}_1\right\|_{\mathrm{F}}^2 \\&+ \frac{\mu}{2}\left\|\mathbf{L}_{1}\mathbf{X}_{1} -  \mathbf{U}_{1} + \mathbf{V}_{1}\right\|_{\mathrm{F}}^{2} + \frac{\mu}{2}\left\|\mathbf{X}_{1}\mathbf{R}_{2} -  \mathbf{U}_{2} + \mathbf{V}_{2} \right\|_{\mathrm{F}}^{2}.
	\end{aligned}
\end{equation}
Closed-form expressions of the minimizers of \eqref{eq:ADMM_Sen10} w.r.t. $\mathbf{X}_{1}$, $\mathbf{U}_{1}$ and $\mathbf{U}_{2}$ can be derived as proposed in \cite{weifast2015-2}, \cite{zhaofast2016} and following a least-square formulation, respectively.
\subsubsection{Correction: optimization w.r.t. \texorpdfstring{$\Delta\mathbf{X}$}{DX}}
		\label{subsubsec:corr_sen10}
For this scenario, given the current state $\mathbf{X}_{1}^{(k)}$ of the latent image, the predicted image that would be observed by the sensor $\textsf{S}_{2}$ at time $t_1$ can be defined as in \eqref{eq:pseudoObs_dX} leading to the predicted change image \eqref{eq:pseudoObs_dX_2}. Then, the correction step consists in minimizing the objective function $\mathcal{J}_{\mathcal{S}_{10},1}(\Delta\mathbf{X})$ in \eqref{eq:objective_dX}.
%
%\begin{equation}
	%\label{eq:pseudoObs_Sen10_2}
  %\check{\mathbf{Y}}_{2}^{(k)} = \mathbf{L}_{2}\mathbf{X}_{1}^{(k)}\mathbf{R}_{2}
%\end{equation}
%
%\begin{equation}	
%\label{eq:pseudoObs_Sen10_3}
	%\Delta{\check{\mathbf{Y}}}_{2}^{(k)} = \mathbf{Y}_{2} - \check{\mathbf{Y}}_{2}^{(k)}.
%\end{equation}
%
%\begin{equation}
%\label{eq:object_Sen10_2}
  %\begin{aligned}
  	%\mathcal{J}_{\mathcal{S}_{10},1}(\Delta\mathbf{X})&\triangleq \mathcal{J}_{\mathcal{S}_{10},1}(\mathbf{X}_{1}^{(k)},\Delta\mathbf{X}) \\
  %&=  \left\|\mathbf{\Lambda}_{2}^{-\frac{1}{2}} \left(\Delta{\check{\mathbf{Y}}}_{2}^{(k)} - \mathbf{L}_{2}\Delta\mathbf{X}\mathbf{R}_{2}\right)\right\|_{\mathrm{F}}^{2} +  \gamma \left\|\Delta\mathbf{X}\right\|_{2,1}.
   %\end{aligned}
%\end{equation}
%
It consists in conducting a spectral deblurring and spatial super-resolution jointly. This problem has no closed-form solution. Therefore, the objective function is iteratively minimized using an ADMM with two splitting variables $\mathbf{W}_{1} \in \mathbb{R}^{m_{\lambda_{1}} \times n} = \mathbf{L}_{1}\Delta\mathbf{X}$ and $\mathbf{W}_{2} \in \mathbb{R}^{n_{\lambda}\times n} = \Delta\mathbf{X}$. The resulting scaled augmented Lagrangian for the problem is expressed as
%\begin{equation}
%\mathbf{W}_{1} = \mathbf{L}_{1}\Delta\mathbf{X} , \mathbf{W}_{2} = \Delta\mathbf{X}.\\
%\label{eq:ADMM_Sen10_split_variable_cor}
%\end{equation}
%
\begin{equation}
		\label{eq:ADMM_Sen10_cor}
		\begin{aligned}
		 &\mathcal{L}_{\mu}(\Delta\mathbf{X},\mathbf{W}_{1},\mathbf{W}_{2},\mathbf{V}_{1},\mathbf{V}_{2}) = \\
		& \left\|\mathbf{\Lambda}_{2}^{-\frac{1}{2}} \left(\Delta{\check{\mathbf{Y}}}_{2}^{(k)} - \mathbf{W}_{1}\mathbf{R}_{2}\right)\right\|_{\mathrm{F}}^{2} + \gamma\left\|\mathbf{W}_{2}\right\|_{\mathrm{2,1}}	 \\&+ \frac{\mu}{2}\left\|\mathbf{L}_{1}\Delta\mathbf{X} -  \mathbf{W}_{1} + \mathbf{V}_{1}\right\|_{\mathrm{F}}^{2} + \frac{\mu}{2}\left\|\Delta\mathbf{X} -  \mathbf{W}_{2} + \mathbf{V}_{2} \right\|_{\mathrm{F}}^{2}.
			\end{aligned}
	\end{equation}
Closed-form expression of the minimizers of \eqref{eq:ADMM_Sen10_cor} w.r.t. $\Delta\mathbf{X}$, $\mathbf{W}_{1}$ and $\mathbf{W}_{2}$ can be derived, following a least-square formulation, the computation proposed in \cite{zhaofast2016} and a group soft-thresholding, respectively.
\section{Experiments}
\label{sec:experiments}

\subsection{Reference images}
\label{subsec:reference_images}
To illustrate the performance of the proposed algorithmic framework using real multi-band optical data on each specific scenario discussed in paragraph \ref{implementation}, observed images from $4$ largely studied open access multi-band sensors have been chosen, namely Landsat-8 from \cite{landsat8}, Sentinel-2 from \cite{sentinel2}, Earth observing-1 Advanced Land Imager (EO-1 ALI) \citep{ali1} and Airborne Visible Infrared Imaging Spectrometer (AVIRIS) from \cite{aviris}. These images have been acquired over the same geographical location, i.e., the Mud Lake region in Lake Tahoe, CA, USA between June 8th, 2011 and October 29th, 2016. Unfortunately, no ground truth information is available for the chosen image pairs, as experienced in numerous experimental situations \citep{bovoloframework2012}. However, this region is characterized by interesting natural meteorological changes, e.g., drought of the Mud Lake, snow falls and vegetation growth, occurring along the seasons which help to visually infer the major changes between two dates and to assess the relevance of the detected changes. All considered images have been manually geographically and geometrically aligned to fulfill the requirements imposed by the considered CD setup.

\begin{figure}[h!]
			\centering	
			\includegraphics[width=\columnwidth]{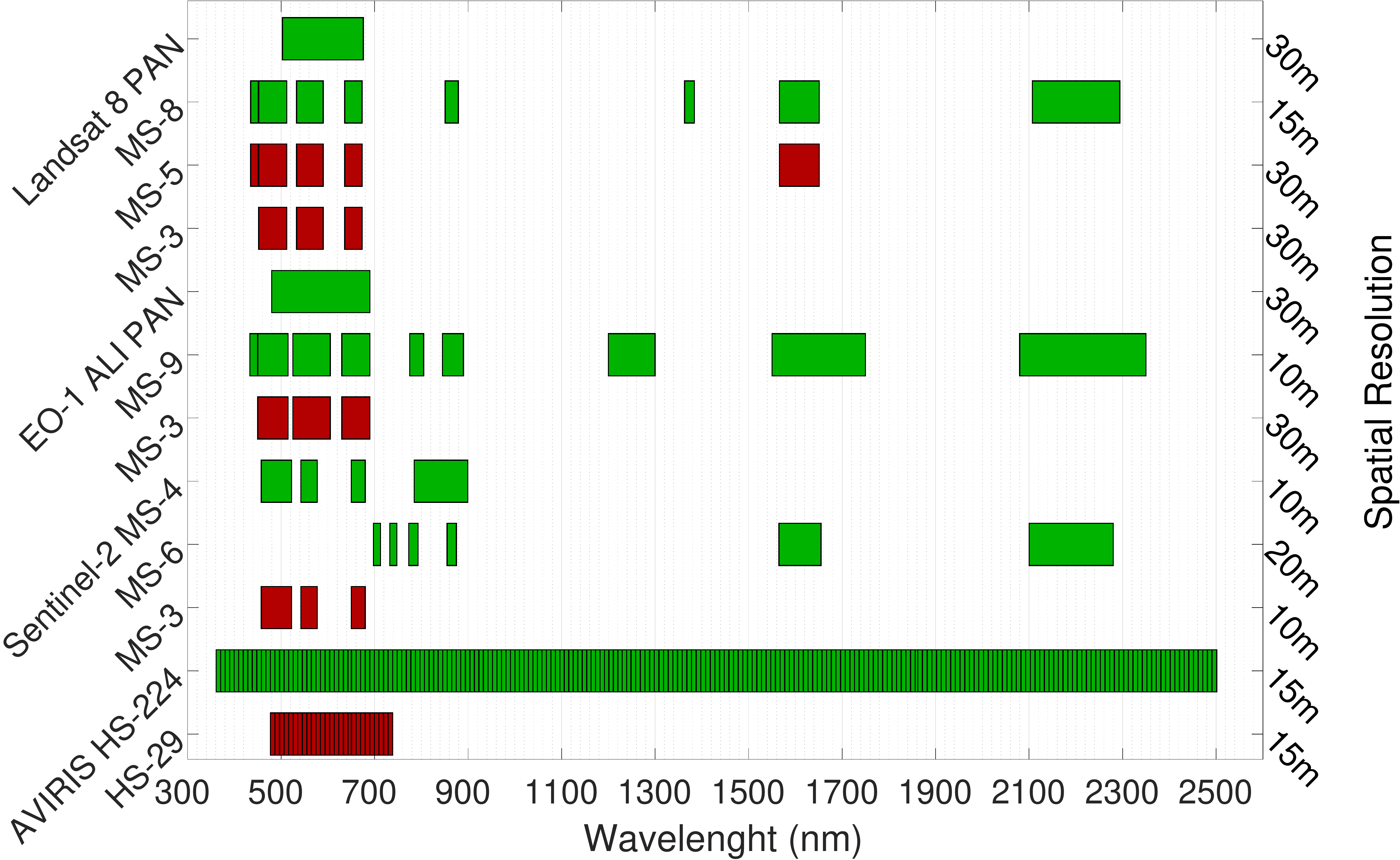}
	\caption{Spectral and spatial characteristics of real (green) and virtual (red) sensors.}%
	\label{fig:respSensors}%
\end{figure}

In addition to the data provided by these sensors, complementary images have been synthetically generated by considering so-called \emph{virtual} sensors derived from the real ones. The specifications of these virtual sensors, summarized in Figure \ref{fig:respSensors}, are chosen such that all applicative scenarios previously discussed can be diversely represented. They are met by selecting a subset of the initial spectral bands or by artificially degrading the spatial resolution of the real sensors.

\subsubsection{Landsat-8 images}
\label{subsubsec:landsat}
%Landsat-8 is the eighth Earth observation satellite series of the US LANDSAT Program \citep{landsat8}, launched on February 11th, 2013 with 16-days revisiting period. It is equipped with the Operational Land Imager (OLI) and the Thermal InfraRed Sensor (TIRS). The satellite sensor ensemble disposes of $11$ spectral bands with three different spatial resolutions, one wide spectrum (0.503$\mu$m - 0.676$\mu$m) 15m spatial resolution panchromatic band and 8 multi-spectral narrow bands 30m spatial resolution, Bands 1 to 7 and 9. Particularly, Bands 2 to 4 represent the visible R (0.452$\mu$m - 0.512$\mu$m), G (0.533$\mu$m - 0.590$\mu$m), B (0.636$\mu$m - 0.673$\mu$m) spectrum also covered by the panchromatic image represented by Band 8. For this sensor, the dataset contain 3 acquisition dates 18/10/2013, 15/04/2015 and 22/09/2015. The spectral responses used for testing are: 15m PAN (Band 8), 30m MS (Bands 1 to 7 and 9), 30m MS-3 (Bands 2 to 4) and 30m MS-5 (Bands 1 to 4 and 7).

Landsat-8 is the eighth Earth observation satellite series of the US LANDSAT Program \citep{landsat8}, launched on February 11th, 2013 with 16-days revisiting period. It is equipped with the Operational Land Imager (OLI) and the Thermal InfraRed Sensor (TIRS). In the conducted experiments, $3$ sets of real images acquired at the dates 10/18/2013, 04/15/2015 and 09/22/2015 have been considered. For each acquisition, Landsat-8 provides
\begin{itemize}
  \item one panchromatic image over the spectral range $0.503$--$0.676\mu$m (band $\sharp8$) at a $15$m spatial resolution (denoted PAN),
  \item one multispectral image of $8$ spectral bands (bands $\sharp1$--$\sharp7$ and $\sharp9$) at a $30$m resolution (denoted MS-$8$).
\end{itemize}
For experimental purpose, as explained above, these real images are complemented with the following virtually acquired images
\begin{itemize}
  \item one multispectral image of $5$ spectral bands (bands $\sharp1$--$\sharp4$ and $\sharp7$) at a $30$m spatial resolution (denoted MS-$5$),
  \item one  red-green-blue (RGB) multispectral image of $3$ spectral bands (bands $\sharp2$--$\sharp4$) at a $30$m spatial resolution (denoted MS-$3$).
\end{itemize}

\subsubsection{Sentinel-2 images}
\label{subsubsec:Sentinel}
Sentinel-2 is a series of two identical satellites for Earth observation missions developed by ESA \citep{sentinel2} as part of the Copernicus Program launched in 2015 and 2017 with 5-days revisiting period. The multi-spectral instrument embedded on each platform is composed of two different sensors for acquisition in the visible and infrared spectral domains, respectively. The actual dataset used in the experiments is composed of two images acquired on 04/12/2016 and 10/29/2016 and, for each real scene, among all available spectral bands, one considers
\begin{itemize}
  \item one multispectral image of $4$ visible/near infrared (VNIR) spectral bands (bands $\sharp2$--$\sharp4$ and $\sharp8$) at a $10$m spatial resolution (denoted MS-4)
  \item one multispectral image of $6$ short wave infrared spectral range (SWIR) spectral bands (bands $\sharp5$--$\sharp8$a and $\sharp11$--$\sharp12$) at a $20$m spatial resolution (denotes MS-$6$)
\end{itemize}
and one additional virtually image, namely,
\begin{itemize}
  \item one RGB multispectral image of $3$ spectral bands (bands $\sharp2$--$\sharp4$) at a $10$m spatial resolution (denoted MS-$3$).
\end{itemize}

%Presenting 13 spectral channels whose 4 bands in visible/near infrared (VNIR) named Bands 2 to 4 and 8, 6 bands in short wave infrared spectral range (SWR) named Bands 5 to 8a and 11 to 12, and 3 bands design for atmospherical corrections namely Bands 1 and 9 to 10. They have respectively 10m, 20m and 60m spatial resolution. Among all bands, Bands 2 to 4 represent the visible RGB, with respective central frequencies 0.490$\mu$m, 0.560$\mu$m and 0.665$\mu$m. This dataset contains 2 acquisition dates, 12/04/2016 and 29/10/2016. Also, the spectral responses used for testing are: 10m MS (Bands 2 to 4 and 8), 10m MS-$3$ (Bands 2 to 4) and 20m MS-$6$ (5 to 8a and 11 to 12).

\subsubsection{EO-1 ALI images}
\label{subsubsec:ALI}
Operated by NASA, EO-1 ALI is a Earth observation satellite part of the New Millennium Program launched in 2000 with 16-days repeat cycle and decommissioned in 2017 \citep{ali1}. The main embedded sensor Advanced Land Imager (ALI) is complemented with the Hyperion spectrometer and the Linear Etalon Imaging Spectrometer Array (LEISA) for atmospheric correction. The considered dataset corresponds to 2 acquisition dates, 06/08/2011 and 08/04/2011, for
\begin{itemize}
  \item one panchromatic image over the spectral range $0.48$--$0.69\mu$m (band $\sharp1$) at a $10$m spatial resolution (denoted PAN),
  \item one multispectral image of $9$ spectral bands (bands $\sharp2$--$\sharp10$) at a $30$m resolution (denoted MS-$9$),
\end{itemize}
in addition to the virtual acquisition of
\begin{itemize}
  \item one RGB multispectral image of $3$ spectral bands (bands $\sharp3$--$\sharp5$) at a $30$m spatial resolution (denoted MS-$3$).
\end{itemize}

%Operated by NASA, EO-1 ALI is a Earth observation satellite part of  New Millennium Program launched in 2000 with 16-days repeat cycle and decommissioned in 2017 \citep{ali1}. It's main mission accounts for validate new technologies used in the scientific instruments. That is why the satellite is embedded with three optical sensors: the main sensor Advanced Land Imager (ALI), the Hyperion spectrometer and the Linear Etalon Imaging Spectrometer Array (LEISA) for atmospheric correction. The ALI sensor measures 10 different spectral channels whose one wide spectrum (0.48 - 0.69 $\mu$m) 10m spatial resolution panchromatic (Band 1) and nine multi-spectral 30m spatial resolution narrow spectum bands (Bands 2 to 10). Similarly as LANDSAT 7 mission the visible spectrum present in Bands 3 to 5 accounts for R (0.45$\mu$m - 0.515$\mu$m), G (0.525$\mu$m - 0.605$\mu$m) and B (0.63$\mu$m - 0.69$\mu$m). This dataset is composed of 2 acquisition dates, 08/06/2011 and 04/08/2011, and by 3 spectral responses: 10m PAN (Band 1), 30m MS (Bands 2 to 10) and 30m MS-$3$ (Bands 3 to 5).

\subsubsection{AVIRIS images}
\label{subsubsec:AVIRIS}
AVIRIS is the second aircraft embedding an image spectrometer developed by Jet Propulsion Laboratory (JPL) for Earth remote sensing \cite{aviris}. It delivers calibrated images in $224$ contiguous $10$nm-width spectral channels ranging from $0.4\mu$m to $2.5\mu$m. Since it is an airborne-dependent system, the spatial resolution is not a priori fixed and is designed for each individual acquisition. The dataset considered in the conducted experiments is composed by two real images acquired on 04/10/2014 and 09/19/2014. For each scene, one considers
\begin{itemize}
  \item the original hyperspectral image of $224$ spectral bands at a $15$m spatial resolution (denoted HS-$224$)
  \item one virtual hyperspectral image of $29$ spectral bands (corresponding to the RGB domain) at a $15$m spatial resolution (denoted HS-$29$)
\end{itemize}

\subsection{Design of the spatial and spectral degradations}

The proposed model requires the prior knowledge of spectral and spatial degradation matrices $\mathbf{L}$ and $\mathbf{R}=\mathbf{BS}$, respectively. Regarding the spectral degradation matrices required in each simulation scenario, they can be easily derived from the intrinsic sensor characteristics freely available by averaging the spectral bands corresponding to the prescribed response. Conversely, the spatial degradation is not a sensor specification. It depends not only on the considered systems as well as external factors but also on the targeted resolution of the fused image. This work relies on commonly adopted assumptions by considering $\mathbf{R}$ as a Gaussian blur and by adjusting the downsampling factor in $\mathbf{S}$ as an integer value corresponding to the relative ratio between spatial resolution of both observed images.

%, as intrinsic sensor's characteristic, each sensor has open acess into design specifications about spectral response which can be use to construct the spectral degradation matrices. The spectral information, generally, correspond to the spectral response for each band in some wavelength. In all cases, the degradation matrix is built by averaging all positive values within the corresponding spectral band that spectrally overlaps the spectral response between sensors. On the other hand, the spatial degradation is not a sensor specification. It depends on the entirely system as well as external factors. Nevertheless, a commonly adopted model considers $\mathbf{R}$ as an gaussian blur and the downsampling factor of $\mathbf{S}$ an integer value corresponding to the relative ratio between spatial resolution of both observed images. Note that, other optical blurs (PSF) can be adopted considering that it stands for a spatially invariant circular convolution.

\subsection{Compared methods}	
	
As previously exposed, the proposed robust fusion-based CD framework (referred to as RF) is able to deal with all combinations of mono- and multi-band optical images of different spatial and spectral resolutions. However, up to author's knowledge, there is no technique in the literature with such a versatility, i.e., able to address all these scenarios. For this reason, the technique referred to as the worst-case (WC) and also considered in \cite{ferrarisrobust2017} has been used as a baseline and state-of-the-art CD technique. This WC method consists in preprocessing the observed images by spatially and/or spectrally degrading them in order to reach a set of observed images of the same spectral and spatial resolutions. Then, when handling images of same resolutions, classical CD technique, e.g., CVA  proposed in \cite{johnsonchange1998}, can be easily conducted to build a low spatial resolution change mask denoted $\hat{\mathbf{d}}_{\mathrm{WC}}$.

\setlength{\tabcolsep}{3pt}
\renewcommand{\arraystretch}{1.1}
\begin{table}[h!]
\begin{footnotesize}
    \centering
    \begin{tabular}{c|c|c|c|c|c|c|}
 \cline{2-7}
 \multicolumn{1}{c|}{} & \multicolumn{3}{c|}{\multirow{2}{*}{\textbf{Image $\sharp{1}$}}} & \multicolumn{3}{c|}{\multirow{2}{*}{\textbf{Image $\sharp{2}$}}} \\ & \multicolumn{3}{c|}{} & \multicolumn{3}{c|}{} \\
\cline{2-7}
\multicolumn{1}{c|}{} & \multirow{2}{*}{\textbf{Sensor}} & \multirow{2}{*}{\shortstack{\textbf{Spatial} \\ \textbf{resol.}}} & \multirow{2}{*}{\shortstack{\textbf{Spectral} \\ \textbf{resol.}}} & \multirow{2}{*}{\textbf{Sensor}}  & \multirow{2}{*}{\shortstack{\textbf{Spatial} \\ \textbf{resol.}}}  & \multirow{2}{*}{\shortstack{\textbf{Spectral} \\ \textbf{resol.}}}\\
\multicolumn{1}{c|}{} & & & & & & \\

    \hline
	\hline
		\multicolumn{1}{|c|}{\multirow{3}{*}{\rotatebox{00}{$\mathcal{S}_1$}}}
        & Landsat-8 & $15$ & PAN & Landsat-8 & $15$ & PAN \\
        \multicolumn{1}{|c|}{} & Landsat-8 & $30$ & MS-$3$ & Landsat-8 & $30$ & MS-$3$ \\
        \multicolumn{1}{|c|}{} & AVIRIS& $15$ & HS-$224$ & AVIRIS & $15$ & HS-$224$ \\
		\hline
    	\multicolumn{1}{|c|}{\multirow{2}{*}{\rotatebox{00}{$\mathcal{S}_2$}}}
    	& EO-1 ALI & $10$ & PAN & Sentinel-2 & $10$ & MS-$3$ \\
    	\multicolumn{1}{|c|}{} & Landsat-8 & $15$ & PAN & AVIRIS & $15$ & HS-$29$ \\
		\hline
        \multicolumn{1}{|c|}{\multirow{2}{*}{\rotatebox{00}{$\mathcal{S}_3$}}}
    	& Sentinel-2 & $10$ & MS-$3$ & EO-1 ALI & $30$ & MS-$3$ \\
    	\multicolumn{1}{|c|}{} & Sentinel-2 & $10$ & MS-$3$ & Landsat-8 & $30$ & MS-$3$ \\
		\hline
		\multicolumn{1}{|c|}{\multirow{3}{*}{\rotatebox{00}{$\mathcal{S}_4$}}}
        & Landsat-8 & $15$ & PAN & Landsat-8 & $30$ & MS-$3$ \\
        \multicolumn{1}{|c|}{} & EO-1 ALI  & $10$ & PAN & Landsat-8 & $30$ & MS-$3$ \\
        \multicolumn{1}{|c|}{} & Landsat-8 & $15$ & PAN & EO-1 ALI & $30$ & MS-$3$ \\
		\hline
		\multicolumn{1}{|c|}{\multirow{2}{*}{\rotatebox{00}{$\mathcal{S}_5$}}}    	
        & EO-1 ALI  & $30$ & MS-$3$ & AVIRIS & $15$ & HS-$29$ \\
    	\multicolumn{1}{|c|}{} & Landsat-8 & $30$ & MS-$3$ & AVIRIS & $15$ & HS-$29$ \\
		\hline
		\multicolumn{1}{|c|}{\multirow{1}{*}{\rotatebox{00}{$\mathcal{S}_6$}}}
		& EO-1 ALI  & $10$ & PAN & Landsat-8 & $15$ & PAN \\
		\hline
		\multicolumn{1}{|c|}{\multirow{1}{*}{\rotatebox{00}{$\mathcal{S}_7$}}}
		& Sentinel-2  & $10$ & MS-$3$ & Landsat-8 & $15$ & PAN \\
		\hline
        \multicolumn{1}{|c|}{\multirow{1}{*}{\rotatebox{00}{$\mathcal{S}_8$}}}
        & Landsat-8 & $30$ & MS-$8$ & EO-1 ALI & $30$ & MS-$9$\\
		\hline
		\multicolumn{1}{|c|}{\multirow{1}{*}{\rotatebox{00}{$\mathcal{S}_9$}}}
        & Landsat-8 & $30$ & MS-$5$ & Sentinel-2 & $10$ & MS-4\\
		\hline
		\multicolumn{1}{|c|}{\multirow{1}{*}{\rotatebox{00}{$\mathcal{S}_{10}$}}}
        & Sentinel-2 & $20$ & MS-$6$ & EO-1 ALI & $30$ & MS-$9$\\
		\hline
    \hline
    \end{tabular}
    \caption{Pairs of real and/or virtual images, and their spatial and spectral characteristics, used for each applicative scenario.}
  \label{table:sen_res}
  \end{footnotesize}
\end{table}

\subsection{Results}

The following paragraphs discuss the CD performance of the proposed RF method and of the WC approach for each applicative scenario detailed in paragraph \ref{implementation} (see also Table \ref{table:SEN_mat}). Depending on the considered scenario, pairs of real and/or virtual images described in paragraph \ref{subsec:reference_images} are selected, benefiting from different acquisition times but common acquisition location.  Table \ref{table:sen_res} summarizes the pair of observed images provided by the real and/or virtual sensors used in each scenario. Note that several combinations of images can be made for Scenarios $\mathcal{S}_1$--$\mathcal{S}_5$.

\subsubsection{Scenario $\mathcal{S}_1$}

In the first scenario, CD is conducted on a pair of images of same spatial and spectral resolutions, which corresponds to the most favorable and commonly considered CD framework. Figures \ref{fig:sen1_1} to \ref{fig:sen1_3} present the CD binary masks recovered by the proposed RF-based CD method as well as by the WC method for three pairs of panchromatic, multispectral and hyperspectral images, respectively. Note that, in this scenario, the WC boils down to conduct CVA directly on the observed images since they already share the same spatial and spectral resolutions and, thus, do not require to be degraded to be able to conduct pixelwise comparison. These CD maps show that both CD methods detect the most significant changes, in particular the draught of the lake. However, for all configurations, the proposed method visually present CD maps with better detection/false alarm rates when compared with the WC method. This can be explained by the fact that the proposed method denoises the observed image while jointly estimating the change image $\Delta\mathbf{X}$. Conversely, the WC method directly uses the observed images to derive the change image, which may suffer from noise introducing false alarms and misdetections. This is particularly visible in Fig. \ref{fig:sen1_3} depicting the results obtained from a hyperspectral image, known to be of lower signal-to-noise ratio.

\ifsinglecolumn
\newcommand{\subfwidth}{0.4\textwidth}
\newcommand{\figsize}{\textwidth}
\else
\newcommand{\subfwidth}{0.23\textwidth}
\newcommand{\figsize}{0.8\textwidth}
\fi

\begin{figure}[h!]
	\centering
	\begin{subfigure}{\subfwidth}
			\centering	
			\includegraphics[width=\figsize]{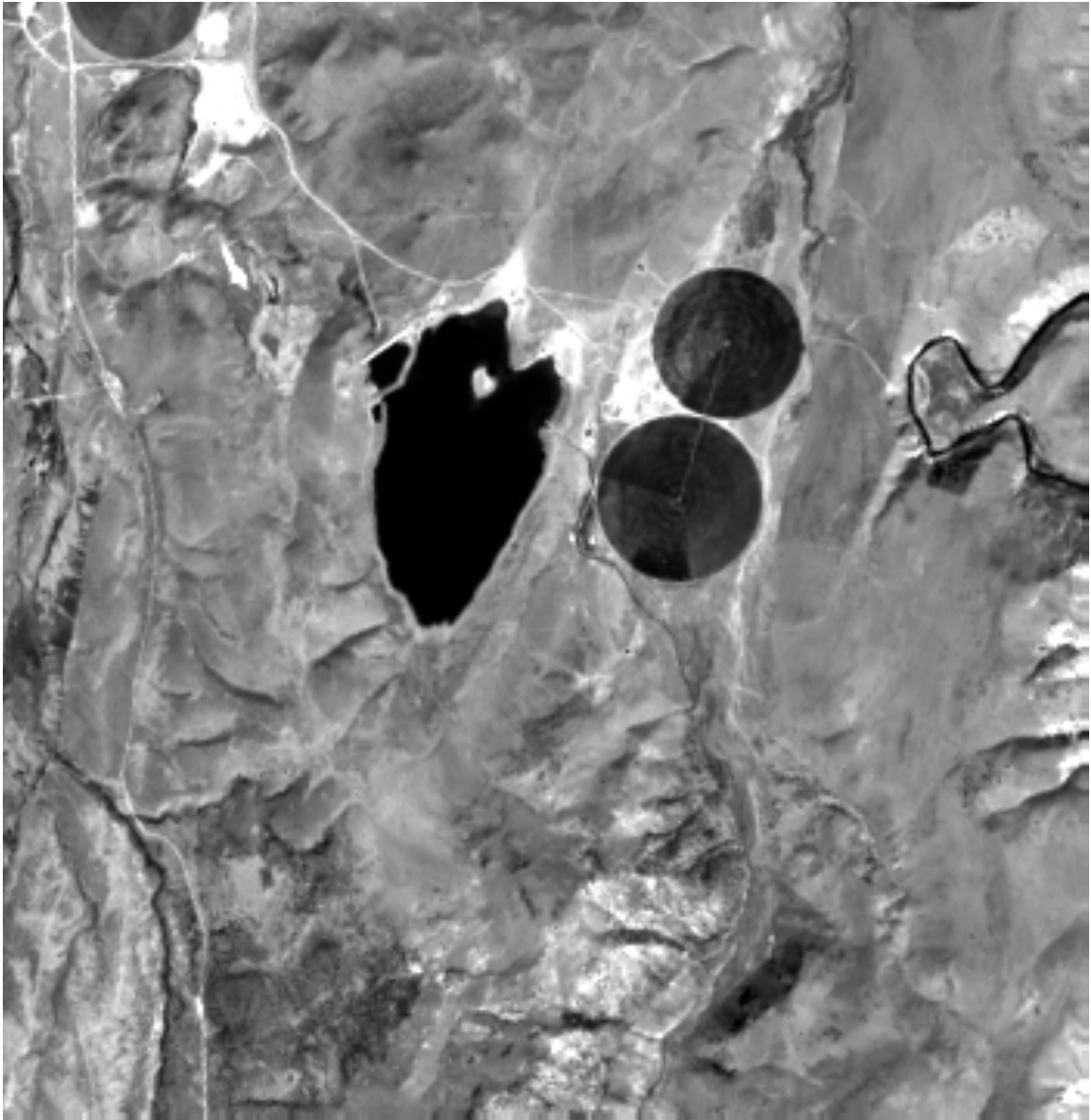}
			\caption{$\mathbf{Y}_{1}$}
			\label{fig:imT1_sen1_1}
	\end{subfigure}
	\begin{subfigure}{\subfwidth}
		\centering	
			\includegraphics[width=\figsize]{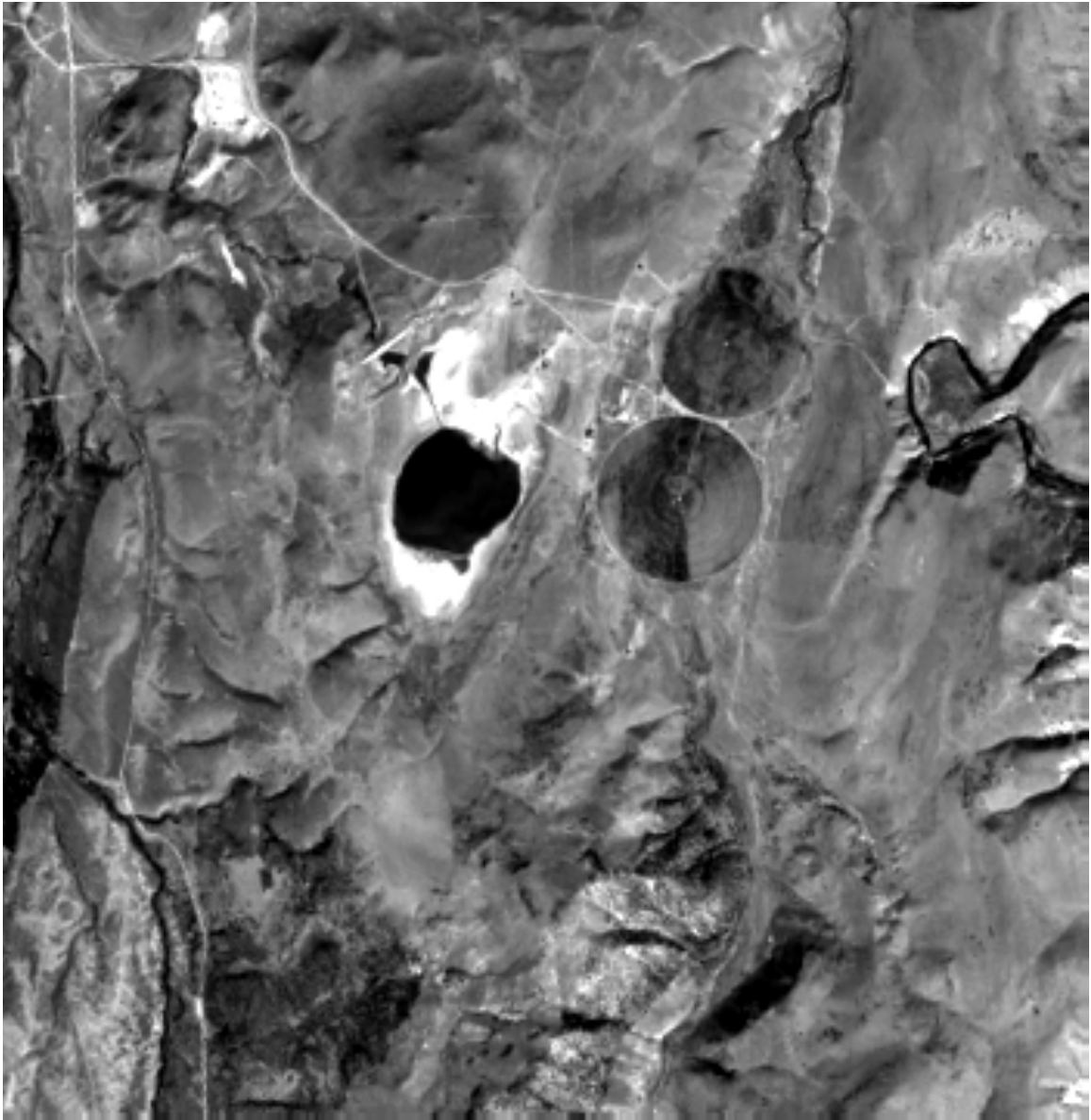}
			\caption{$\mathbf{Y}_{2}$}
			\label{fig:imT2_sen1_1}
	\end{subfigure}
	\begin{subfigure}{\subfwidth}
		\centering
			\includegraphics[width=\figsize]{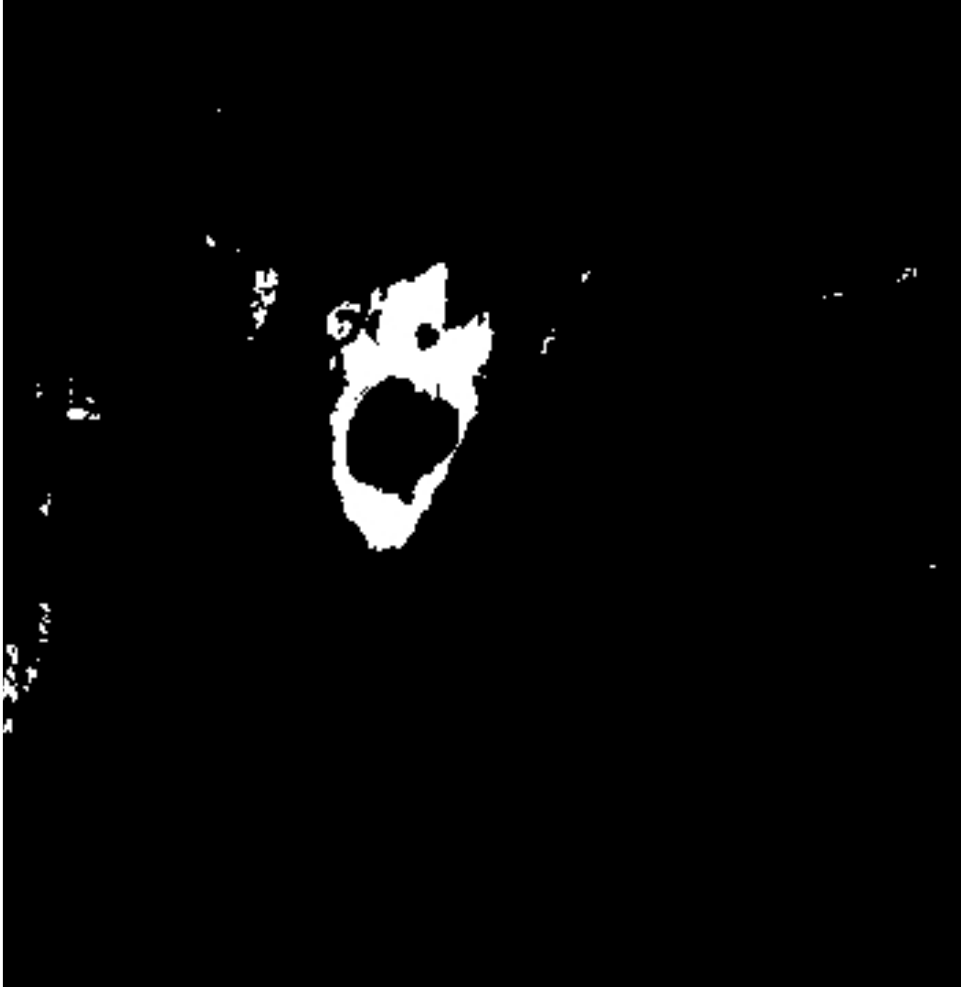}
			\caption{$\hat{\mathbf{d}}_{\mathrm{WC}}$}
			\label{fig:wcCD_sen1_1}
	\end{subfigure}
	\begin{subfigure}{\subfwidth}
			\centering	
			\includegraphics[width=\figsize]{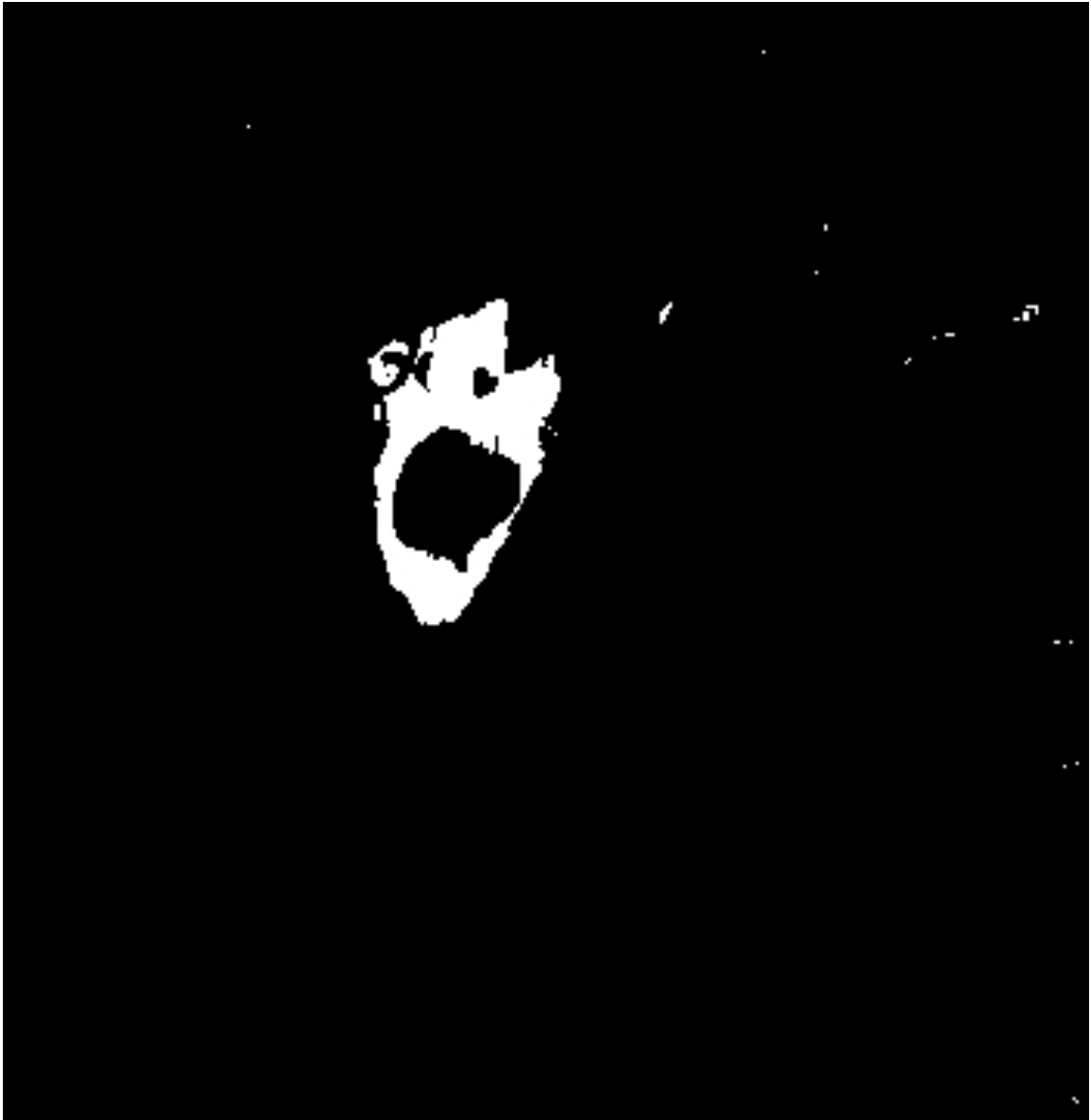}
			\caption{$\hat{\mathbf{d}}_{\mathrm{RF}}$}
			\label{fig:reCD_sen1_1}
	\end{subfigure}
	\caption{Scenario $\mathcal{S}_1$: \protect\subref{fig:imT1_sen1_1} Landsat-8 $15$m PAN observed image $\mathbf{Y}_{1}$ acquired on 04/15/2015, \protect\subref{fig:imT2_sen1_1} Landsat-8 $15$m PAN observed image $\mathbf{Y}_{2}$ acquired on 09/22/2015, \protect\subref{fig:wcCD_sen1_1} change mask $\hat{\mathbf{d}}_{\mathrm{WC}}$ estimated by the WC method and \protect\subref{fig:reCD_sen1_1} change mask $\hat{\mathbf{d}}_{\mathrm{RF}}$ estimated by the proposed approach.}%
	\label{fig:sen1_1}%
\end{figure}

\begin{figure}[h!]
	\centering
	\begin{subfigure}{\subfwidth}
			\centering	
			\includegraphics[width=\figsize]{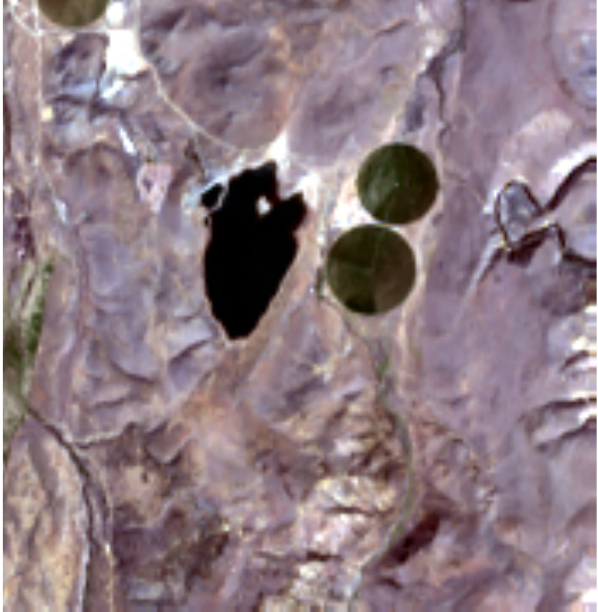}
			\caption{$\mathbf{Y}_{1}$}
			\label{fig:imT1_sen1_2}
	\end{subfigure}
	\begin{subfigure}{\subfwidth}
			\centering	
			\includegraphics[width=\figsize]{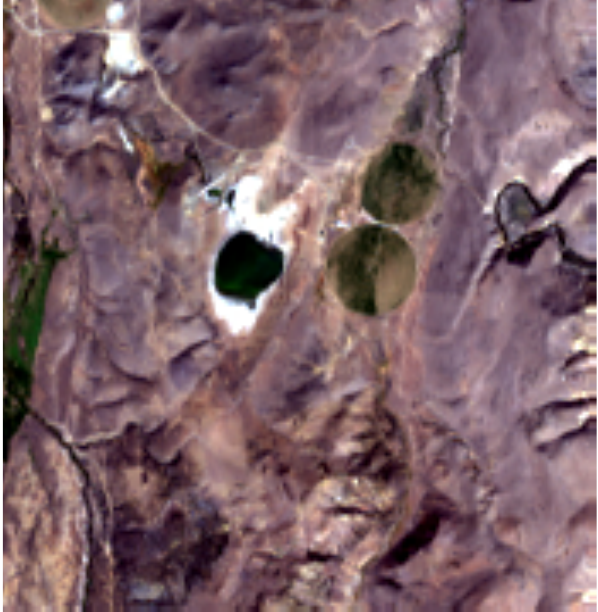}
			\caption{$\mathbf{Y}_{2}$}
			\label{fig:imT2_sen1_2}
	\end{subfigure}
	\begin{subfigure}{\subfwidth}
			\centering
			\includegraphics[width=\figsize]{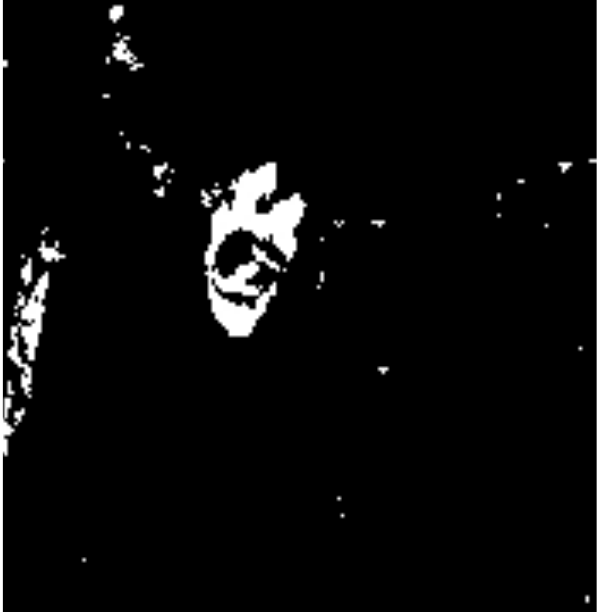}
			\caption{$\hat{\mathbf{d}}_{\mathrm{WC}}$}
			\label{fig:wcCD_sen1_2}
	\end{subfigure}
	\begin{subfigure}{\subfwidth}
			\centering	
			\includegraphics[width=\figsize]{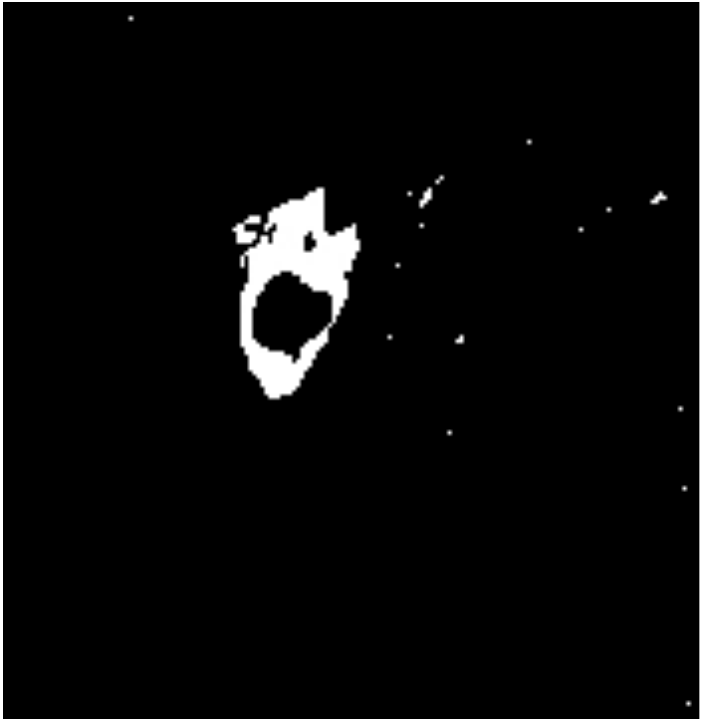}
			\caption{$\hat{\mathbf{d}}_{\mathrm{RF}}$}
			\label{fig:reCD_sen1_2}
	\end{subfigure}
	\caption{Scenario $\mathcal{S}_1$: \protect\subref{fig:imT1_sen1_2} Landsat-8 $30$m MS-$3$ observed image $\mathbf{Y}_{1}$ acquired on 04/15/2015, \protect\subref{fig:imT2_sen1_2} Landsat-8 $30$m MS-$3$ observed image $\mathbf{Y}_{2}$ acquired on 09/22/2015, \protect\subref{fig:wcCD_sen1_2} change mask $\hat{\mathbf{d}}_{\mathrm{WC}}$ estimated by the WC method and  \protect\subref{fig:reCD_sen1_2} change mask $\hat{\mathbf{d}}_{\mathrm{RF}}$ estimated by the proposed approach.}%
	\label{fig:sen1_2}%
\end{figure}

\begin{figure}[h!]
	\centering
	\begin{subfigure}{\subfwidth}
			\centering	
			\includegraphics[width=\figsize]{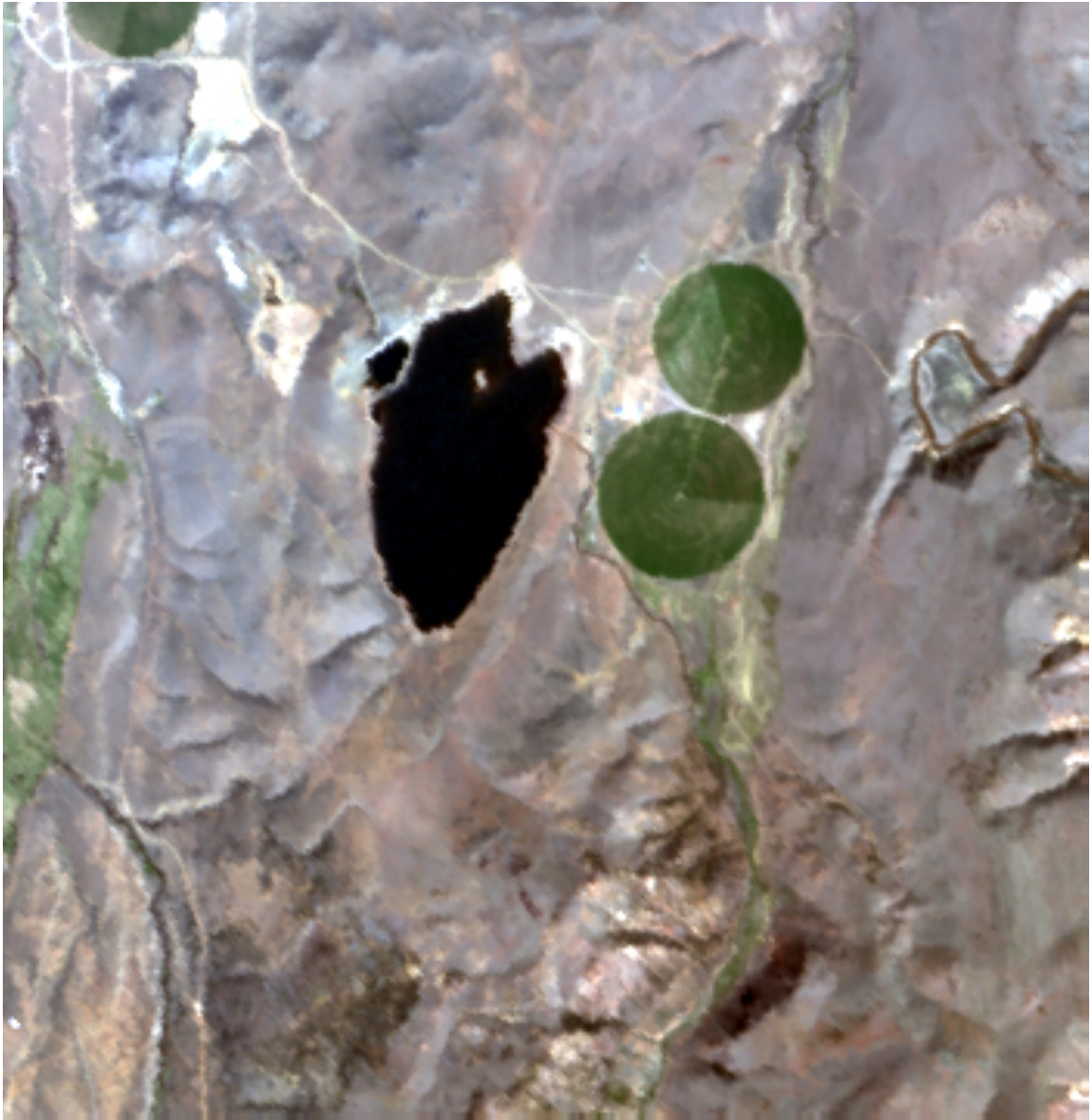}
			\caption{$\mathbf{Y}_{1}$}
			\label{fig:imT1_sen1_3}
	\end{subfigure}
	\begin{subfigure}{\subfwidth}
			\centering	
			\includegraphics[width=\figsize]{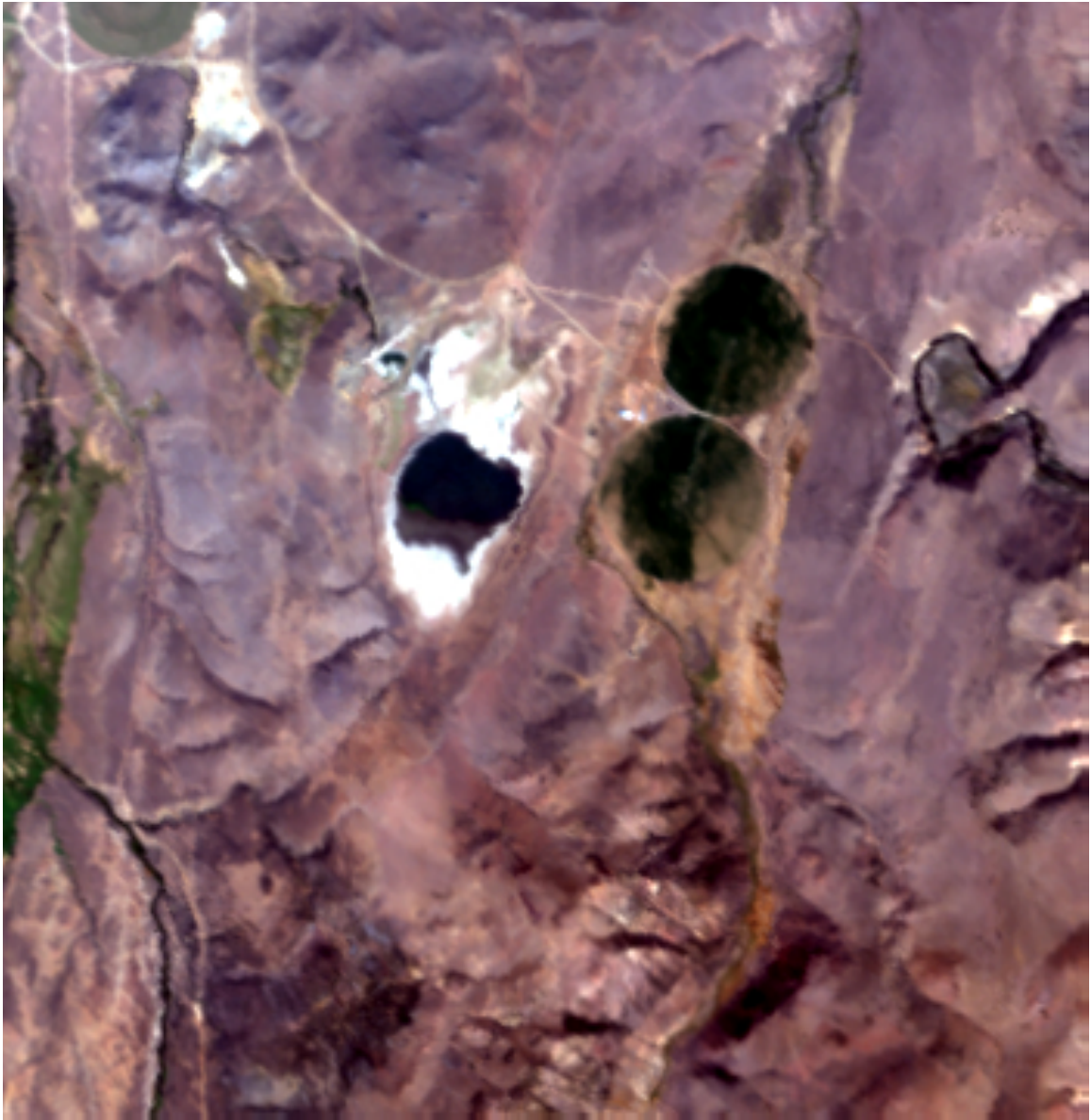}
			\caption{$\mathbf{Y}_{2}$}
			\label{fig:imT2_sen1_3}
	\end{subfigure}
	\begin{subfigure}{\subfwidth}
			\centering
			\includegraphics[width=\figsize]{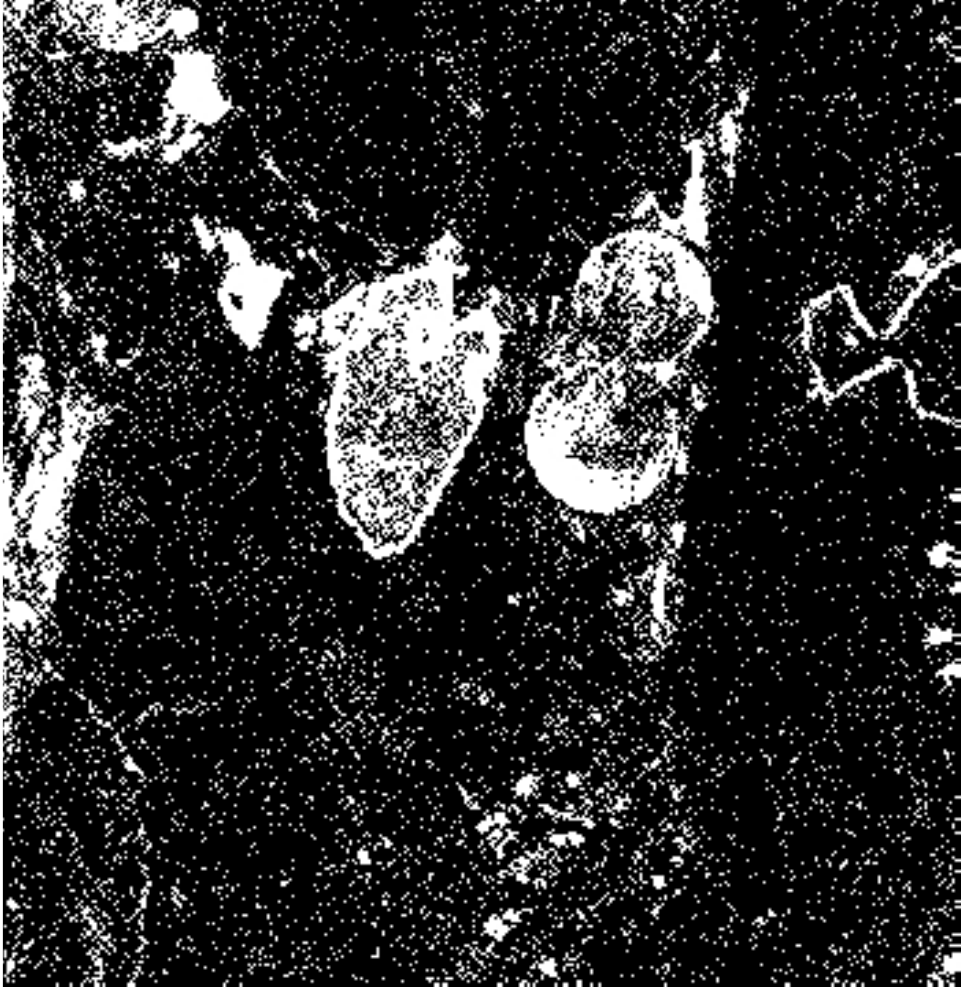}
			\caption{$\hat{\mathbf{d}}_{\mathrm{WC}}$}
			\label{fig:wcCD_sen1_3}
	\end{subfigure}
	\begin{subfigure}{\subfwidth}
			\centering	
			\includegraphics[width=\figsize]{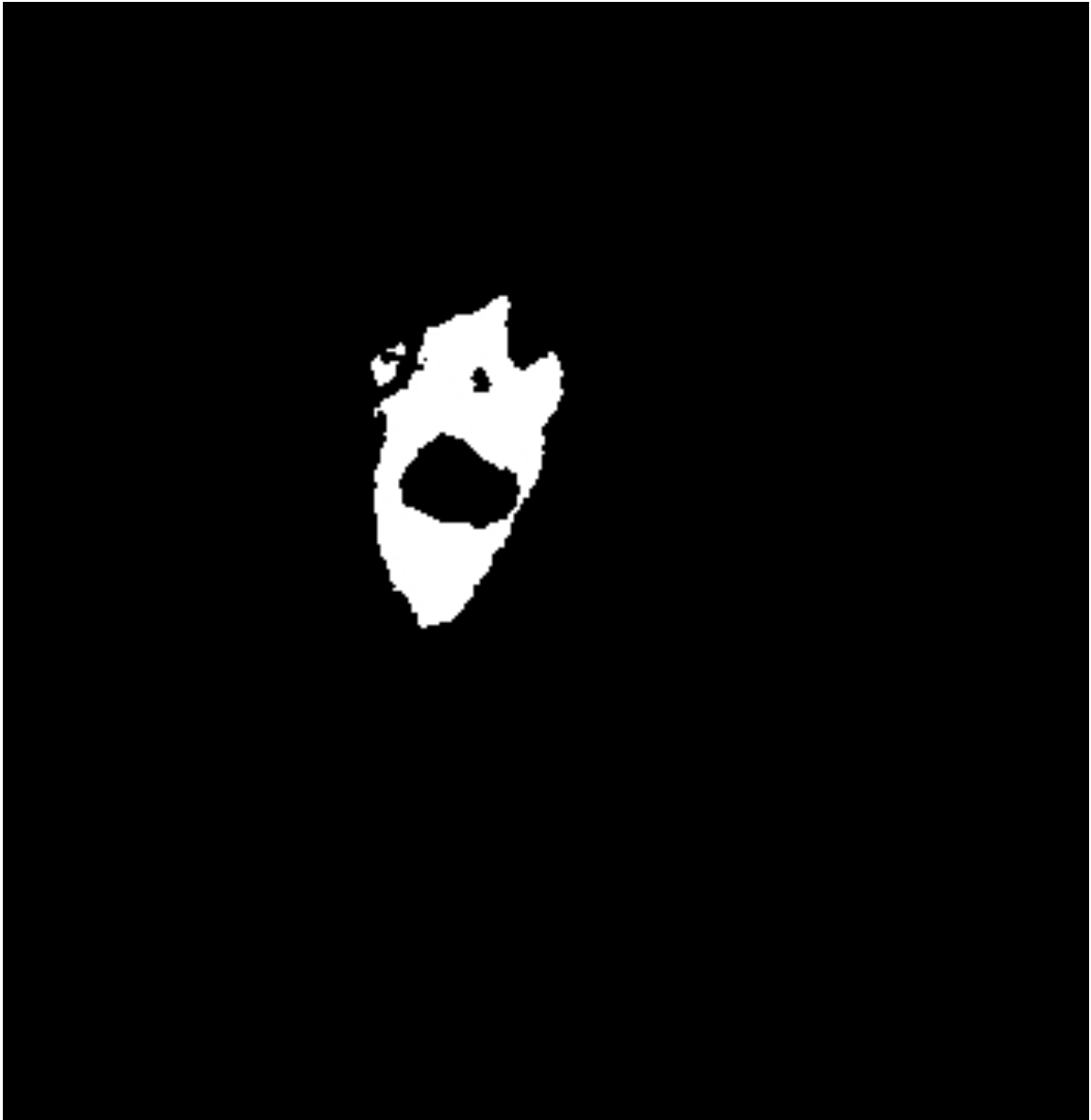}
			\caption{$\hat{\mathbf{d}}_{\mathrm{RF}}$}
			\label{fig:reCD_sen1_3}
	\end{subfigure}
	\caption{Scenario $\mathcal{S}_1$: \protect\subref{fig:imT1_sen1_3} AVIRIS $15$m HS-$224$ observed image $\mathbf{Y}_{1}$ acquired on 04/10/2014, \protect\subref{fig:imT2_sen1_3} AVIRIS $15$m HS-$224$ observed image $\mathbf{Y}_{2}$ acquired on 09/19/2014, \protect\subref{fig:wcCD_sen1_3} change mask $\hat{\mathbf{d}}_{\mathrm{WC}}$ estimated by the WC method and \protect\subref{fig:reCD_sen1_3} change mask $\hat{\mathbf{d}}_{\mathrm{RF}}$ estimated by the proposed approach.}%
	\label{fig:sen1_3}%
\end{figure}

\subsubsection{Scenario $\mathcal{S}_{2}$}

\begin{figure}[h!]
	\centering
	\begin{subfigure}{\subfwidth}
			\centering	
			\includegraphics[width=\figsize]{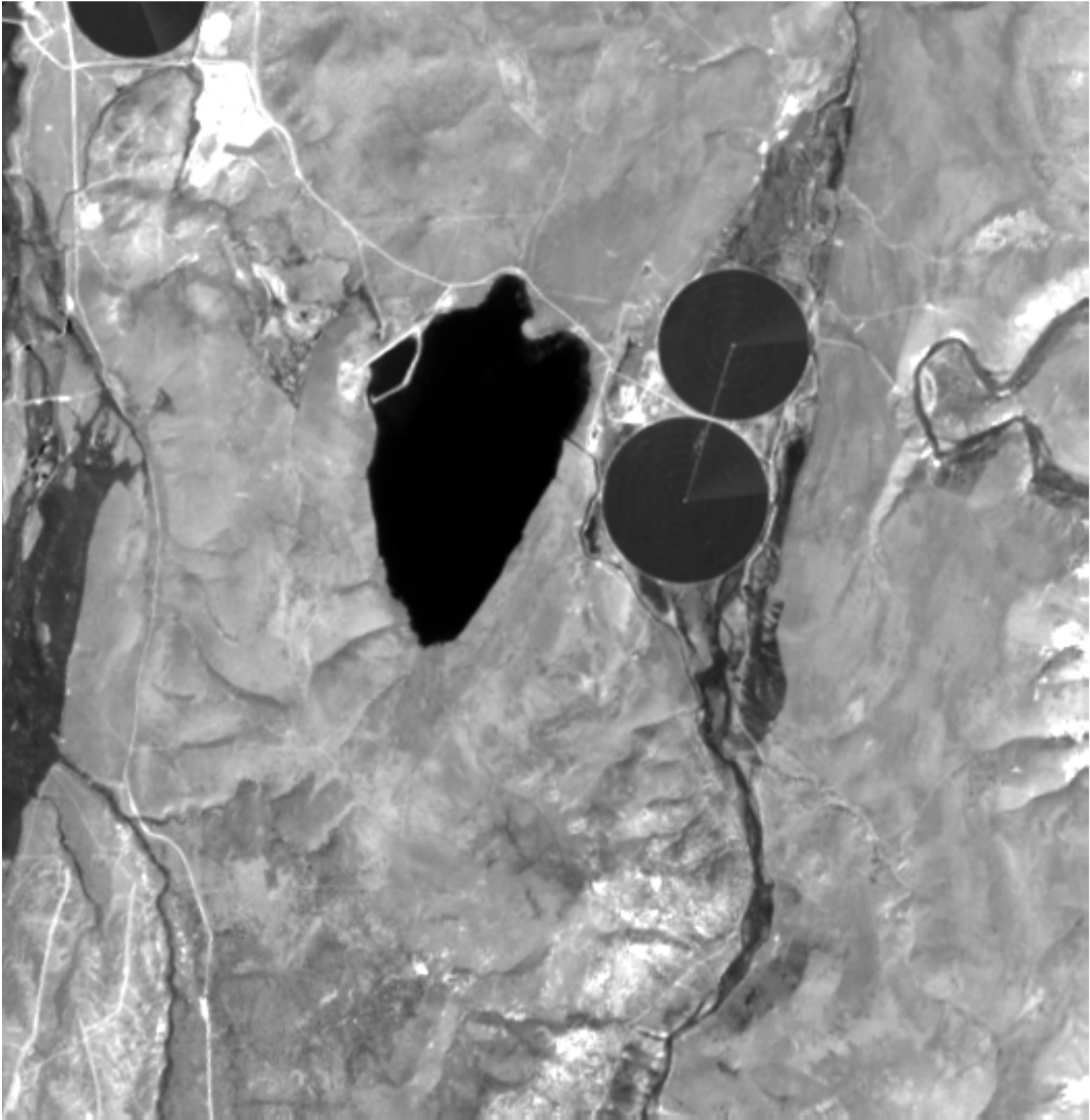}
			\caption{$\mathbf{Y}_{1}$}
			\label{fig:imT1_sen2_1}
	\end{subfigure}
	\begin{subfigure}{\subfwidth}
			\centering	
			\includegraphics[width=\figsize]{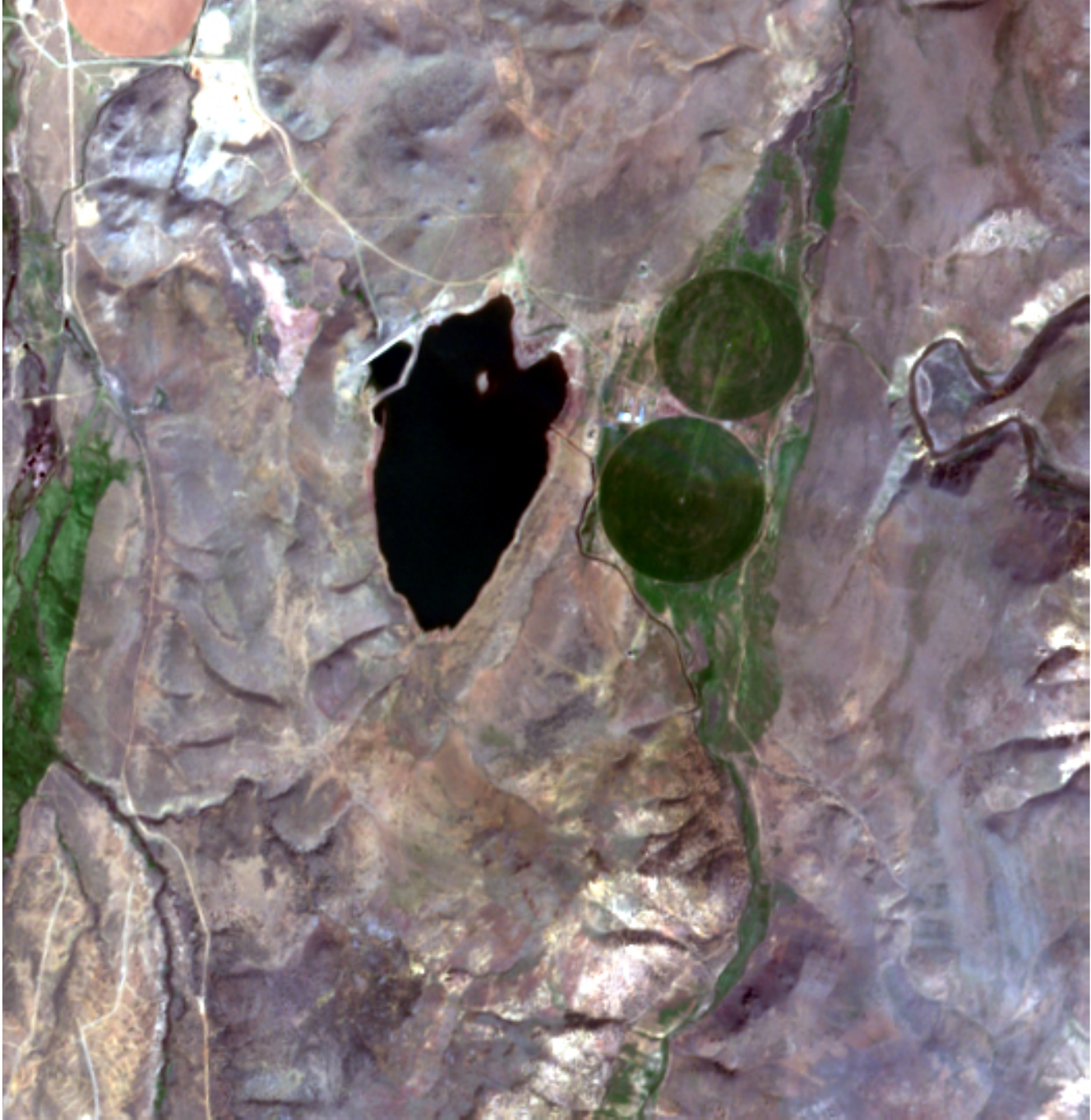}
			\caption{$\mathbf{Y}_{2}$}
			\label{fig:imT2_sen2_1}
	\end{subfigure}
	\begin{subfigure}{\subfwidth}
			\centering
			\includegraphics[width=\figsize]{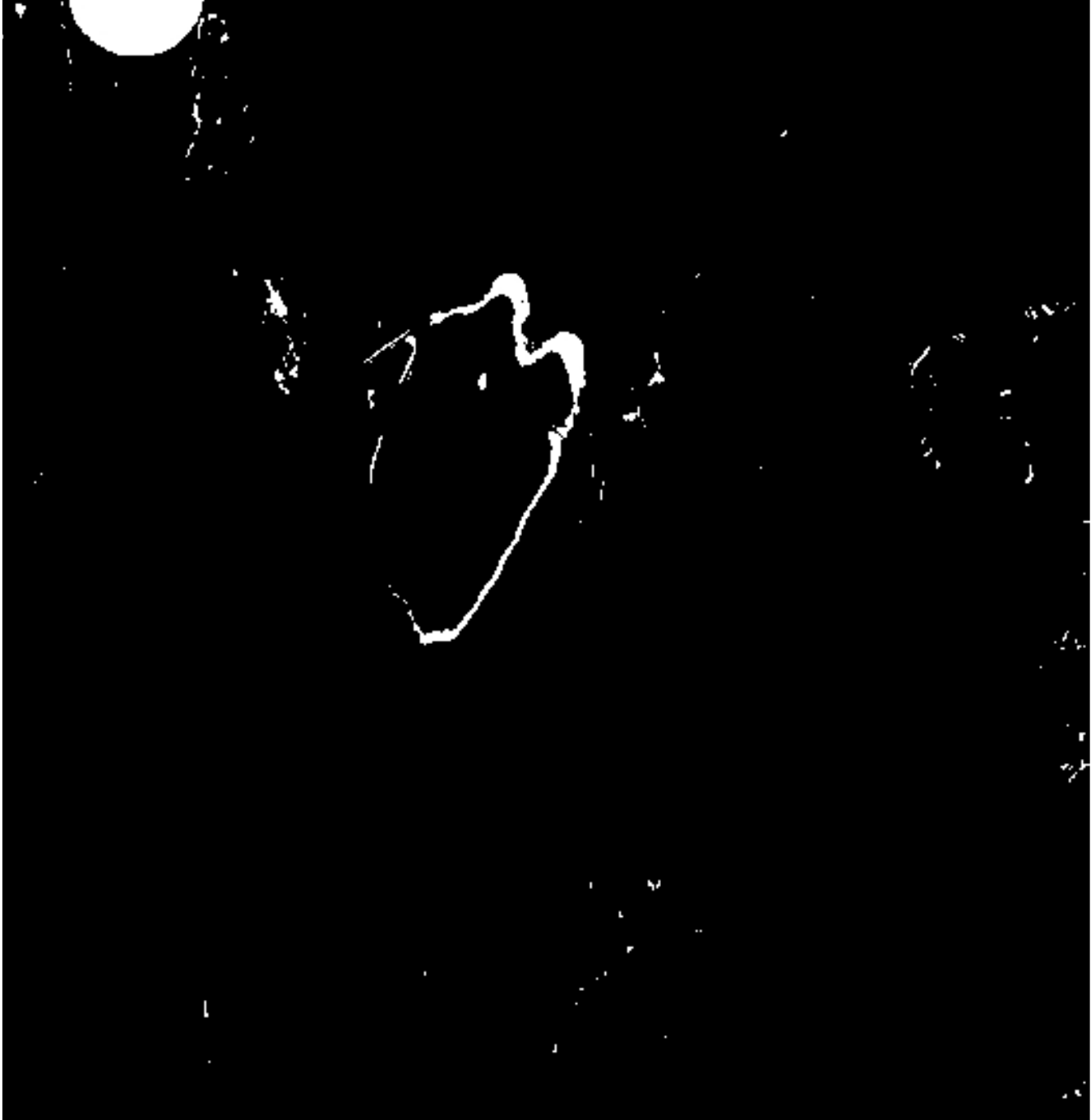}
			\caption{$\hat{\mathbf{d}}_{\mathrm{WC}}$}
			\label{fig:wcCD_sen2_1}
	\end{subfigure}
	\begin{subfigure}{\subfwidth}
			\centering	
			\includegraphics[width=\figsize]{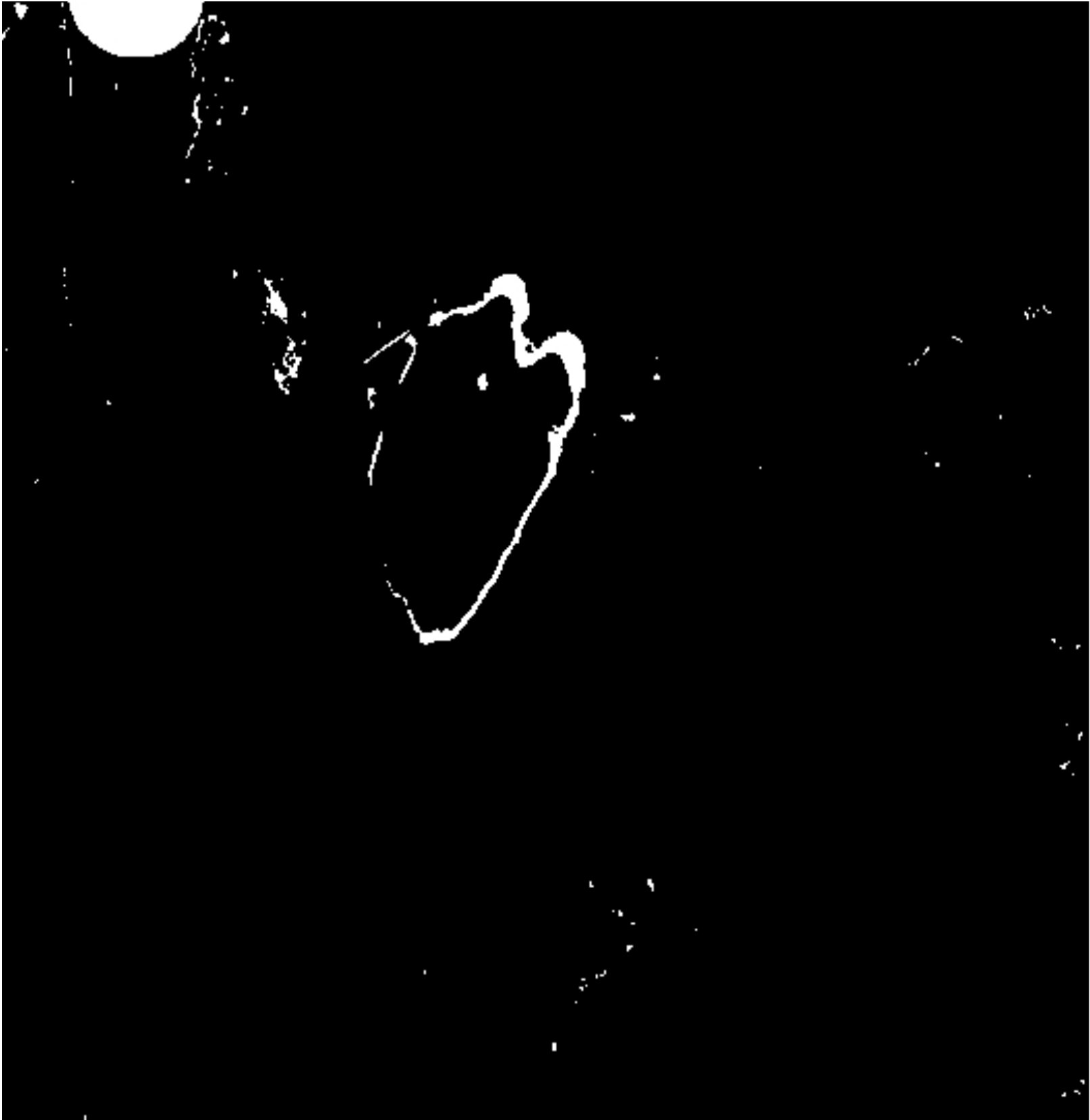}
			\caption{$\hat{\mathbf{d}}_{\mathrm{RF}}$}
			\label{fig:reCD_sen2_1}
	\end{subfigure}
	\caption{Scenario $\mathcal{S}_{2}$: \protect\subref{fig:imT1_sen2_1} EO-1 ALI $10$m PAN observed image $\mathbf{Y}_{1}$ acquired on 06/08/2011, \protect\subref{fig:imT2_sen2_1} Sentinel-2 $10$m MS-$3$ observed image $\mathbf{Y}_{2}$ acquired on 04/12/2016, \protect\subref{fig:wcCD_sen2_1} change mask $\hat{\mathbf{d}}_{\mathrm{WC}}$ estimated by the WC approach from a pair of $10$m PAN degraded images and \protect\subref{fig:reCD_sen2_1} change mask $\hat{\mathbf{d}}_{\mathrm{RF}}$ estimated by the proposed approach from a $10$m MS-$3$ change image $\Delta\hat{\mathbf{X}}$.}%
	\label{fig:sen2_1}%
\end{figure}

\begin{figure}[h!]
	\centering
	\begin{subfigure}{\subfwidth}
			\centering	
			\includegraphics[width=\figsize]{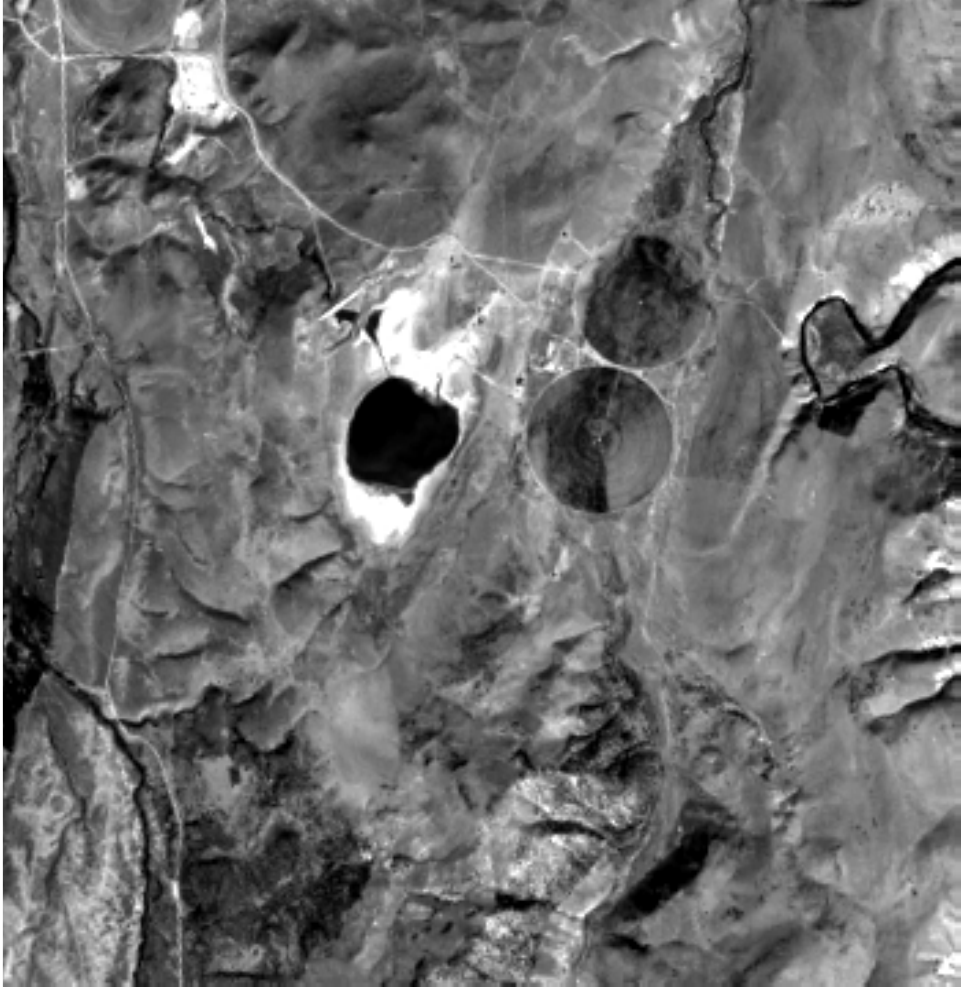}
			\caption{$\mathbf{Y}_{1}$}
			\label{fig:imT1_sen2_2}
	\end{subfigure}
	\begin{subfigure}{\subfwidth}
			\centering	
			\includegraphics[width=\figsize]{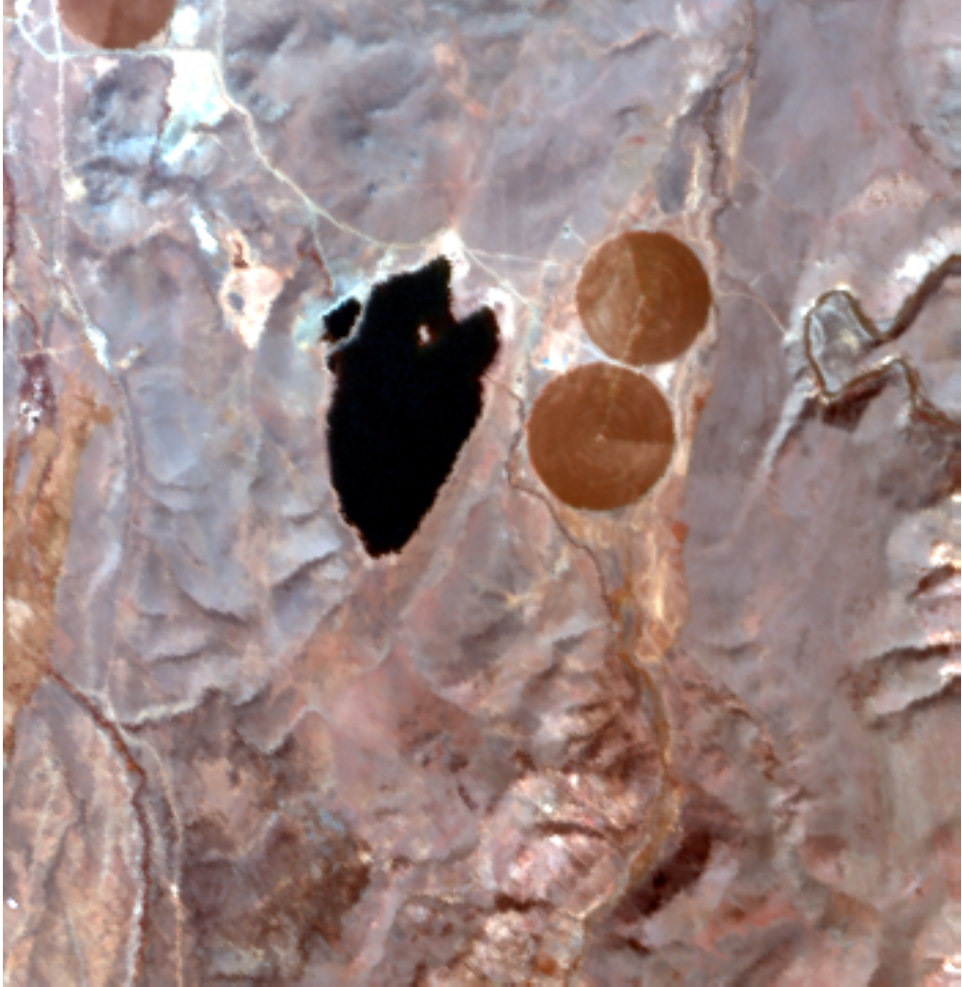}
			\caption{$\mathbf{Y}_{2}$}
			\label{fig:imT2_sen2_2}
	\end{subfigure}
	\begin{subfigure}{\subfwidth}
			\centering
			\includegraphics[width=\figsize]{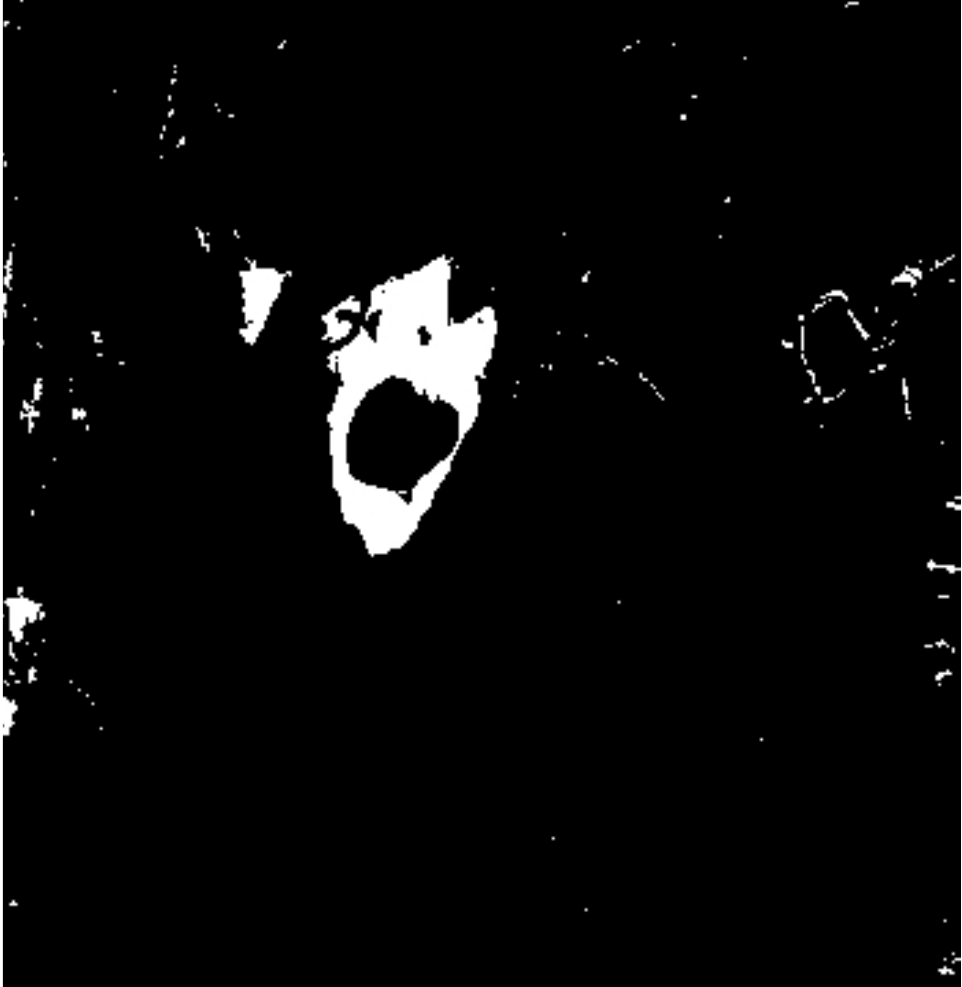}
			\caption{$\hat{\mathbf{d}}_{\mathrm{WC}}$}
			\label{fig:wcCD_sen2_2}
	\end{subfigure}
	\begin{subfigure}{\subfwidth}
			\centering	
			\includegraphics[width=\figsize]{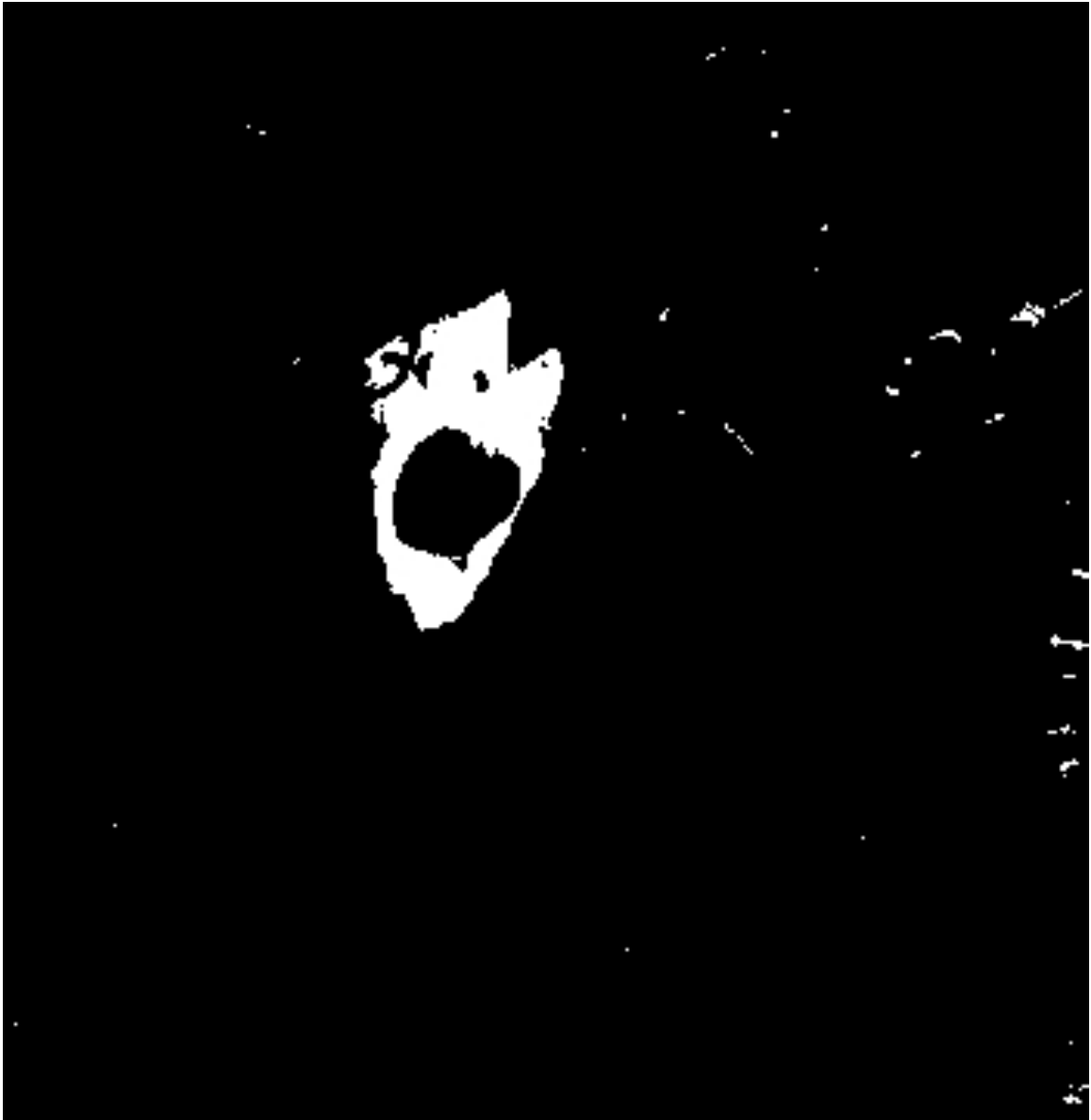}
			\caption{$\hat{\mathbf{d}}_{\mathrm{RF}}$}
			\label{fig:reCD_sen2_2}
	\end{subfigure}
	\caption{Scenario $\mathcal{S}_{2}$: \protect\subref{fig:imT1_sen2_2} Landsat-8 $15$m PAN observed image $\mathbf{Y}_{1}$ acquired on 09/22/2015, \protect\subref{fig:imT2_sen2_2} AVIRIS $15$m HS-$29$ observed image $\mathbf{Y}_{2}$ acquired on 04/10/2014, \protect\subref{fig:wcCD_sen2_2} change mask $\hat{\mathbf{d}}_{\mathrm{WC}}$ estimated by the WC approach from a pair of $15$m PAN degraded images and \protect\subref{fig:reCD_sen2_2} change mask $\hat{\mathbf{d}}_{\mathrm{RF}}$ estimated by the proposed approach from a $15$m HS-$29$ change image $\Delta\hat{\mathbf{X}}$.}%
	\label{fig:sen2_2}%
\end{figure}

This CD scenario deals with observed images of same spatial resolution but different spectral resolutions. Figures \ref{fig:sen2_1} and \ref{fig:sen2_2} illustrate two possible situations and show the CD results of the proposed RF-based CD method compared with the WC method. In this scenario, similarly to scenario $\mathcal{S}_{1}$, both estimated CD maps have the same spatial resolution as the observed image pair, which means that there is not any loss of spatial resolution. On the other hand, the proposed method delivers a change map estimated from $\Delta\mathbf{X}$ of same spectral resolution as the highest spectral resolution among the two observed images. Conversely, the WC method conducts CVA on a pair of images after spectral degradation to reach the lowest spectral resolution, which possibly results in loss of significant information. The consequent impact on the change/no-change decision is the visual reduction of false alarm rate for the proposed RF method, even if both CD maps have the same spatial resolution.

\subsubsection{Scenario $\mathcal{S}_{3}$}

\begin{figure}[h!]
	\centering
	\begin{subfigure}{\subfwidth}
			\centering	
			\includegraphics[width=\figsize]{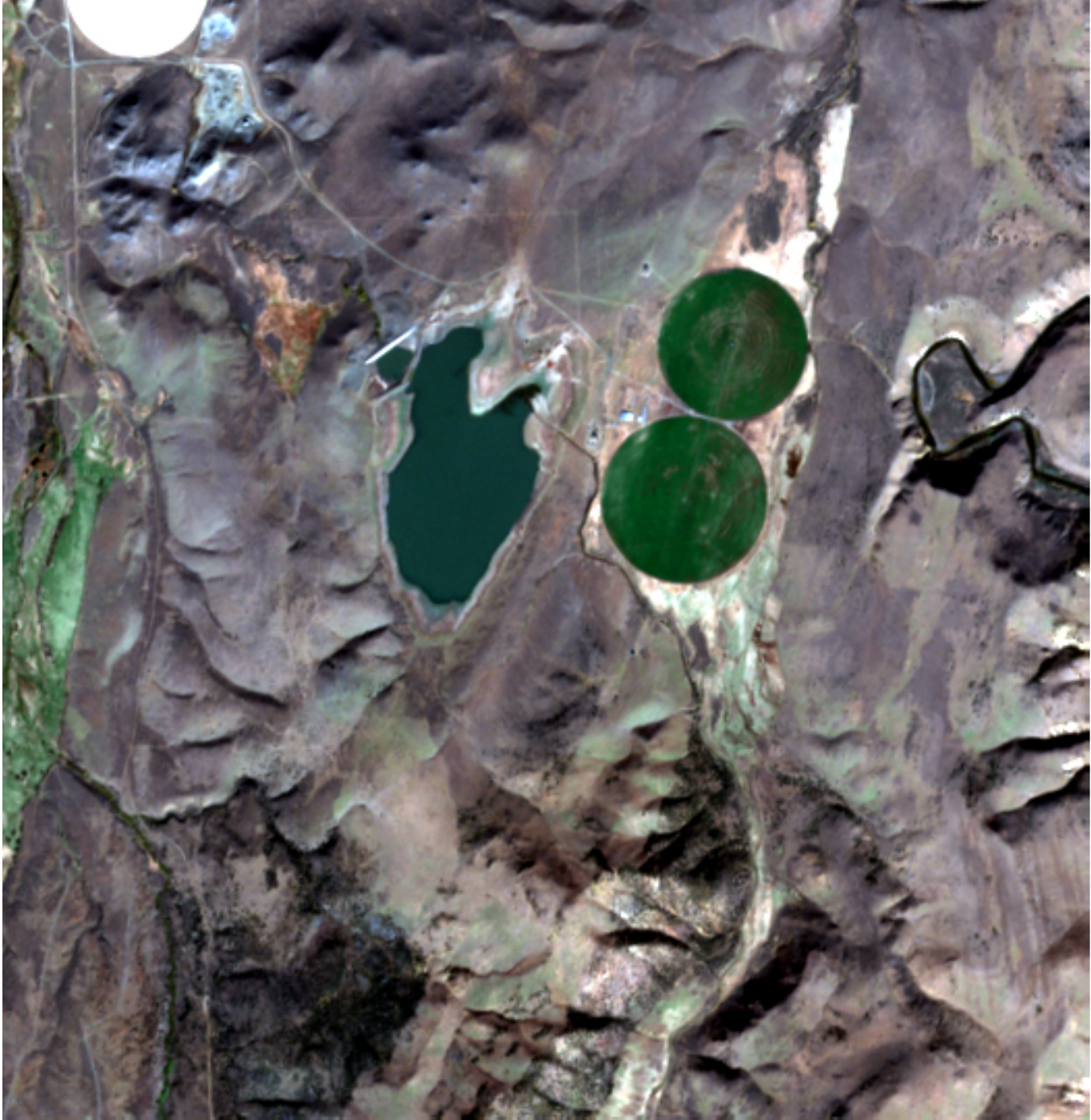}
			\caption{$\mathbf{Y}_{1}$}
			\label{fig:imT1_sen3_1}
	\end{subfigure}
	\begin{subfigure}{\subfwidth}
			\centering	
			\includegraphics[width=\figsize]{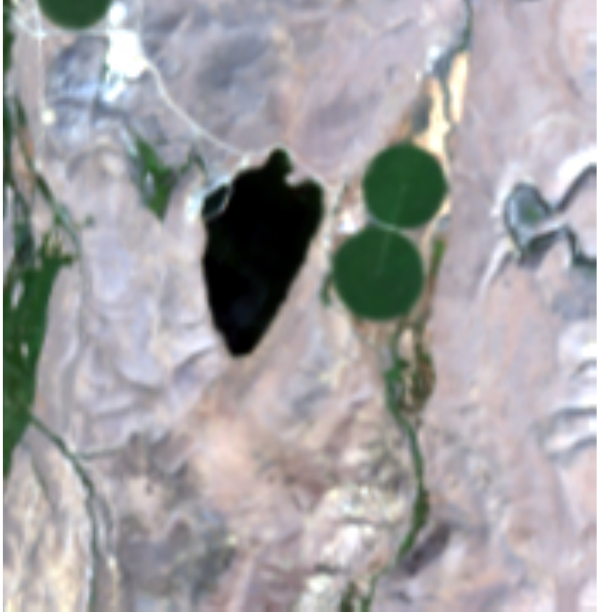}
			\caption{$\mathbf{Y}_{2}$}
			\label{fig:imT2_sen3_1}
	\end{subfigure}
		\begin{subfigure}{\subfwidth}
			\centering
			\includegraphics[width=\figsize]{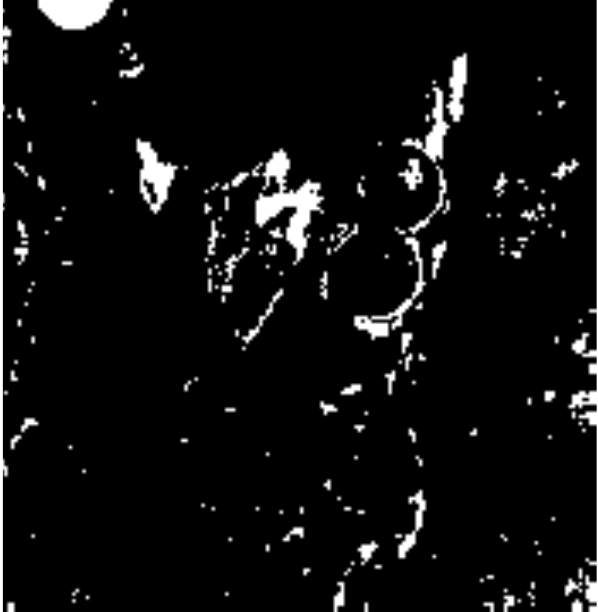}
			\caption{$\hat{\mathbf{d}}_{\mathrm{WC}}$}
			\label{fig:wcCD_sen3_1}
	\end{subfigure}
	\begin{subfigure}{\subfwidth}
			\centering	
			\includegraphics[width=\figsize]{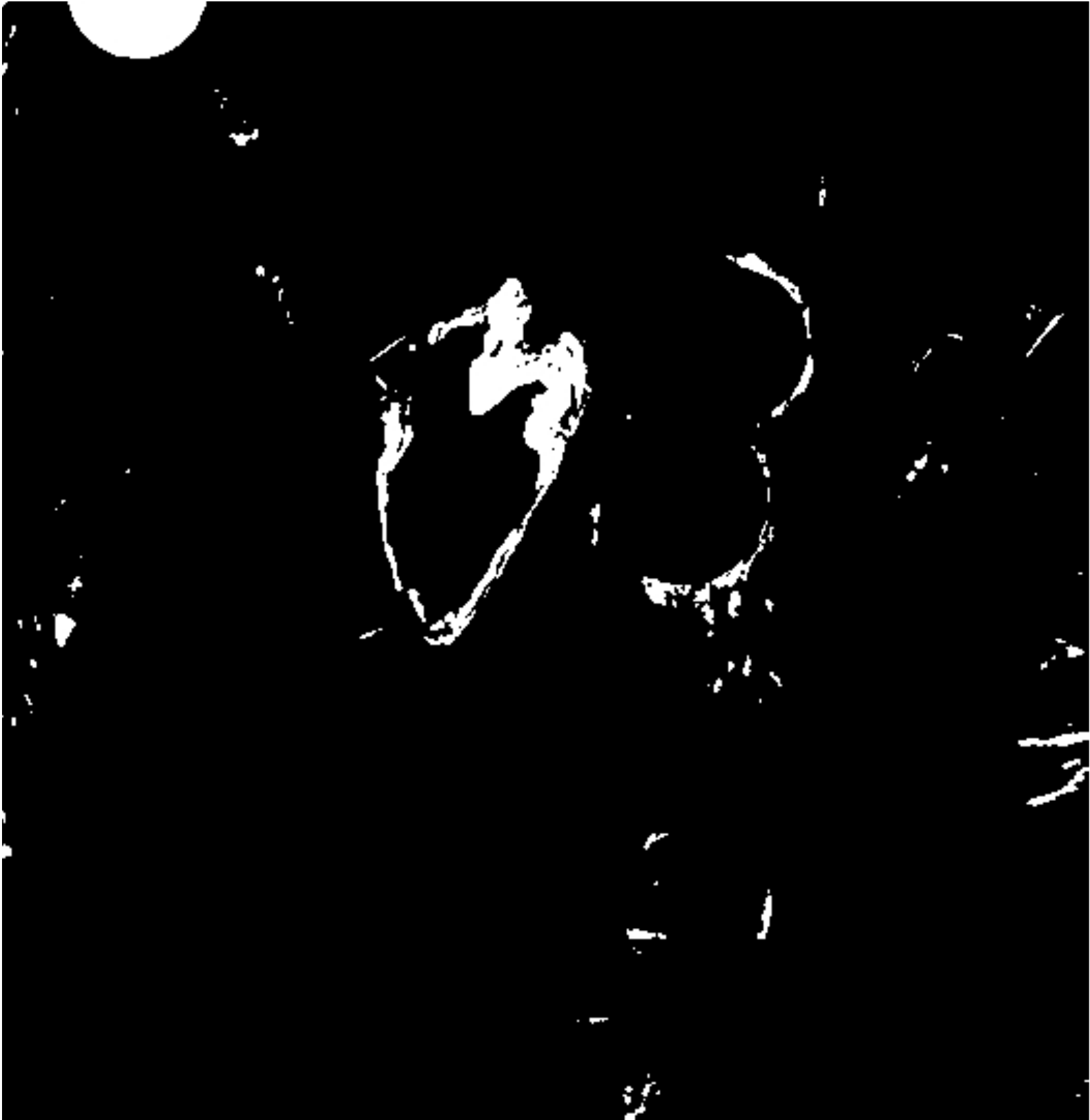}
			\caption{$\hat{\mathbf{d}}_{\mathrm{RF}}$}
			\label{fig:reCD_sen3_1}
	\end{subfigure}
	\caption{Scenario $\mathcal{S}_{3}$: \protect\subref{fig:imT1_sen3_1} Sentinel-2 $10$m MS-$3$ observed image $\mathbf{Y}_{1}$ acquired on 10/29/2016, \protect\subref{fig:imT2_sen3_1} EO-1 ALI $30$m MS-$3$ observed image $\mathbf{Y}_{2}$ acquired on 08/04/2011, \protect\subref{fig:wcCD_sen3_1} change mask $\hat{\mathbf{d}}_{\mathrm{WC}}$ estimated by the WC approach from a pair of $30$m MS-$3$ degraded images and \protect\subref{fig:reCD_sen3_1} change mask $\hat{\mathbf{d}}_{\mathrm{RF}}$ estimated by the proposed approach from a $10$m MS-$3$ change image $\Delta\hat{\mathbf{X}}$.}%
	\label{fig:sen3_1}%
\end{figure}

\begin{figure}[h!]
	\centering
	\begin{subfigure}{\subfwidth}
			\centering	
			\includegraphics[width=\figsize]{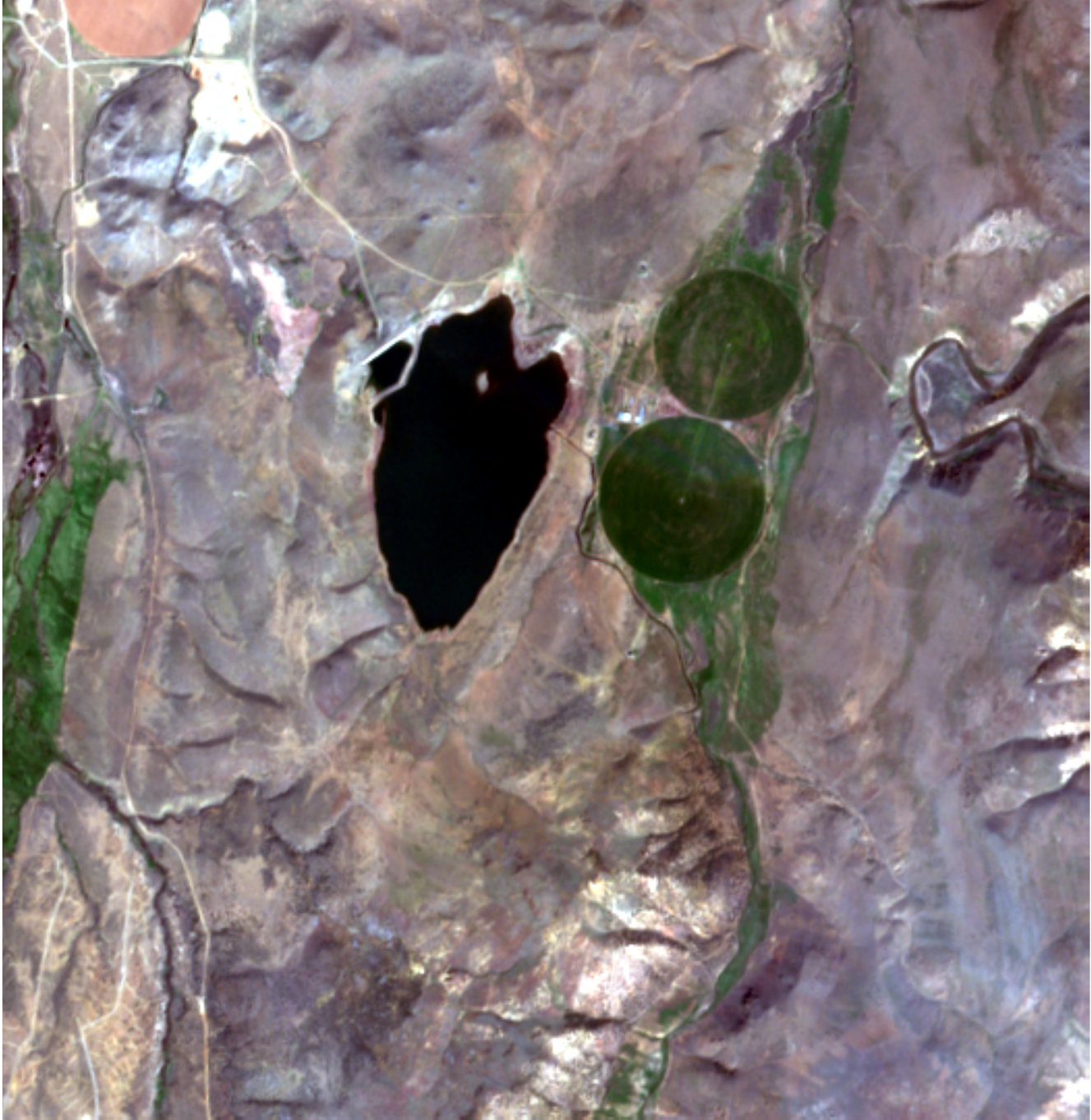}
			\caption{$\mathbf{Y}_{1}$}
			\label{fig:imT1_sen3_2}
	\end{subfigure}
	\begin{subfigure}{\subfwidth}
			\centering	
			\includegraphics[width=\figsize]{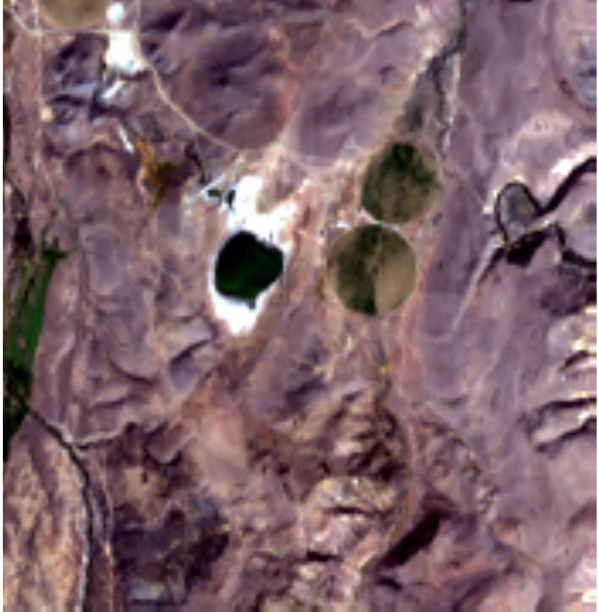}
			\caption{$\mathbf{Y}_{2}$}
			\label{fig:imT2_sen3_2}
	\end{subfigure}
	\begin{subfigure}{\subfwidth}
			\centering
			\includegraphics[width=\figsize]{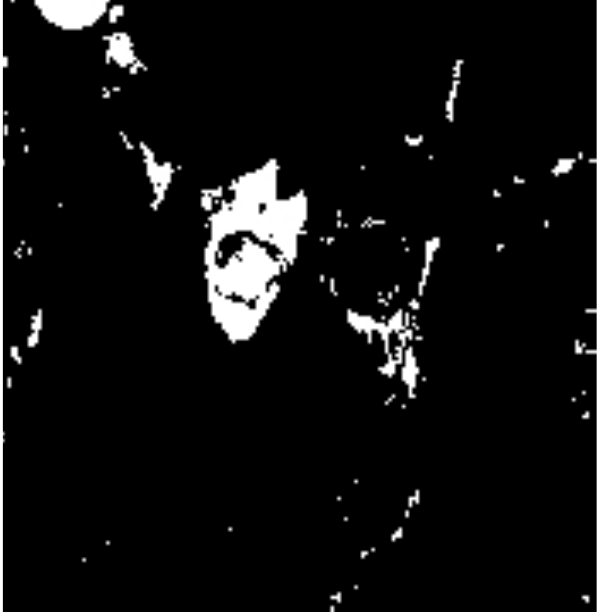}
			\caption{$\hat{\mathbf{d}}_{\mathrm{WC}}$}
			\label{fig:wcCD_sen3_2}
	\end{subfigure}
	\begin{subfigure}{\subfwidth}
			\centering	
			\includegraphics[width=\figsize]{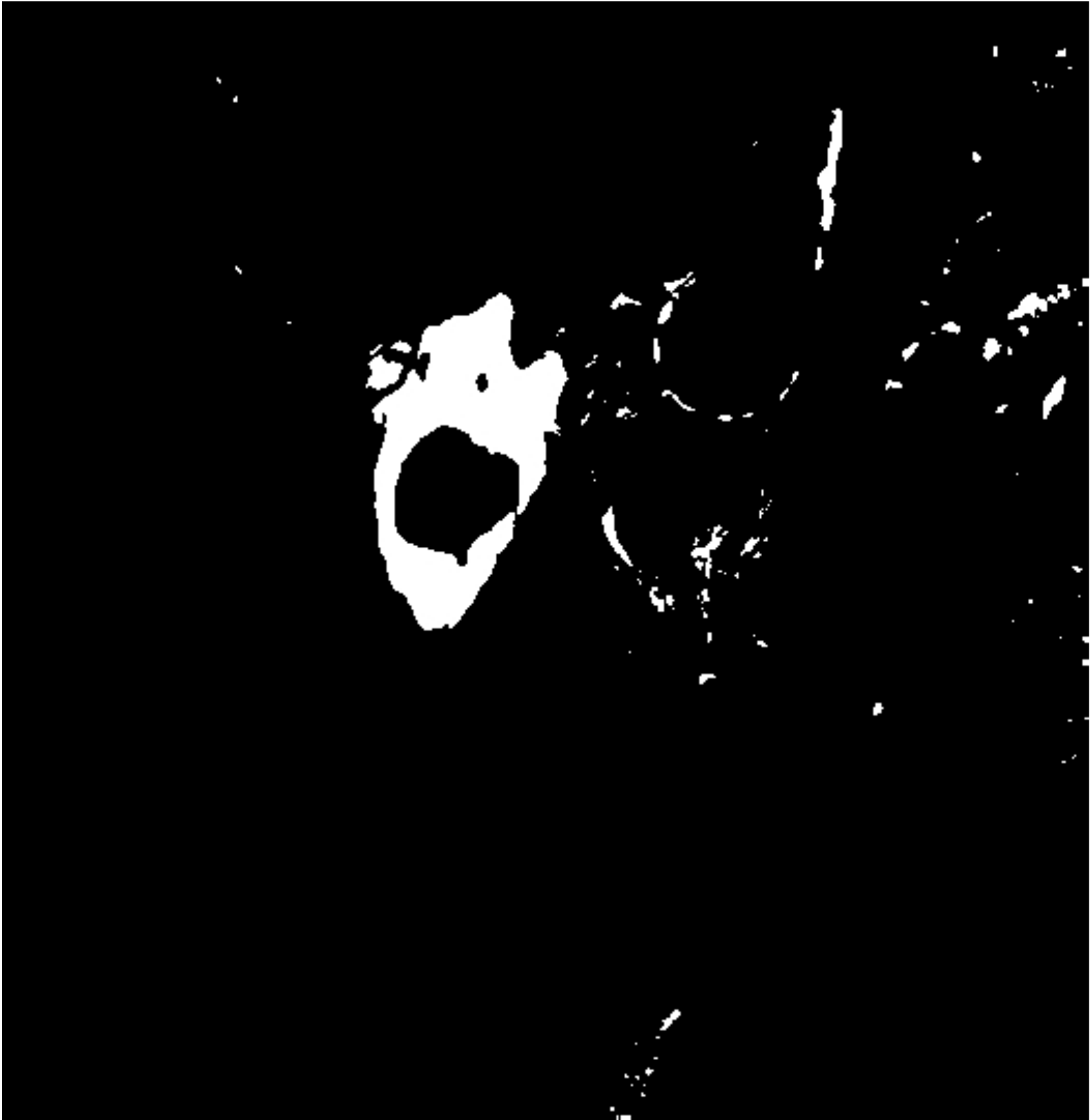}
			\caption{$\hat{\mathbf{d}}_{\mathrm{RF}}$}
			\label{fig:reCD_sen3_2}
	\end{subfigure}
	\caption{Scenario $\mathcal{S}_{3}$: \protect\subref{fig:imT1_sen3_2} Sentinel-2 $10$m MS-$3$ observed image $\mathbf{Y}_{1}$ acquired on 04/12/2016, \protect\subref{fig:imT2_sen3_2} Landsat-8 $30$m MS-$3$ observed image $\mathbf{Y}_{2}$ acquired on 09/22/2015, \protect\subref{fig:wcCD_sen3_2} change mask $\hat{\mathbf{d}}_{\mathrm{WC}}$ estimated by the WC approach from a pair of $30$m MS-$3$ degraded images and \protect\subref{fig:reCD_sen3_2} change mask $\hat{\mathbf{d}}_{\mathrm{RF}}$ estimated by the proposed approach from a $10$m MS-$3$ change image $\Delta\hat{\mathbf{X}}$.}%
	\label{fig:sen3_2}%
\end{figure}

In scenario $\mathcal{S}_{3}$, corresponding to the reverse situation encountered in scenario $\mathcal{S}_2$, observed images share the same spectral resolution but with different spatial resolution. Figures \ref{fig:sen3_1} and \ref{fig:sen3_2} present the results obtained for two possible real situations. Note that CD maps obtained by the proposed RF method are of higher spatial resolutions than the ones estimated by the WC approach. Thus, this scenario is the first to illustrate the most important differences between both approaches, i.e., the difference in spatial resolutions of the CD maps. In scenario $\mathcal{S}_{2}$, the results have already shown that the loss of spectral information inherent to the WC approach leads to an increase of false alarms and misdetections. Here, the loss of spatial information when conducting the WC approach results in an inaccurate localization of the possible changes.

\subsubsection{Scenario $\mathcal{S}_{4}$}

\begin{figure}[h!]
	\centering
	\begin{subfigure}{\subfwidth}
			\centering
			\includegraphics[width=\figsize]{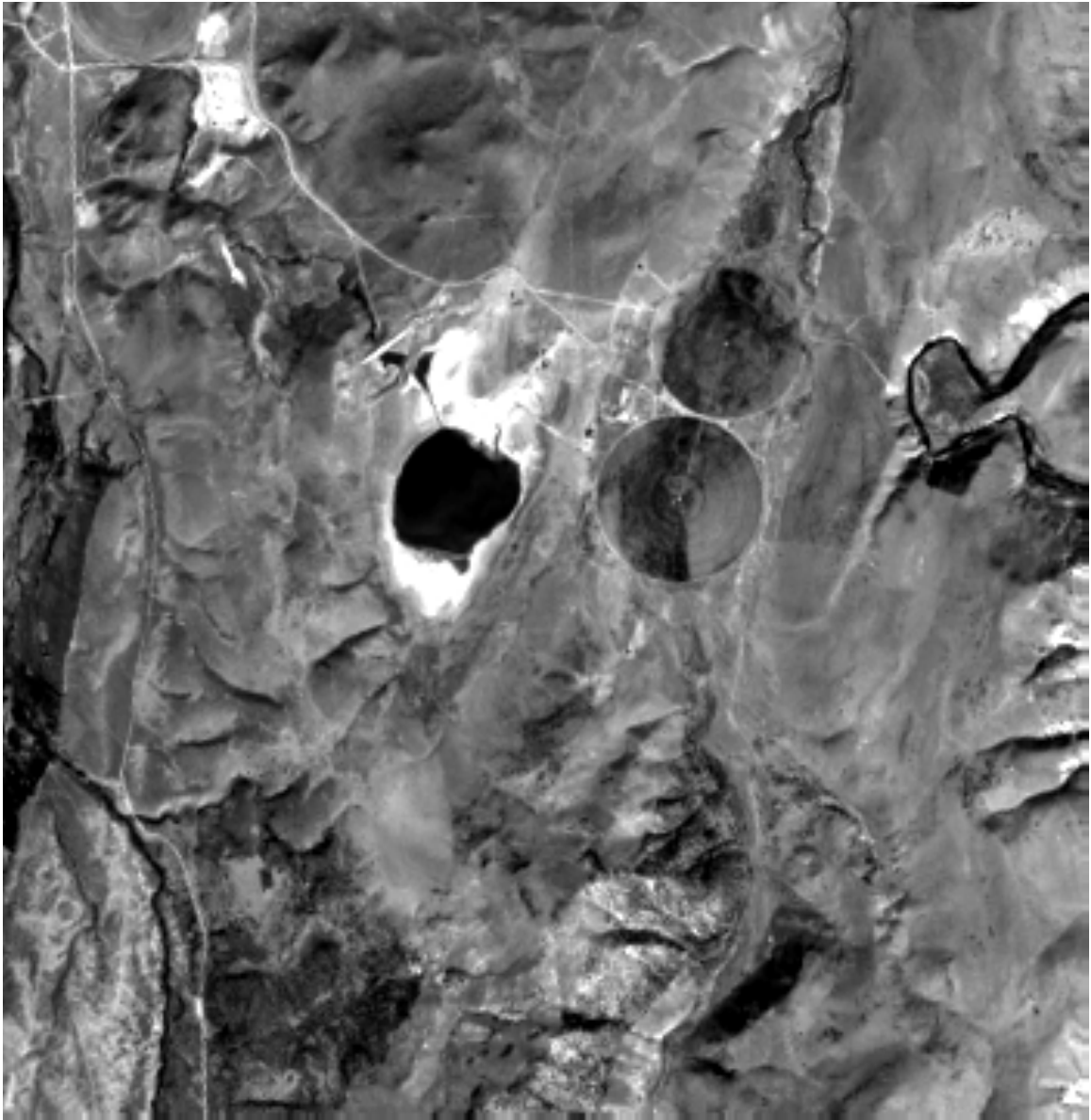}
			\caption{$\mathbf{Y}_{1}$}
			\label{fig:imT1_sen4_1}
	\end{subfigure}
		\begin{subfigure}{\subfwidth}
			\centering
			\includegraphics[width=\figsize]{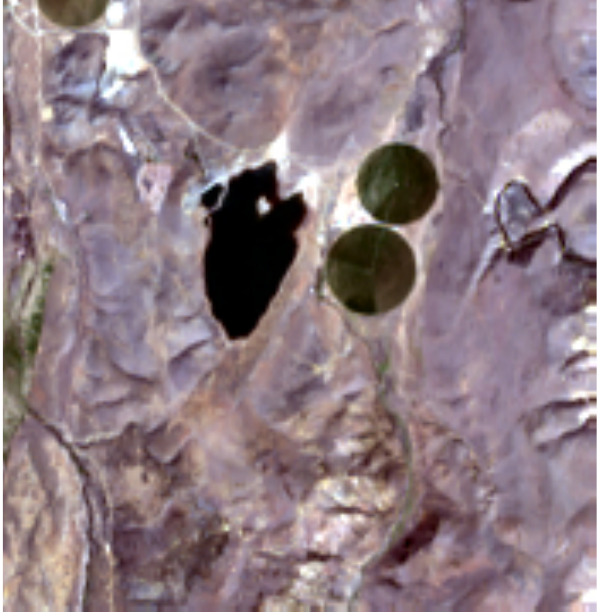}
			\caption{$\mathbf{Y}_{2}$}
			\label{fig:imT2_sen4_1}
	\end{subfigure}
		\begin{subfigure}{\subfwidth}
			\centering
			\includegraphics[width=\figsize]{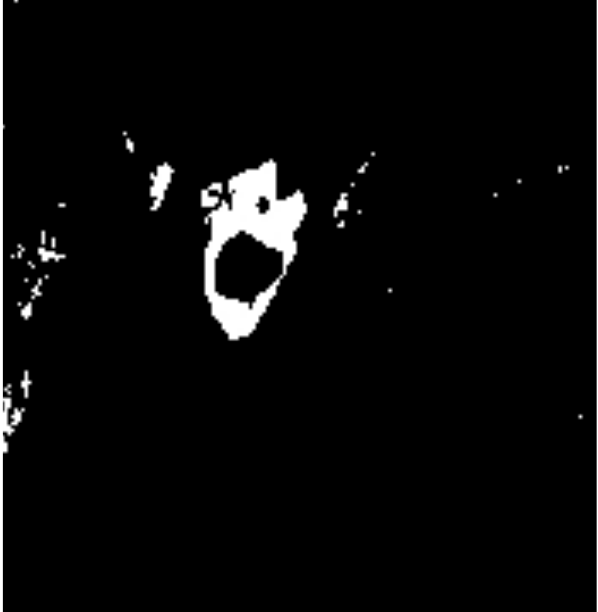}
			\caption{$\hat{\mathbf{d}}_{\mathrm{WC}}$}
			\label{fig:wcCD_sen4_1}
	\end{subfigure}
	\begin{subfigure}{\subfwidth}
			\centering
			\includegraphics[width=\figsize]{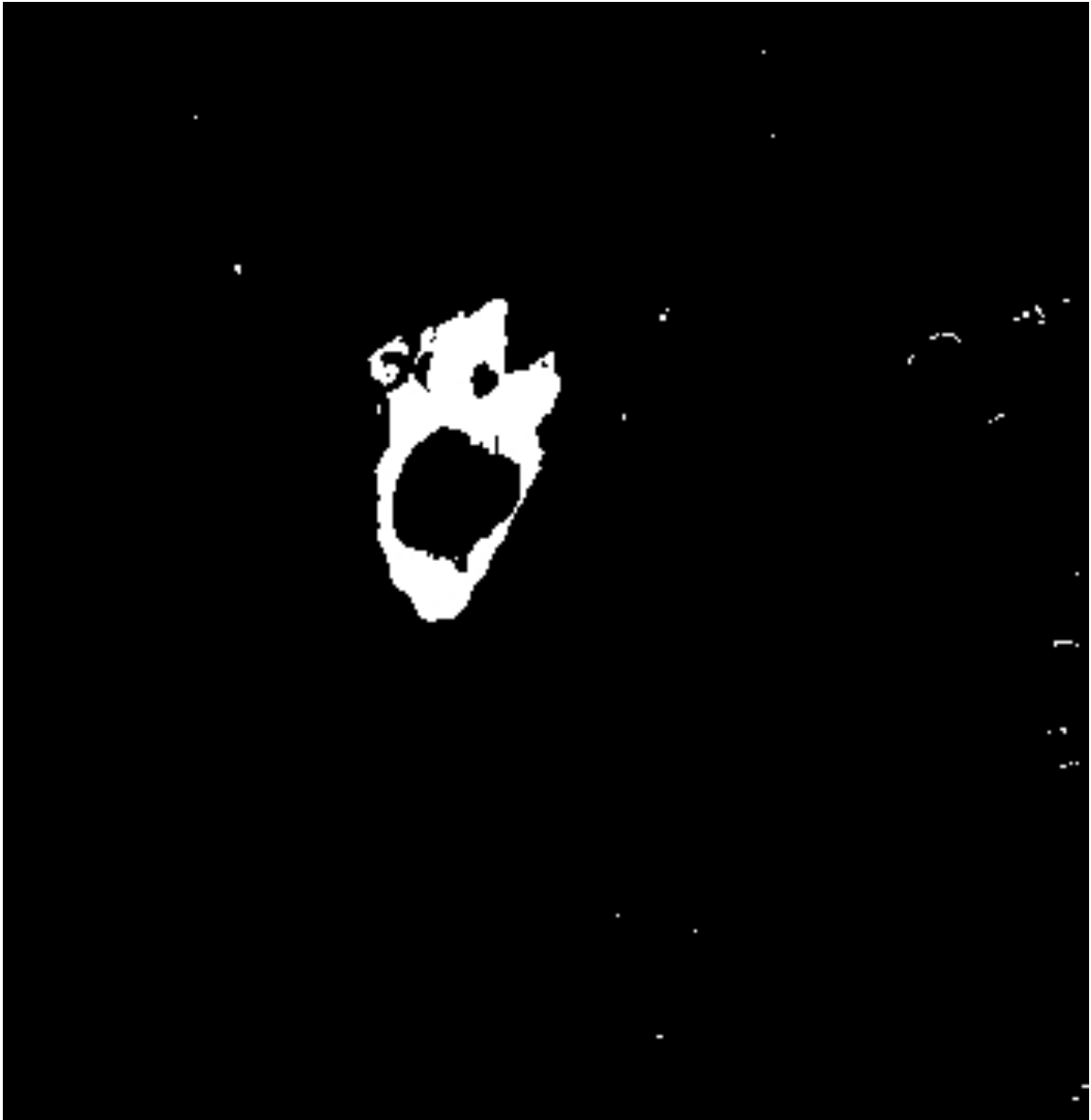}
			\caption{$\hat{\mathbf{d}}_{\mathrm{RF}}$}
			\label{fig:reCD_sen4_1}
	\end{subfigure}
	\caption{Scenario $\mathcal{S}_{4}$: \protect\subref{fig:imT1_sen4_1} Landsat-8 $15$m PAN observed image $\mathbf{Y}_{1}$ acquired on 09/22/2015, \protect\subref{fig:imT2_sen4_1} Landsat-8 $30$m MS-$3$ observed image $\mathbf{Y}_{2}$ acquired on 04/15/2015, \protect\subref{fig:wcCD_sen4_1} change mask $\hat{\mathbf{d}}_{\mathrm{WC}}$ estimated by the WC approach from a pair of $30$m PAN degraded images and \protect\subref{fig:reCD_sen4_1} change mask $\hat{\mathbf{d}}_{\mathrm{RF}}$ estimated by the proposed approach from a $15$m MS-$3$ change image $\Delta\hat{\mathbf{X}}$.}%
	\label{fig:sen4_1}%
\end{figure}

\begin{figure}[h!]
	\centering
	\begin{subfigure}{\subfwidth}
			\centering
			\includegraphics[width=\figsize]{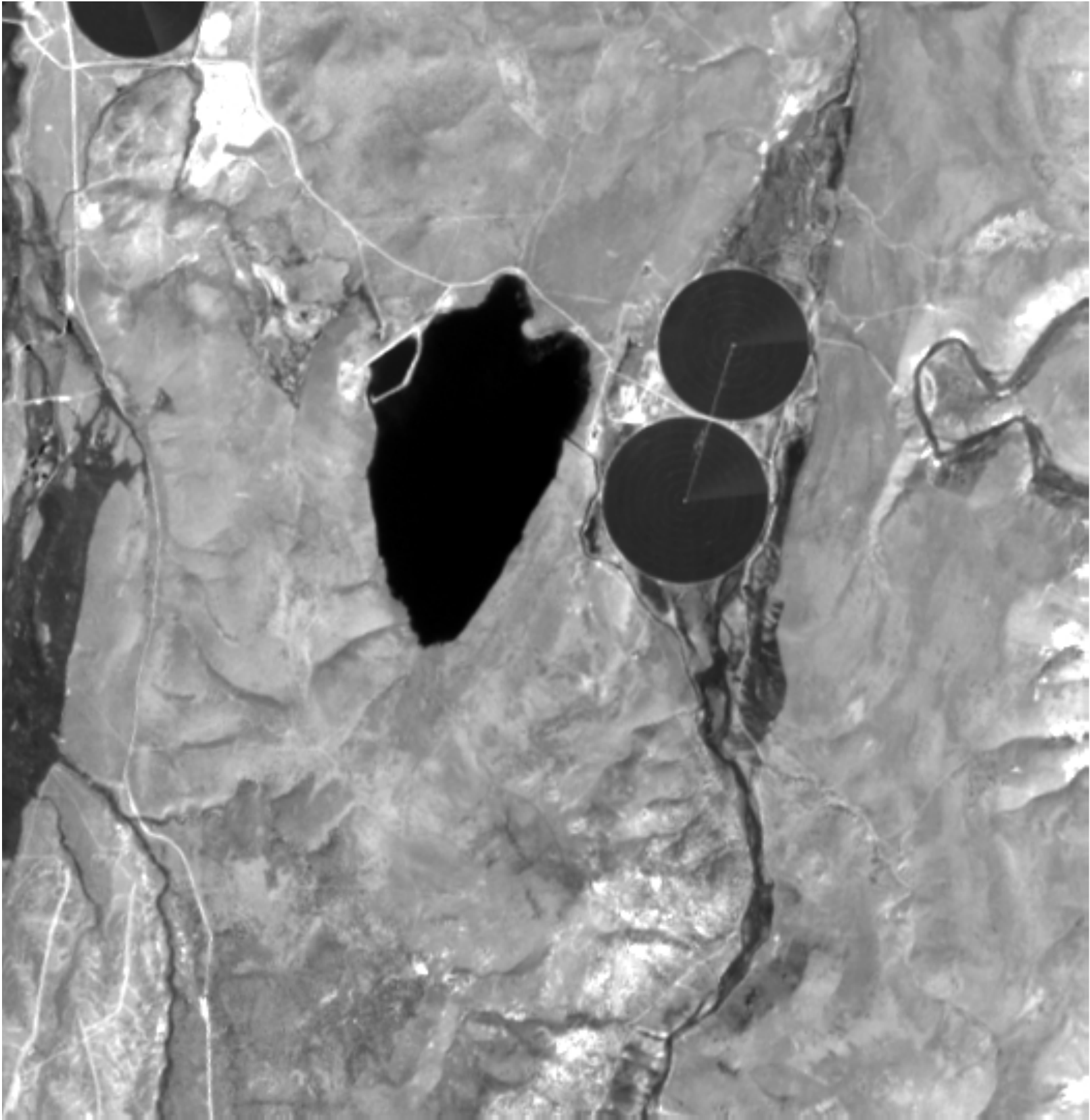}
			\caption{$\mathbf{Y}_{1}$}
			\label{fig:imT1_sen4_2}
	\end{subfigure}
	\begin{subfigure}{\subfwidth}
			\centering
			\includegraphics[width=\figsize]{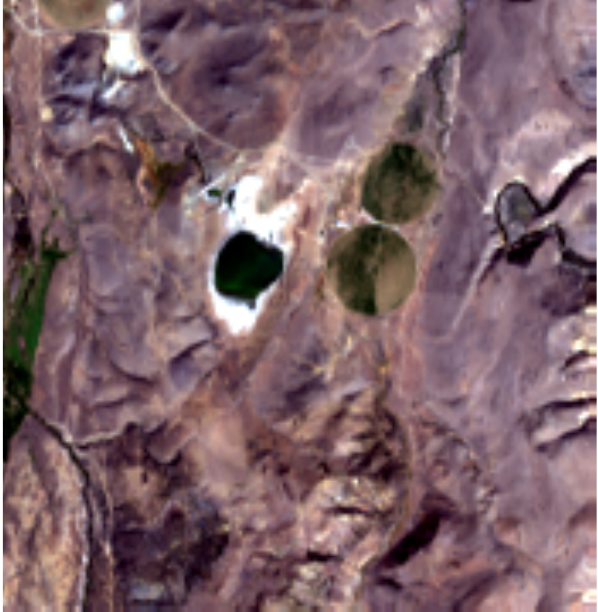}
			\caption{$\mathbf{Y}_{2}$}
			\label{fig:imT2_sen4_2}
	\end{subfigure}
	\begin{subfigure}{\subfwidth}
			\centering
			\includegraphics[width=\figsize]{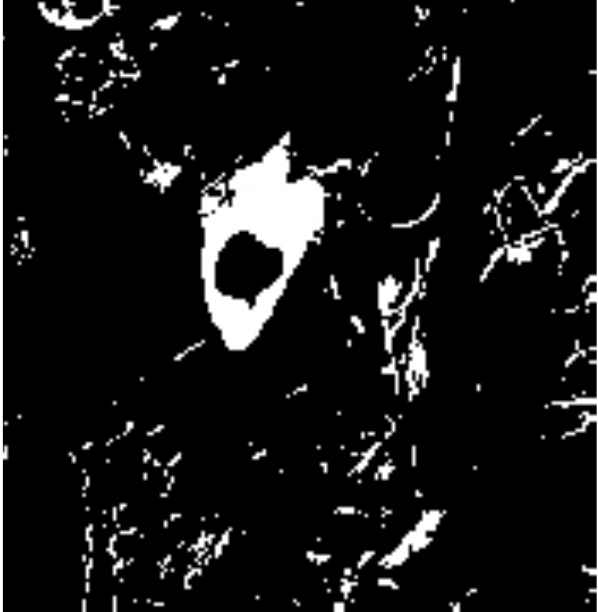}
			\caption{$\hat{\mathbf{d}}_{\mathrm{WC}}$}
			\label{fig:wcCD_sen4_2}
	\end{subfigure}
		\begin{subfigure}{\subfwidth}
			\centering
			\includegraphics[width=\figsize]{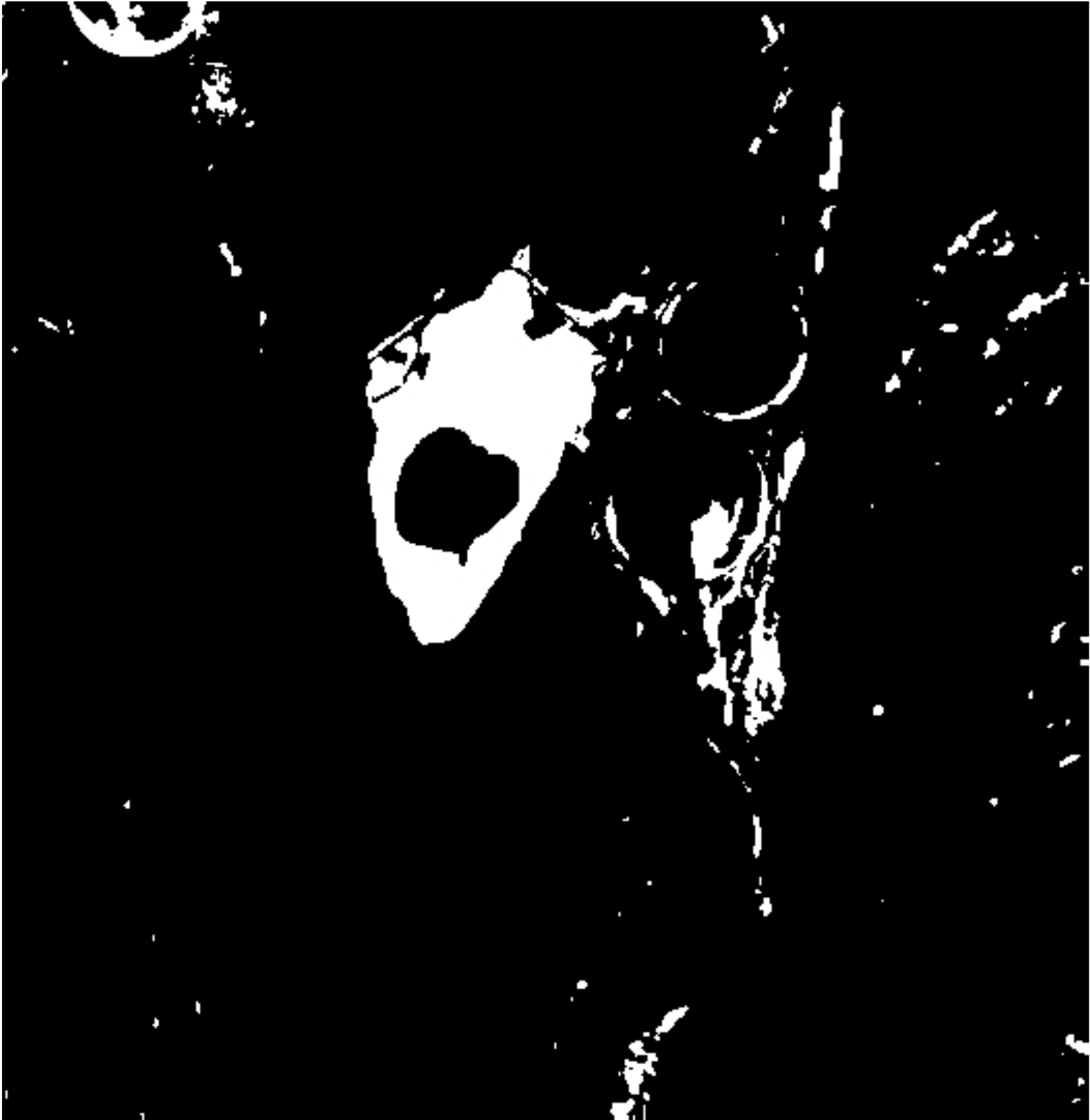}
			\caption{$\hat{\mathbf{d}}_{\mathrm{RF}}$}
			\label{fig:reCD_sen4_2}
	\end{subfigure}
	\caption{Scenario $\mathcal{S}_{4}$: \protect\subref{fig:imT1_sen4_2} EO-1 ALI $10$m PAN observed image $\mathbf{Y}_{1}$ acquired on 06/08/2011, \protect\subref{fig:imT2_sen4_2} Landsat-8 $30$m MS-$3$ observed image $\mathbf{Y}_{2}$ acquired on 09/22/2015, \protect\subref{fig:wcCD_sen4_2} change mask $\hat{\mathbf{d}}_{\mathrm{WC}}$ estimated by the WC approach from a pair of $30$m PAN degraded images \protect\subref{fig:reCD_sen4_2} change mask $\hat{\mathbf{d}}_{\mathrm{RF}}$ estimated by the proposed approach from a $10$m MS-$3$ change image $\Delta\hat{\mathbf{X}}$.}%
	\label{fig:sen4_2}%
\end{figure}

\begin{figure}[h!]
	\centering
	\begin{subfigure}{\subfwidth}
			\centering
			\includegraphics[width=\figsize]{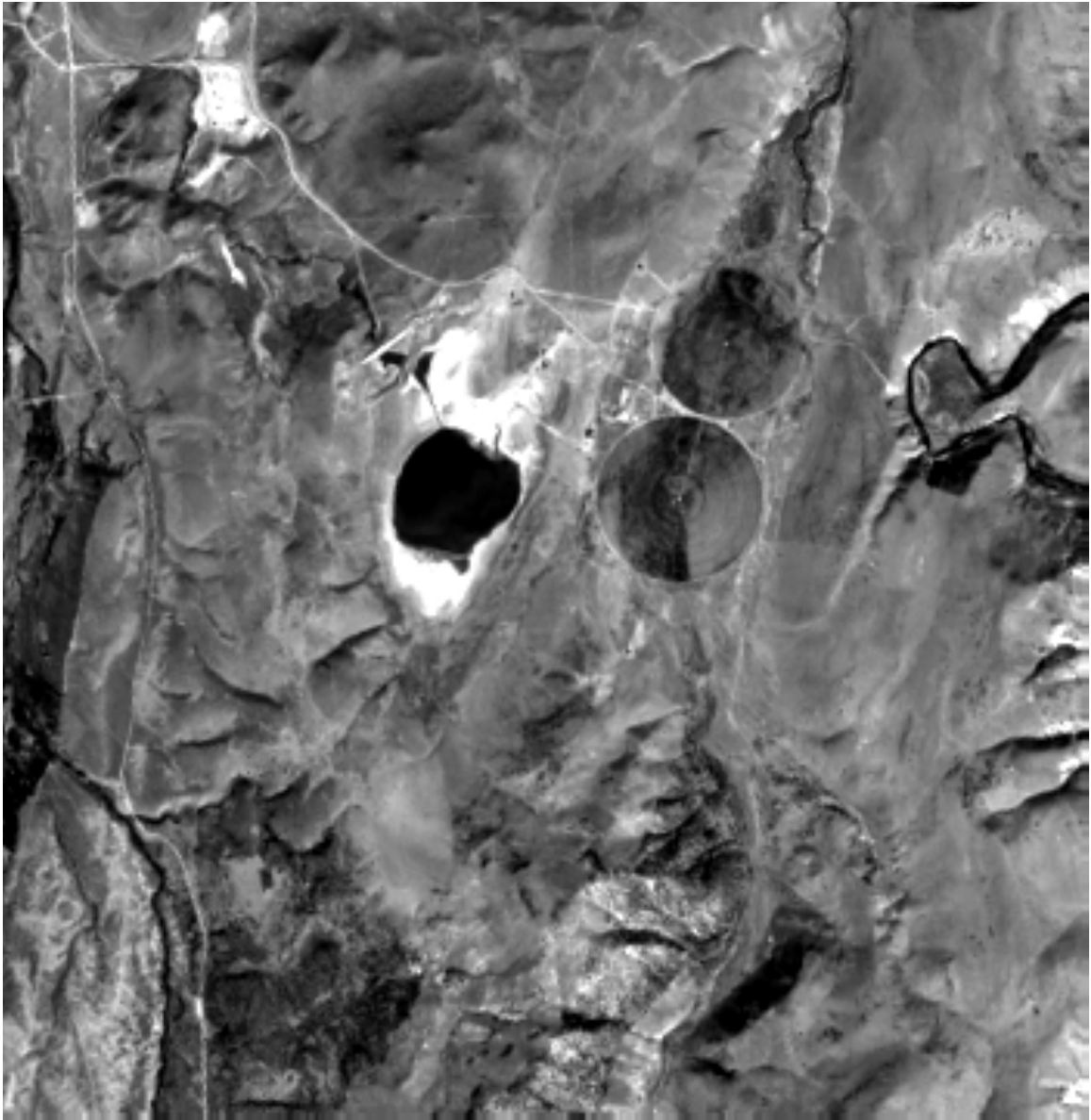}
			\caption{$\mathbf{Y}_{1}$}
			\label{fig:imT1_sen4_3}
	\end{subfigure}
	\begin{subfigure}{\subfwidth}
			\centering
			\includegraphics[width=\figsize]{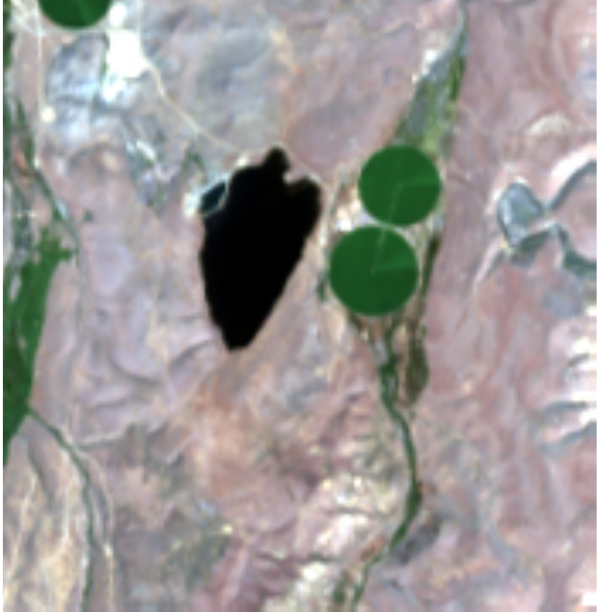}
			\caption{$\mathbf{Y}_{2}$}
			\label{fig:imT2_sen4_3}
	\end{subfigure}
		\begin{subfigure}{\subfwidth}
			\centering
			\includegraphics[width=\figsize]{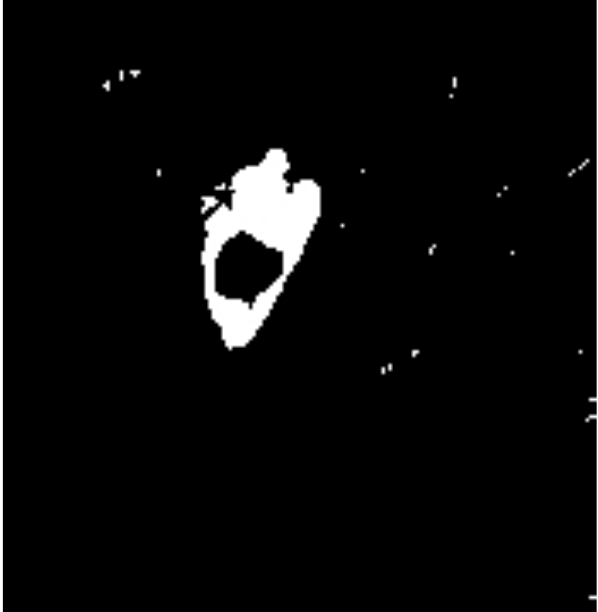}
			\caption{$\hat{\mathbf{d}}_{\mathrm{WC}}$}
			\label{fig:wcCD_sen4_3}
	\end{subfigure}
	\begin{subfigure}{\subfwidth}
			\centering
			\includegraphics[width=\figsize]{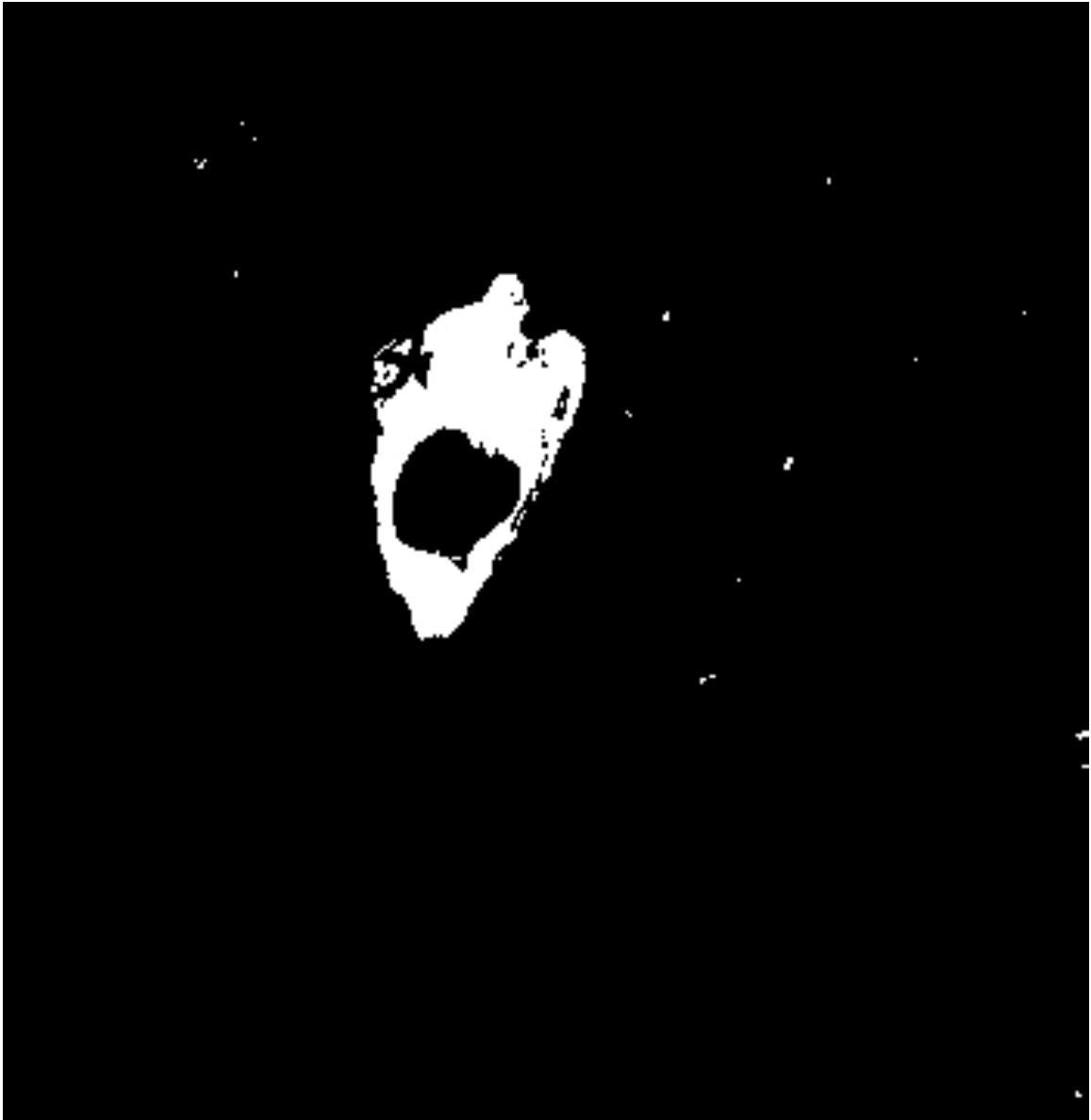}
			\caption{$\hat{\mathbf{d}}_{\mathrm{RF}}$}
			\label{fig:reCD_sen4_3}
	\end{subfigure}
	\caption{Scenario $\mathcal{S}_{4}$: \protect\subref{fig:imT1_sen4_3} Landsat-8 $15$m PAN  observed image $\mathbf{Y}_{1}$ acquired on 09/22/2015, \protect\subref{fig:imT2_sen4_3} EO-1 ALI $30$m MS-$3$ observed image $\mathbf{Y}_{2}$ acquired on 06/08/2011, \protect\subref{fig:wcCD_sen4_3} change mask $\hat{\mathbf{d}}_{\mathrm{WC}}$ estimated by the WC approach from a pair of $30$m PAN degraded images \protect\subref{fig:reCD_sen4_3} change mask $\hat{\mathbf{d}}_{\mathrm{RF}}$ estimated by the proposed approach from a $15$m MS-$3$ change image $\Delta\hat{\mathbf{X}}$.}%
	\label{fig:sen4_3}%
\end{figure}

This scenario has been deeply investigated in \cite{ferrarisrobust2017} who conducted a comprehensive analysis of the performance of the proposed RF-based CD method. This scenario corresponds to a more difficult CD investigation than all previous ones since the pair of observed images have not the same spatial neither spectral resolutions. As a consequence, the conventional WC approach is constrained to compare a spatially degraded version of one observed image with a spectrally degraded version of the other observed image. Irredeemably, these degradations result in a loss of spectral information, essential to assess the presence of change, and a loss of spatial information, required to accurately localize the possible changes. On the contrary, the proposed method is able to derive the CD mask from a change image characterized by the best of the spectral and spatial resolution of the observed images. Figures \ref{fig:sen4_1} to \ref{fig:sen4_3} depict the CD results obtained for three common configurations and illustrate the superiority of the proposed RF-based CD method.

\subsubsection{Scenario $\mathcal{S}_{5}$}

\begin{figure}[h!]
	\centering
	\begin{subfigure}{\subfwidth}
			\centering
			\includegraphics[width=\figsize]{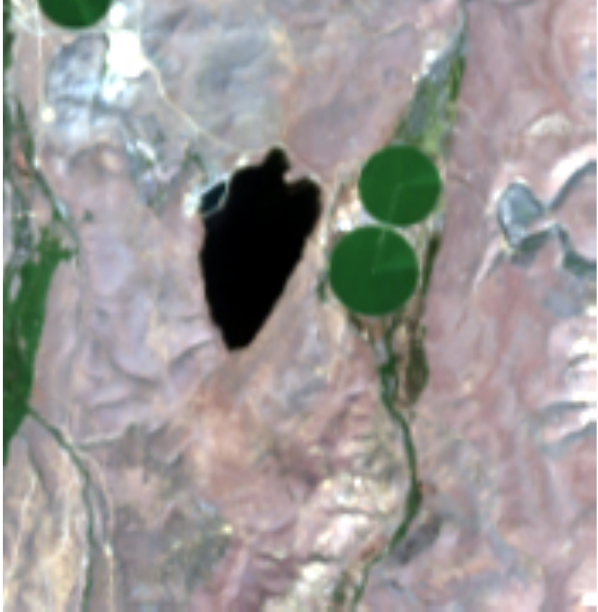}
			\caption{$\mathbf{Y}_{1}$}
			\label{fig:imT1_sen5_1}
	\end{subfigure}
	\begin{subfigure}{\subfwidth}
			\centering
			\includegraphics[width=\figsize]{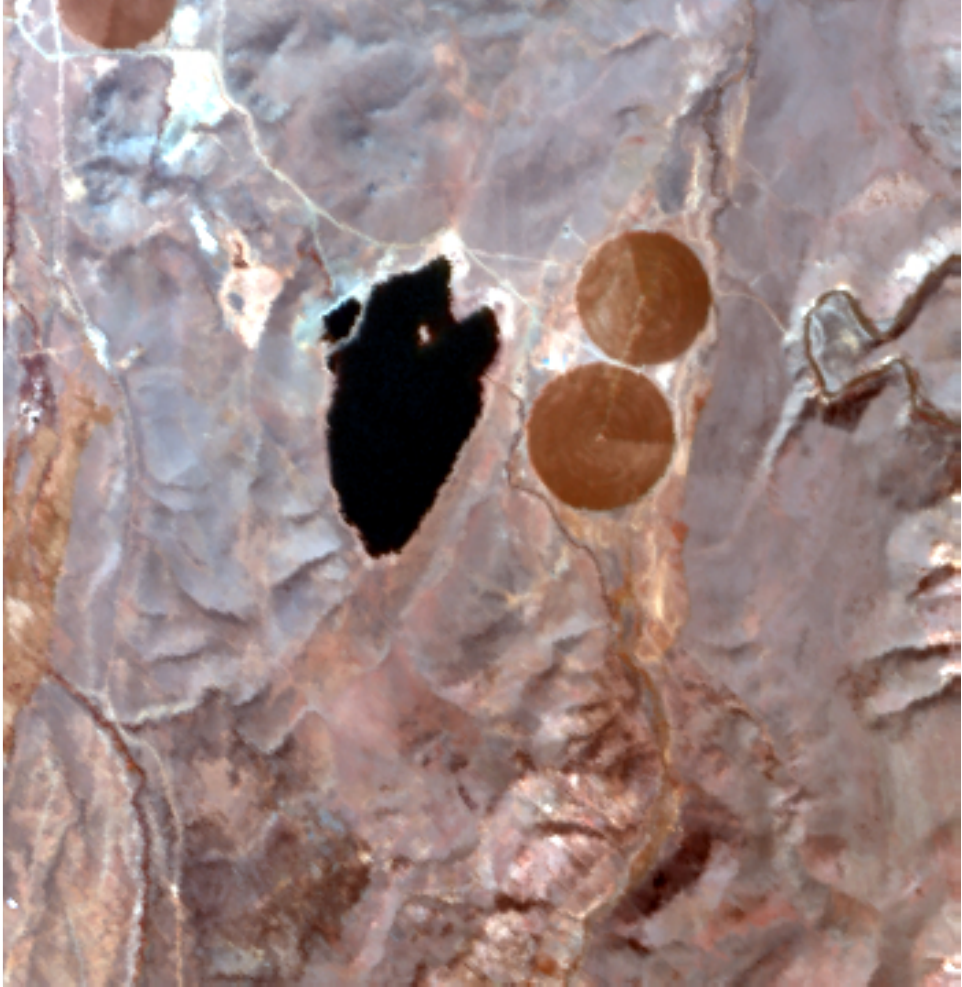}
			\caption{$\mathbf{Y}_{2}$}
			\label{fig:imT2_sen5_1}
	\end{subfigure}
		\begin{subfigure}{\subfwidth}
			\centering
			\includegraphics[width=\figsize]{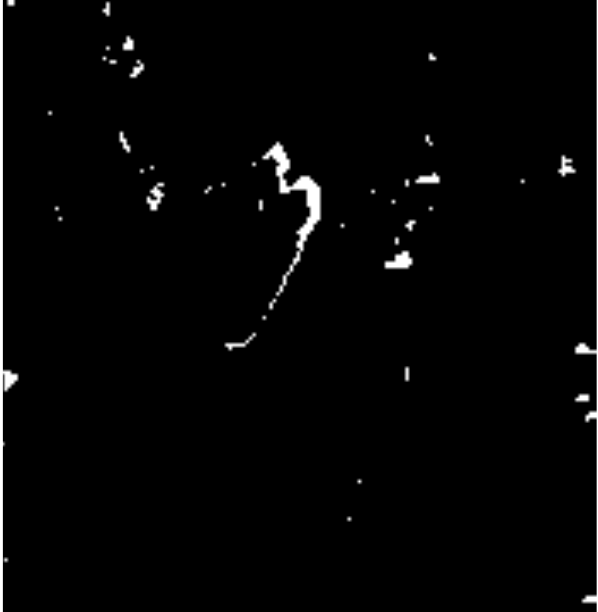}
			\caption{$\hat{\mathbf{d}}_{\mathrm{WC}}$}
			\label{fig:wcCD_sen5_1}
	\end{subfigure}
	\begin{subfigure}{\subfwidth}
			\centering
			\includegraphics[width=\figsize]{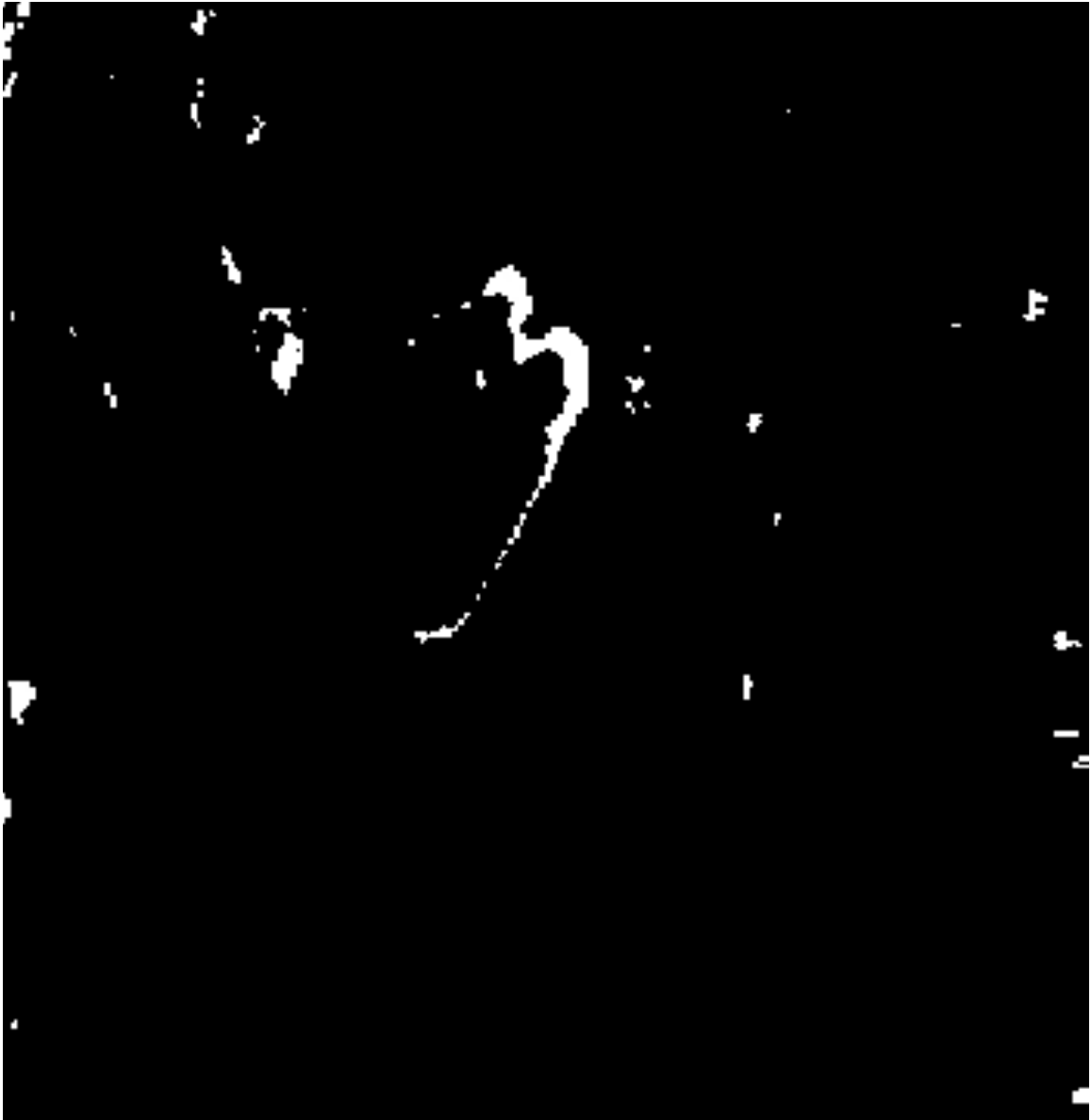}
			\caption{$\hat{\mathbf{d}}_{\mathrm{RF}}$}
			\label{fig:reCD_sen5_1}
	\end{subfigure}
	\caption{Scenario $\mathcal{S}_{5}$: \protect\subref{fig:imT1_sen5_1} EO-1 ALI $30$m MS-$3$ observed image $\mathbf{Y}_{1}$ acquired on 08/04/2011, \protect\subref{fig:imT2_sen5_1} AVIRIS $15$m HS-$29$ observed image $\mathbf{Y}_{2}$ acquired on 04/10/2014, \protect\subref{fig:wcCD_sen5_1} change mask $\hat{\mathbf{d}}_{\mathrm{WC}}$ estimated by the WC approach from a pair of $30$m MS-$3$ degraded images \protect\subref{fig:reCD_sen5_1} change mask $\hat{\mathbf{d}}_{\mathrm{RF}}$ estimated by the proposed approach from a $15$m HS-$29$ change image $\Delta\hat{\mathbf{X}}$.}%
	\label{fig:sen5_1}%
\end{figure}

\begin{figure}[h!]
	\centering
	\begin{subfigure}{\subfwidth}
			\centering
			\includegraphics[width=\figsize]{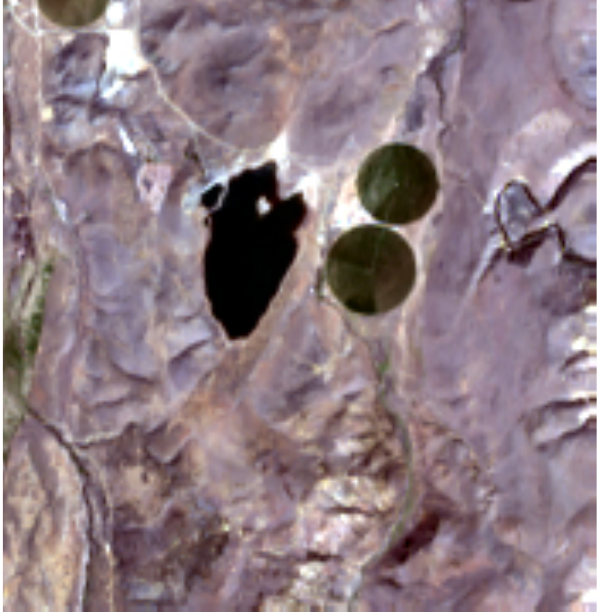}
			\caption{$\mathbf{Y}_{1}$}
			\label{fig:imT1_sen5_2}
	\end{subfigure}
	\begin{subfigure}{\subfwidth}
			\centering
			\includegraphics[width=\figsize]{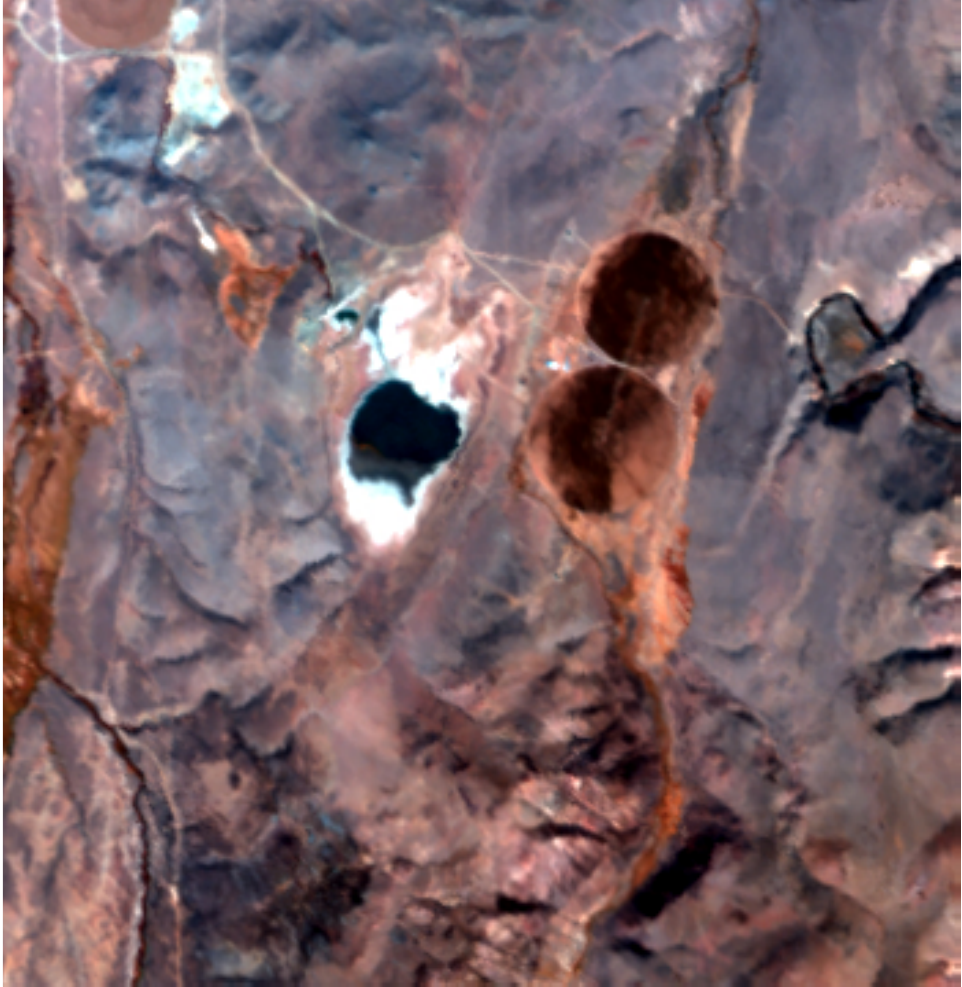}
			\caption{$\mathbf{Y}_{2}$}
			\label{fig:imT2_sen5_2}
	\end{subfigure}
		\begin{subfigure}{\subfwidth}
			\centering
			\includegraphics[width=\figsize]{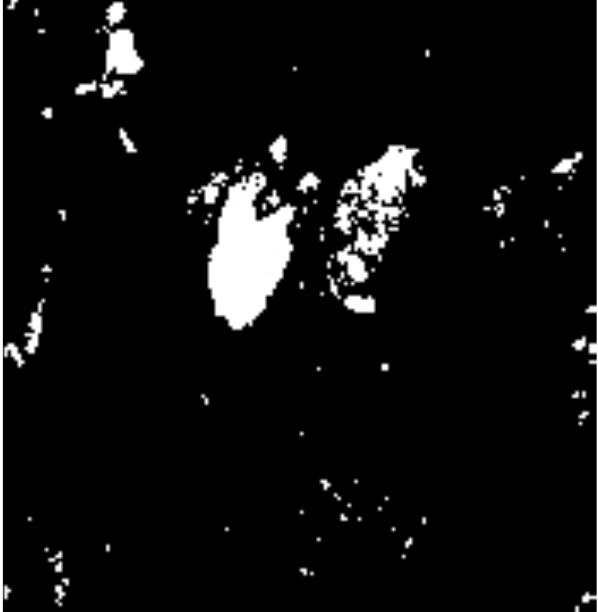}
			\caption{$\hat{\mathbf{d}}_{\mathrm{WC}}$}
			\label{fig:wcCD_sen5_2}
	\end{subfigure}
	\begin{subfigure}{\subfwidth}
			\centering
			\includegraphics[width=\figsize]{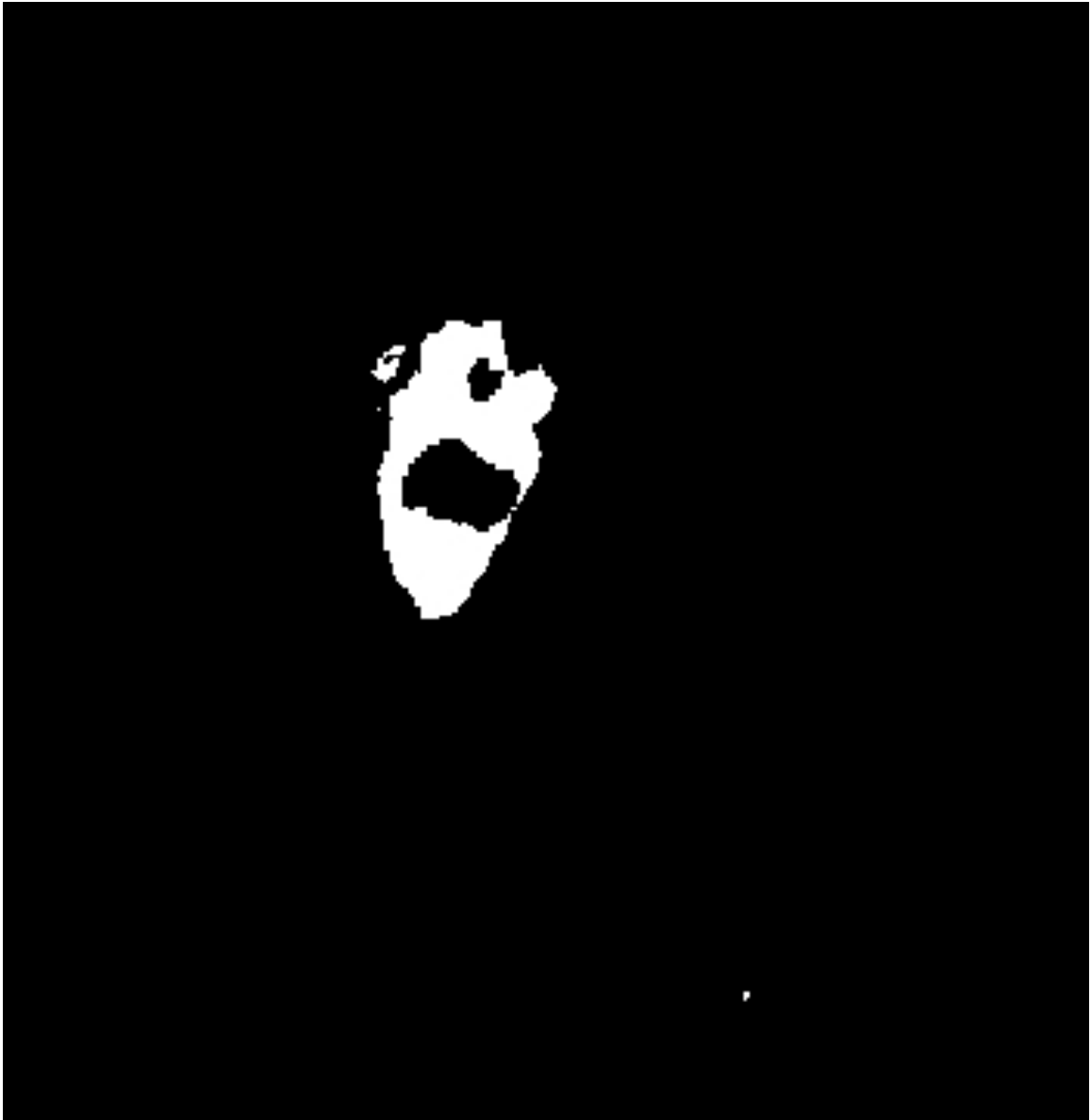}
			\caption{$\hat{\mathbf{d}}_{\mathrm{RF}}$}
			\label{fig:reCD_sen5_2}
	\end{subfigure}
	\caption{Scenario $\mathcal{S}_{5}$: \protect\subref{fig:imT1_sen5_2} Landsat-8 $30$m MS-$3$ observed image $\mathbf{Y}_{1}$ acquired on 04/15/2015, \protect\subref{fig:imT2_sen5_2} AVIRIS $15$m HS-$29$ observed image $\mathbf{Y}_{2}$ acquired on 09/19/2014, \protect\subref{fig:wcCD_sen5_2} change mask $\hat{\mathbf{d}}_{\mathrm{WC}}$ estimated by the WC approach from a pair of $30$m MS-$3$ degraded images and \protect\subref{fig:reCD_sen5_2} change mask $\hat{\mathbf{d}}_{\mathrm{RF}}$ estimated by the proposed approach from a $15$m HS-$29$ change image $\Delta\hat{\mathbf{X}}$.}%
	\label{fig:sen5_2}%
\end{figure}

As in the previous case, this scenario handles images which do not share the same spatial neither spectral resolutions. However, contrary to scenario $\mathcal{S}_{4}$, this scenario considers one of the two images of higher spatial and spectral resolution. Again, the WC is expected to be less reliable (in terms of decision and localization) due to the loss of spectral and spatial information consecutive to the degradations before conducting CVA. Figures \ref{fig:sen5_1} and \ref{fig:sen5_2} present the results obtained from two possible real configurations. As expected the proposed RF-based CD method provides visually more satisfactory results. In particular, as shown in Fig. \ref{fig:sen5_2}, the WC method is unable to accurately localize the change due to lake draught from the pair of multispectral and hyperspectral images.

%is a modified version of \emph{Scenario} 4. Here two observed images with different spatial and spectral resolutions are considered. However, differently from the previous scenario, one image has the lowest spatial and spectral resolution while the other one has the highest resolutions. As previously mentioned, this scenario is a particular instance of CD because, in a change free situation, the observed image that gathers all high spectral and spatial resolution information of the scene will be enough to represent it the entirely. Nevertheless, as changes are supposed to be present multitemporally, one must estimate them. Figures \ref{fig:sen5_1} and \ref{fig:sen5_2} present two possible real configurations for \emph{Scenario} 5. It is worthy to say that, similarly as \emph{Scenario} 4, both WC and the proposed method CD maps are estimated from variables, degraded observed image pair and $\Delta\mathbf{X}$ respectivelly, with different spatial and spectral resolution. The same remark, also, can be observed in this scenario, the proposed method dispose the highest resolutions CD maps reducing the false alarm/detection rate.

\subsubsection{Scenario $\mathcal{S}_{6}$}

\begin{figure}[h!]
	\centering
	\begin{subfigure}{\subfwidth}
			\centering
			\includegraphics[width=\figsize]{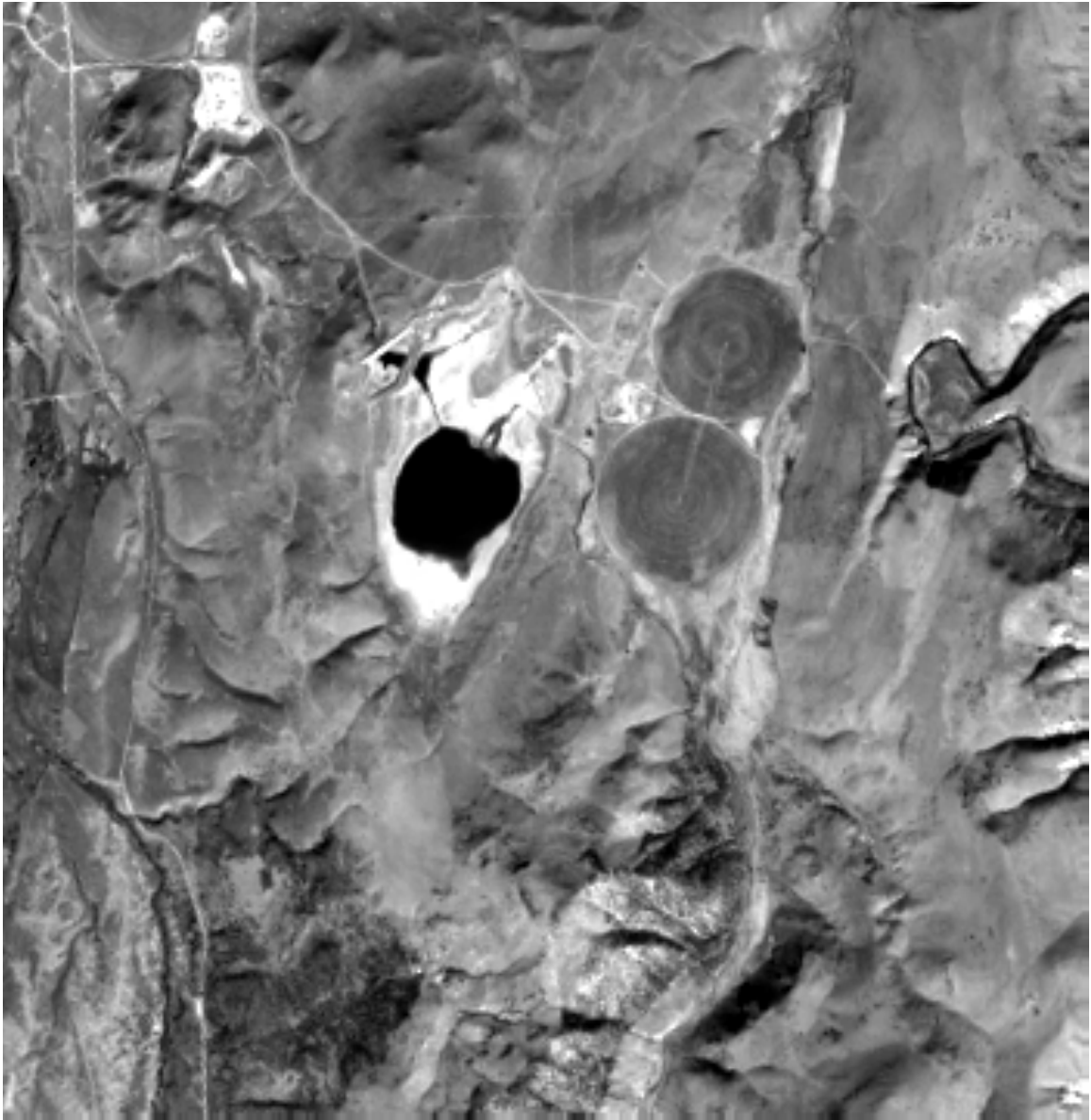}
			\caption{$\mathbf{Y}_{1}$}
			\label{fig:imT1_sen6_1}
	\end{subfigure}
	\begin{subfigure}{\subfwidth}
			\centering
			\includegraphics[width=\figsize]{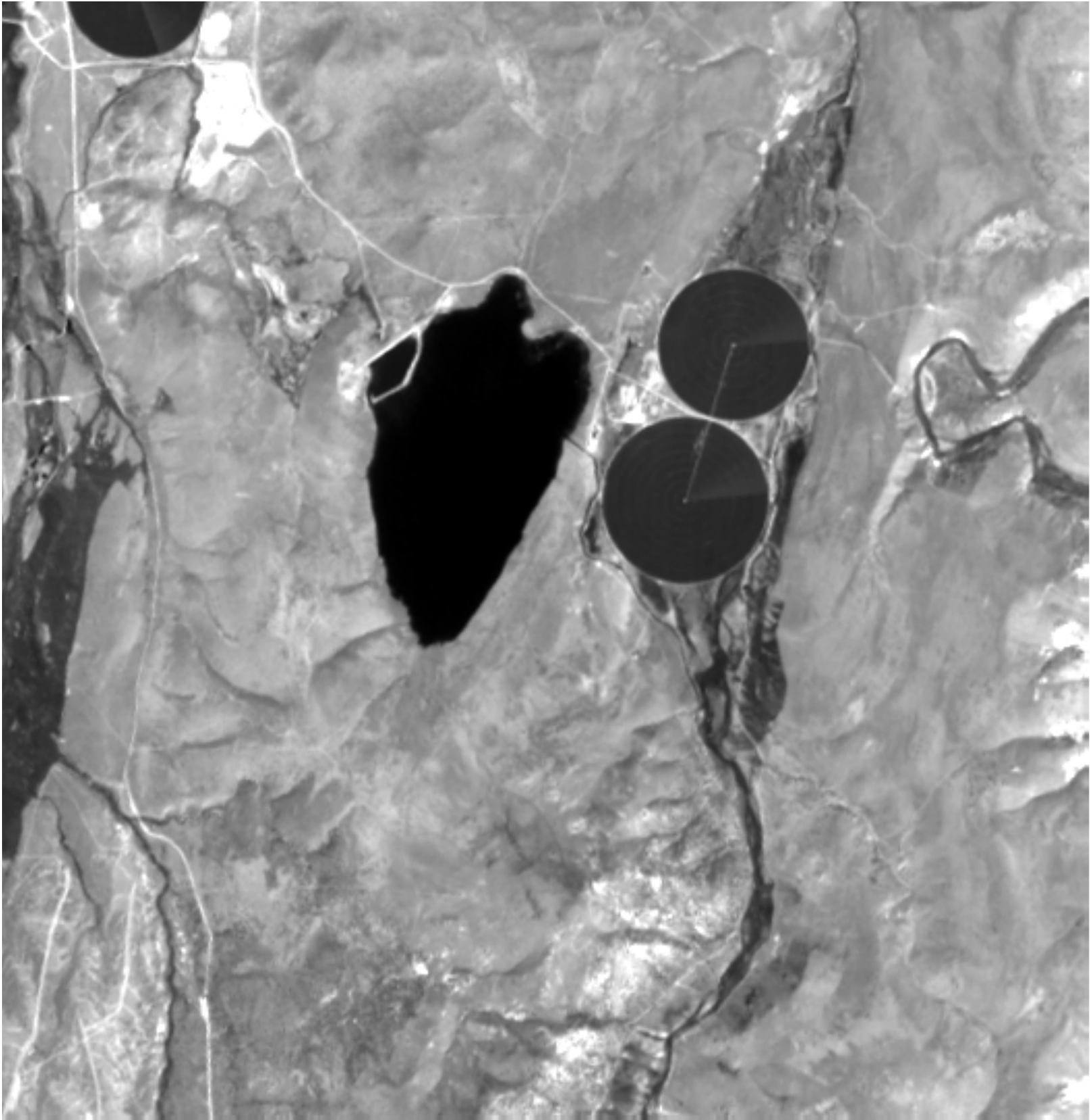}
			\caption{$\mathbf{Y}_{2}$}
			\label{fig:imT2_sen6_1}
	\end{subfigure}
		\begin{subfigure}{\subfwidth}
			\centering
			\includegraphics[width=\figsize]{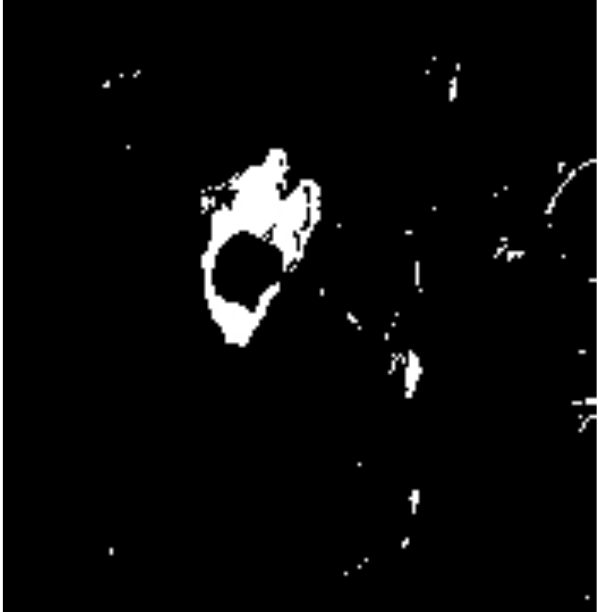}
			\caption{$\hat{\mathbf{d}}_{\mathrm{WC}}$}
			\label{fig:wcCD_sen6_1}
	\end{subfigure}
	\begin{subfigure}{\subfwidth}
			\centering
			\includegraphics[width=\figsize]{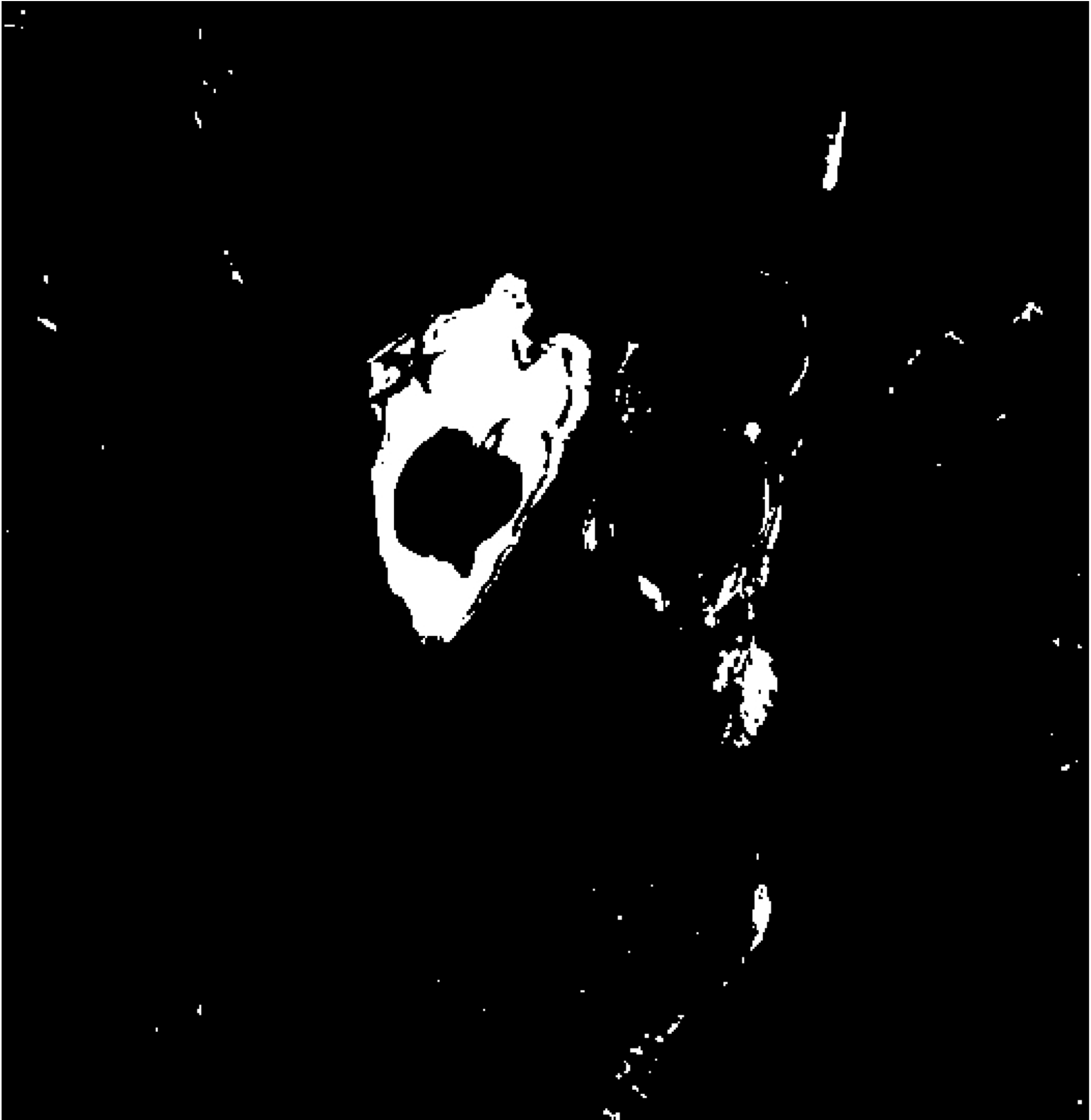}
			\caption{$\hat{\mathbf{d}}_{\mathrm{RF}}$}
			\label{fig:reCD_sen6_1}
	\end{subfigure}
	\caption{Scenario $\mathcal{S}_{6}$: \protect\subref{fig:imT1_sen6_1} Landsat-8 $15$m PAN observed image $\mathbf{Y}_{1}$ acquired on 10/18/2013, \protect\subref{fig:imT2_sen6_1} EO-1 ALI $10$m PAN observed image $\mathbf{Y}_{2}$ acquired on 08/04/2011, \protect\subref{fig:wcCD_sen6_1} change mask $\hat{\mathbf{d}}_{\mathrm{WC}}$ estimated by the WC approach from a pair of $30$m PAN degraded images \protect\subref{fig:reCD_sen6_1} change mask $\hat{\mathbf{d}}_{\mathrm{RF}}$ estimated by the proposed approach from $5$m PAN change image $\Delta\hat{\mathbf{X}}$.}%
	\label{fig:sen6_1}%
\end{figure}

This scenario represents a particular instance of scenario $\mathcal{S}_{3}$, i.e.,  with two observed images of different spatial resolution but same spectral resolution. Nevertheless, here, the two spatial resolutions are related by a non-integer downsampling ratio which precludes the use of a unique spatial degradation matrix in the proposed RF-based CD method. As detailed in paragraph \ref{subsec:sen6}, super-resolutions are conducted during the fusion and correction steps of the AM algorithm, which leads to a change image $\Delta\hat{\mathbf{X}}$ with a spatial resolution higher than the ones of the two observed images (defined as the greatest common divisor of the resolutions). For instance, Fig. \ref{fig:sen6_1} illustrates one possible configuration for which the observed images $\mathbf{Y}_1$ and $\mathbf{Y}_2$, depicted in Fig. \ref{fig:imT1_sen6_1} and \ref{fig:imT2_sen6_1}, are of $15$m and $10$m spatial resolutions, respectively. Thus the  change image $\Delta\hat{\mathbf{X}}$ and change mask $\hat{\mathbf{d}}_{\mathrm{RF}}$ estimated by the proposed method are at a $5$m resolution. Conversely, the WC method provides a CD map at a spatial resolution based on the least common multiple, which is, in this case, $30$m. The significantly higher spatial resolution of the change map is clear in Fig. \ref{fig:reCD_sen6_1}.

\subsubsection{Scenario $\mathcal{S}_{7}$}

\begin{figure}[h!]
	\centering
	\begin{subfigure}{\subfwidth}
			\centering
			\includegraphics[width=\figsize]{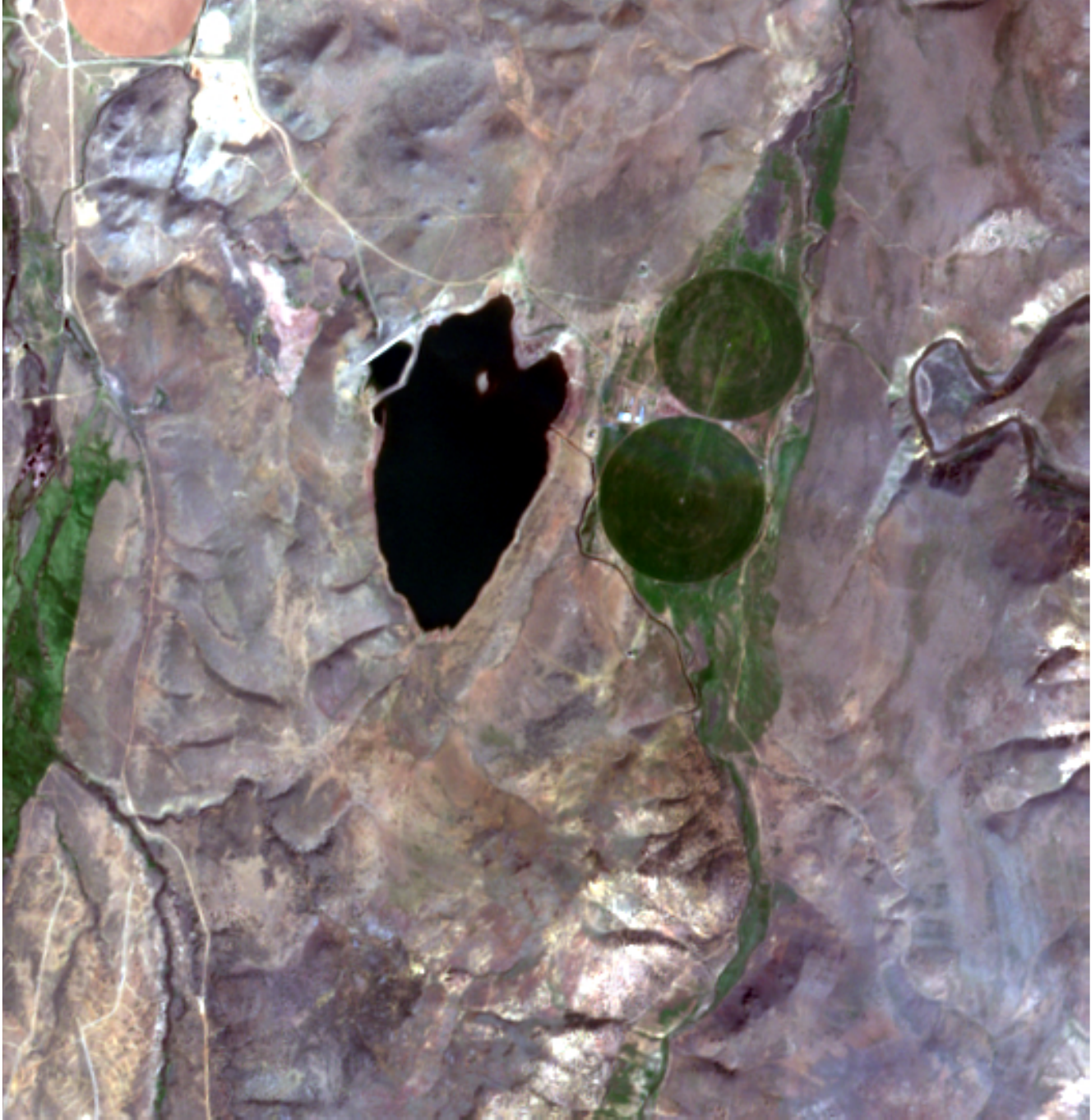}
			\caption{$\mathbf{Y}_{1}$}
			\label{fig:imT1_sen7_1}
	\end{subfigure}
	\begin{subfigure}{\subfwidth}
			\centering
			\includegraphics[width=\figsize]{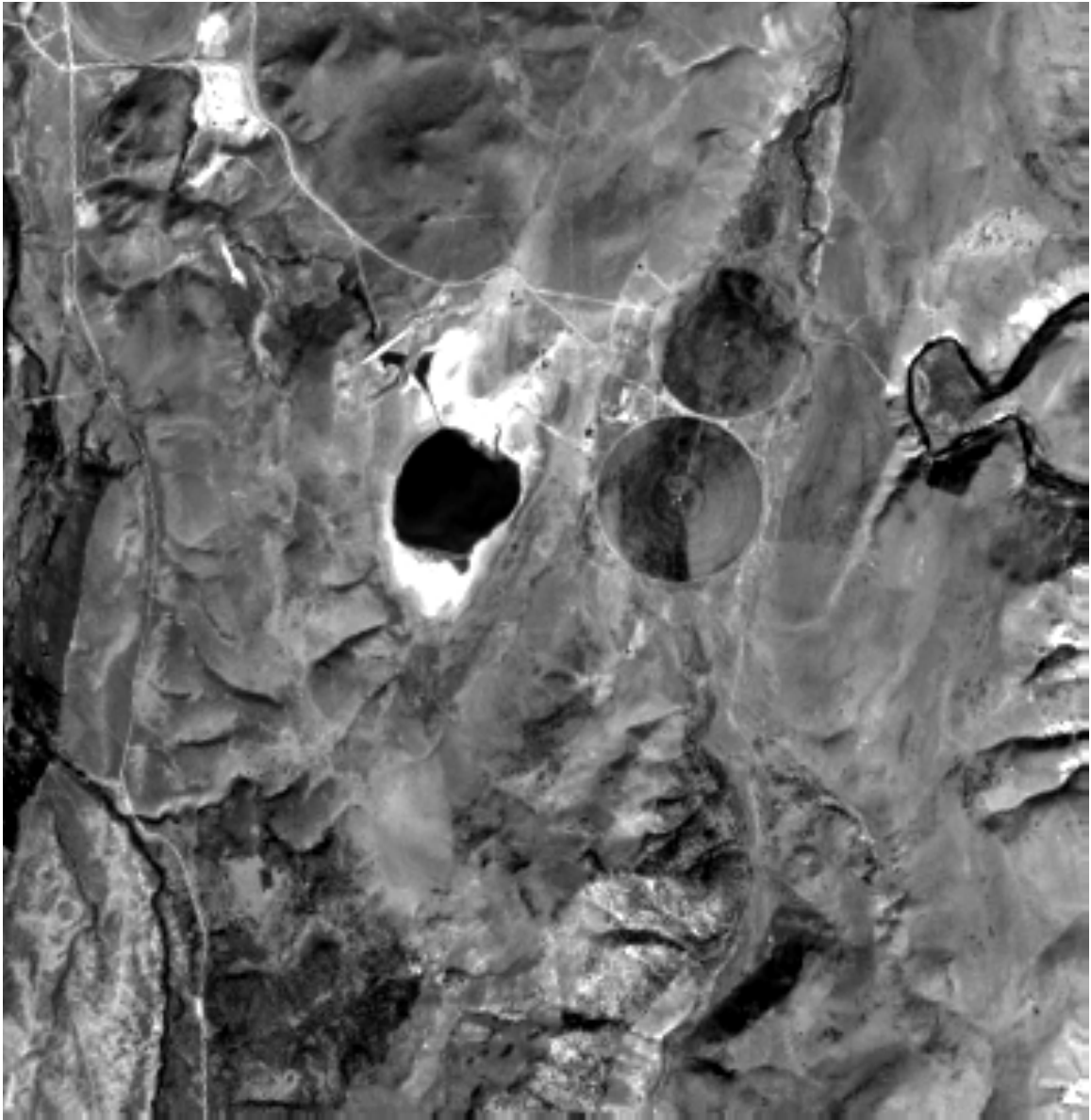}
			\caption{$\mathbf{Y}_{2}$}
			\label{fig:imT2_sen7_1}
	\end{subfigure}
		\begin{subfigure}{\subfwidth}
			\centering
			\includegraphics[width=\figsize]{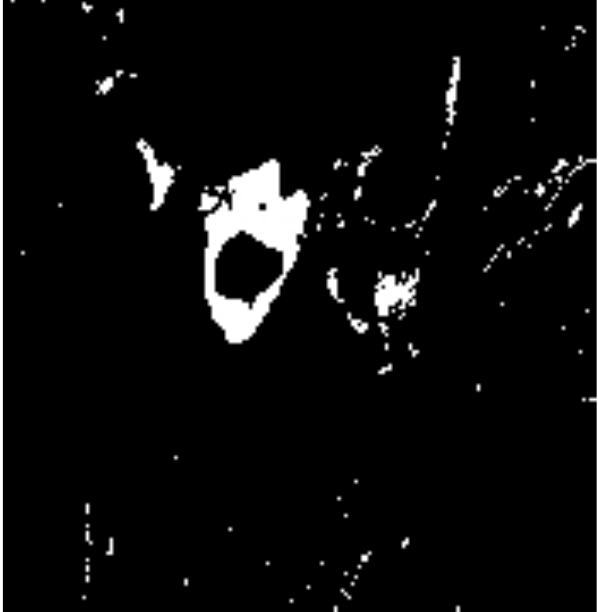}
			\caption{$\hat{\mathbf{d}}_{\mathrm{WC}}$}
			\label{fig:wcCD_sen7_1}
	\end{subfigure}
	\begin{subfigure}{\subfwidth}
			\centering
			\includegraphics[width=\figsize]{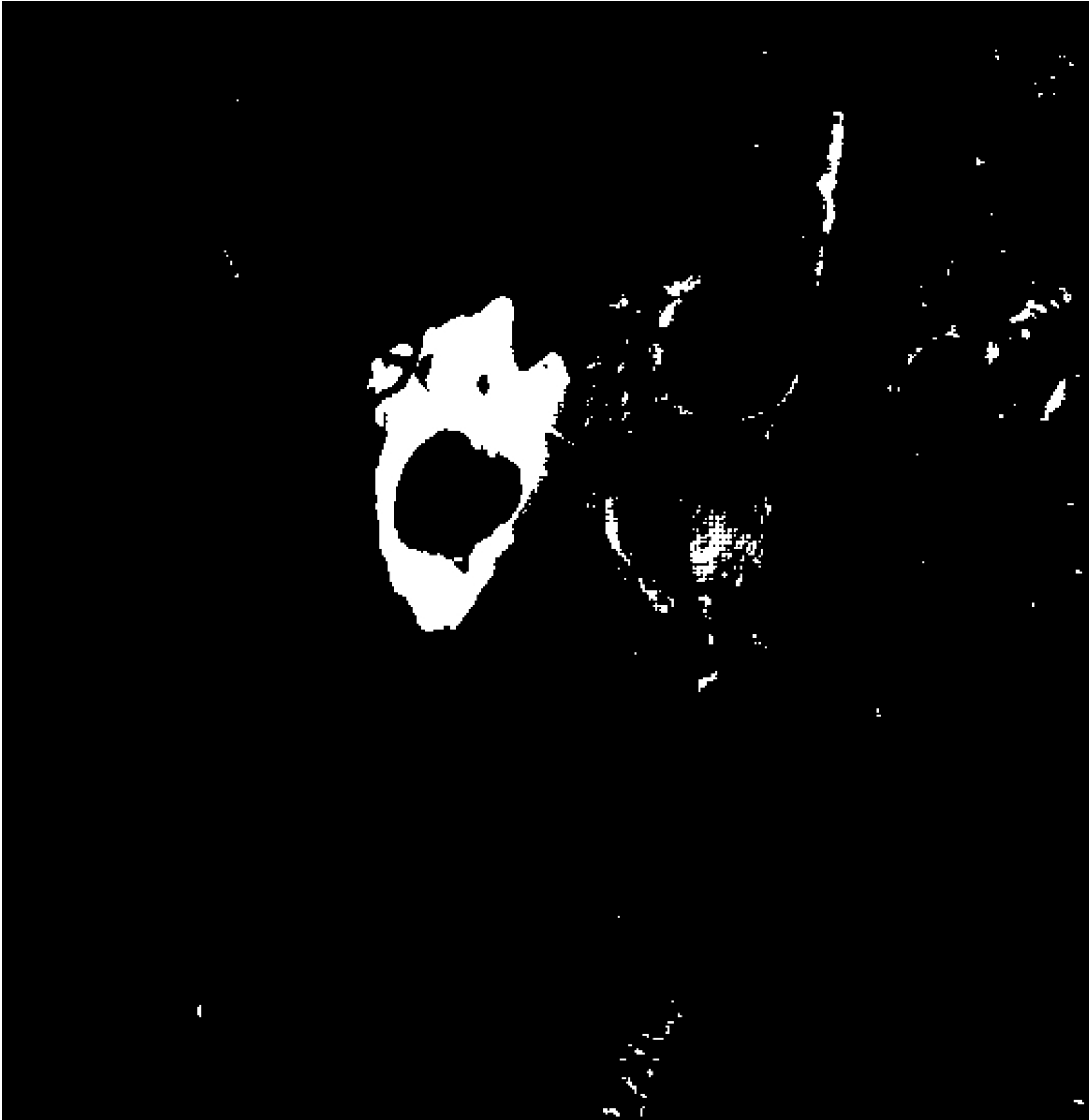}
			\caption{$\hat{\mathbf{d}}_{\mathrm{RF}}$}
			\label{fig:reCD_sen7_1}
	\end{subfigure}
	\caption{Scenario $\mathcal{S}_{7}$: \protect\subref{fig:imT1_sen7_1} Sentinel-2 $10$m MS-$3$ observed image $\mathbf{Y}_{1}$ acquired on 04/12/2016, \protect\subref{fig:imT2_sen7_1} Landsat-8 $15$m PAN observed image $\mathbf{Y}_{2}$ acquired on 09/22/2015, \protect\subref{fig:wcCD_sen7_1} change mask $\hat{\mathbf{d}}_{\mathrm{WC}}$ estimated by the WC approach from a pair of $30$m PAN degraded images and \protect\subref{fig:reCD_sen7_1} change mask $\hat{\mathbf{d}}_{\mathrm{RF}}$ estimated by the proposed approach from $5$m MS-$3$ change image $\Delta\hat{\mathbf{X}}$.}%
	\label{fig:sen7_1}%
\end{figure}

This scenario consists in a more challenging context than scenario $\mathcal{S}_{6}$ since, in addition to the non-integer relative downsampling factor, the two observed images do not share the same spectral resolution. As before, the change image $\Delta\hat{\mathbf{X}}$ and the binary change mask $\hat{\mathbf{d}}_{\mathrm{RF}}$ estimated by the proposed RF-based CD method are defined at a higher spatial resolution than both resolutions of the observed image. Figure \ref{fig:sen7_1} presents one example of this scenario. As expected, the proposed method benefits from the estimated highest spatial and spectral resolution change image $\Delta\mathbf{X}$ to localize the changes, contrary to the WC method which can only exploit a pair of spatially and spectrally degrade images. This important dual resolution gap contributes a lot in the observed differences on the false alarm and good detection rates.

%and $\mathcal{S}_{6}$. It stands for two images with different spectral resolutions and spatial resolutions attached by a non-integer downsampling factor. The same stategy as for \emph{Scenario} 6 is employed here in order to overcome the non-integer downsampling factor problem. Additionaly, a spectral superresolution step is required in the proposed method in order to bring all estimated variables to the same spectral resolution, which increase the difficulty to solve the problem.
%
%Figure \ref{fig:sen7_1} represent on example of this scenario together with CD map results of the proposed and WC method. Like the previous scenario, the proposed method present the highest spatial and spectral resolution CD estimation $\Delta\mathbf{X}$ , in this case, a 5m MS image, againist the 30m PAN of the WC method estimation. This important dual resolution gap contributes a lot in the difference on the false alarm/detection rate between both results proving, once more, the efficiency and accuracy of the proposed method.

\subsubsection{Scenario $\mathcal{S}_{8}$}

\begin{figure}[h!]
	\centering
	\begin{subfigure}{\subfwidth}
			\centering
			\includegraphics[width=\figsize]{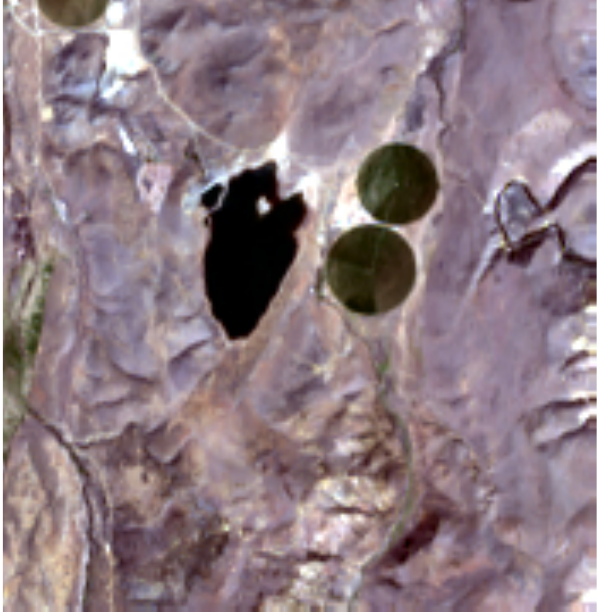}
			\caption{$\mathbf{Y}_{1}$}
			\label{fig:imT1_sen8_1}
	\end{subfigure}
	\begin{subfigure}{\subfwidth}
			\centering
			\includegraphics[width=\figsize]{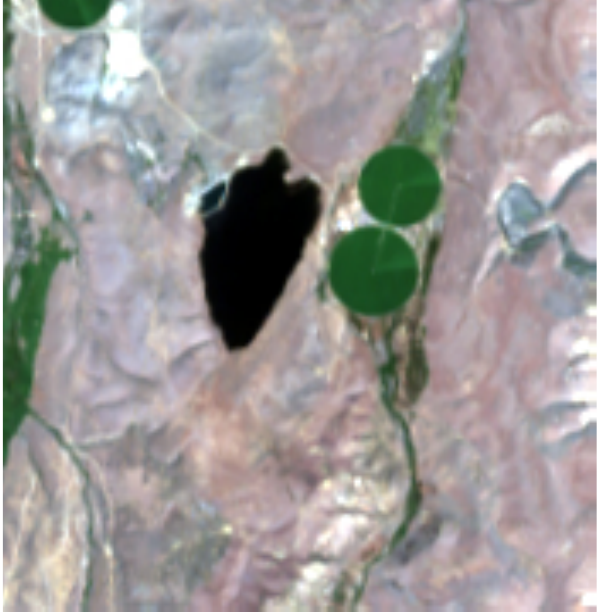}
			\caption{$\mathbf{Y}_{2}$}
			\label{fig:imT2_sen8_1}
	\end{subfigure}
		\begin{subfigure}{\subfwidth}
			\centering
			\includegraphics[width=\figsize]{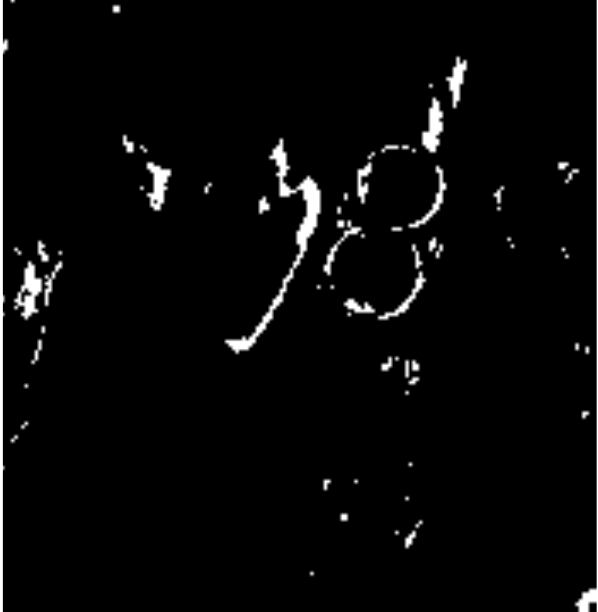}
			\caption{$\hat{\mathbf{d}}_{\mathrm{WC}}$}
			\label{fig:wcCD_sen8_1}
	\end{subfigure}
	\begin{subfigure}{\subfwidth}
			\centering
			\includegraphics[width=\figsize]{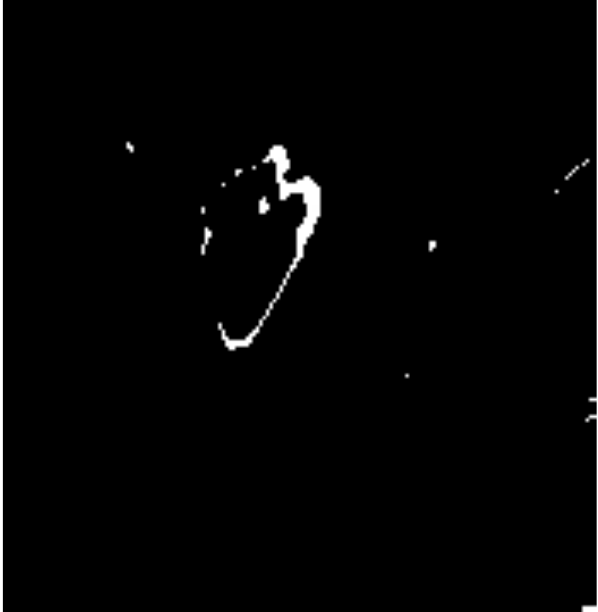}
			\caption{$\hat{\mathbf{d}}_{\mathrm{RF}}$}
			\label{fig:reCD_sen8_1}
	\end{subfigure}
	\caption{Scenario $\mathcal{S}_{8}$: \protect\subref{fig:imT1_sen8_1}  Landsat-8 $30$m MS-$8$ observed image $\mathbf{Y}_{1}$ acquired on 04/15/2015, \protect\subref{fig:imT2_sen8_1} EO-1 ALI $30$m MS-$9$ observed image  $\mathbf{Y}_{2}$ acquired on 06/08/2011, \protect\subref{fig:wcCD_sen8_1} change mask $\hat{\mathbf{d}}_{\mathrm{WC}}$ estimated by the WC approach from a pair of $30$m MS-$7$  degraded images \protect\subref{fig:reCD_sen8_1} change mask $\hat{\mathbf{d}}_{\mathrm{RF}}$ estimated by the proposed approach from $30$m MS-$10$ change image $\Delta\hat{\mathbf{X}}$.}%
	\label{fig:sen8_1}%
\end{figure}

Scenario $\mathcal{S}_{8}$ generalizes scenario $\mathcal{S}_{2}$, with the particular case of presence of non-overlapping bands in the two sensor spectral responses, which requires the simultaneous use of two spectral degradation matrices in the proposed RF method. Figure \ref{fig:sen8_1} provides one instance of this scenario. Due to the presence of non-overlapping bands, before conducting CVA, the WC requires to ignore the spectral bands which are not commonly shared by the two observed images. Conversely, by fully exploiting the whole available spectral information, the proposed method combines the overlapped bands and the non-overlapping bands to estimate a change image $\Delta\hat{\mathbf{X}}$ of higher spectral resolution than the two observed images. This higher amount of information leads to visually more consistent results in Fig. \ref{fig:reCD_sen8_1}.

\subsubsection{Scenario $\mathcal{S}_{9}$}

\begin{figure}[h!]
	\centering
	\begin{subfigure}{\subfwidth}
			\centering
			\includegraphics[width=\figsize]{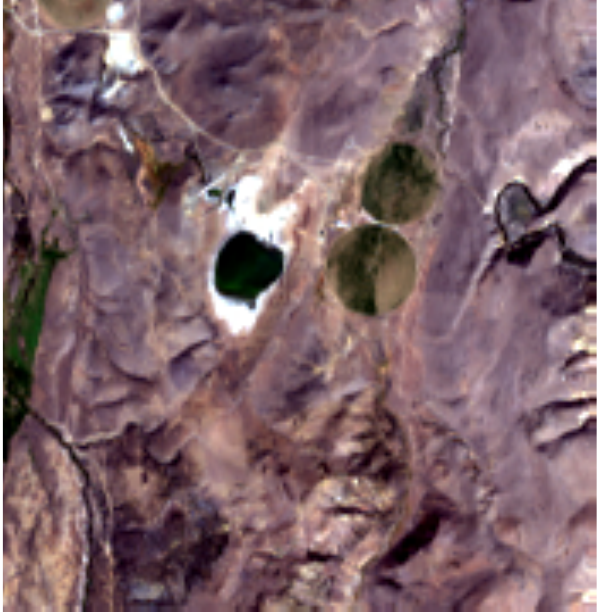}
			\caption{$\mathbf{Y}_{1}$}
			\label{fig:imT1_sen9_1}
	\end{subfigure}
	\begin{subfigure}{\subfwidth}
			\centering
			\includegraphics[width=\figsize]{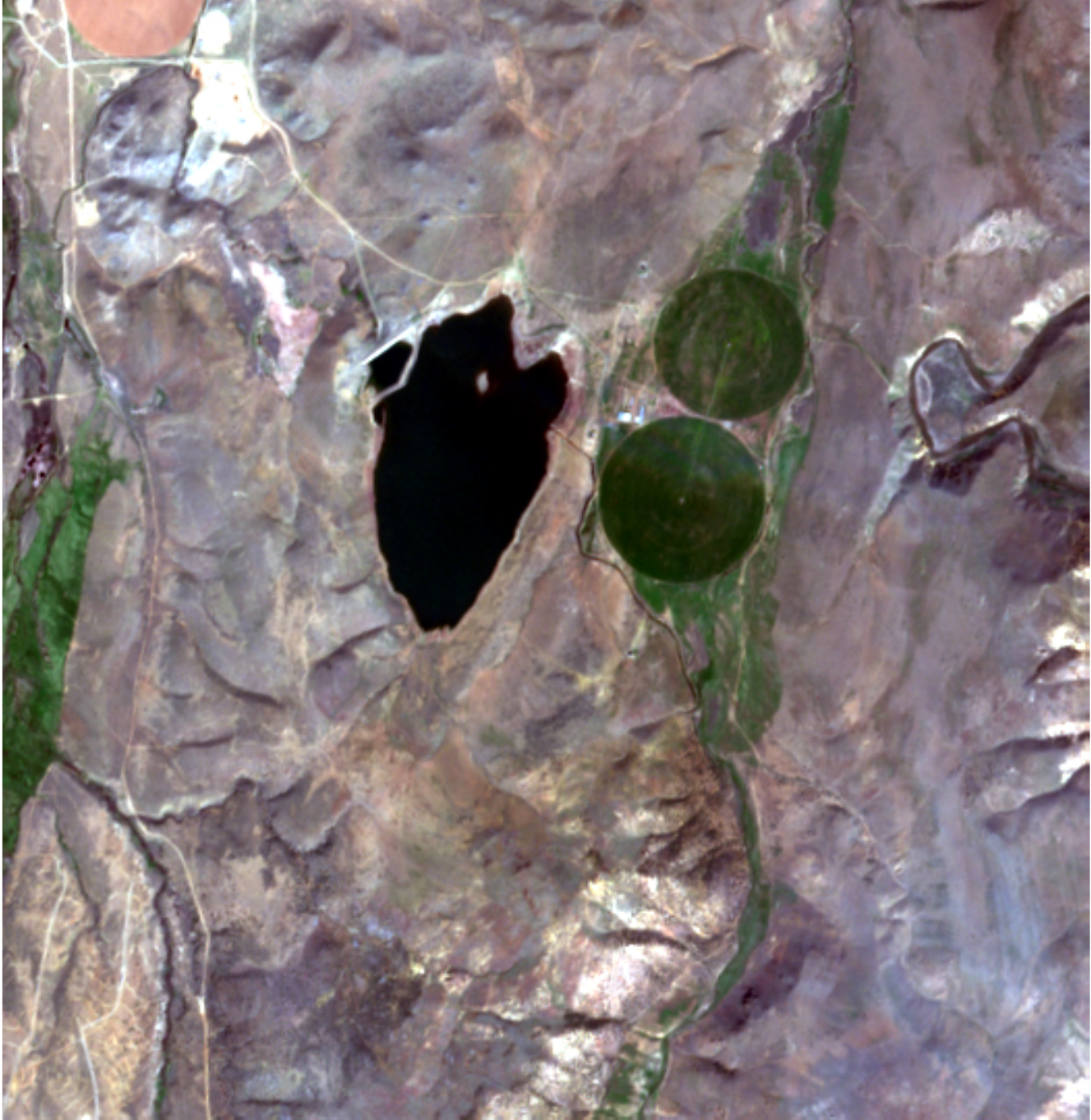}
			\caption{$\mathbf{Y}_{2}$}
			\label{fig:imT2_sen9_1}
	\end{subfigure}
	\begin{subfigure}{\subfwidth}
			\centering
			\includegraphics[width=\figsize]{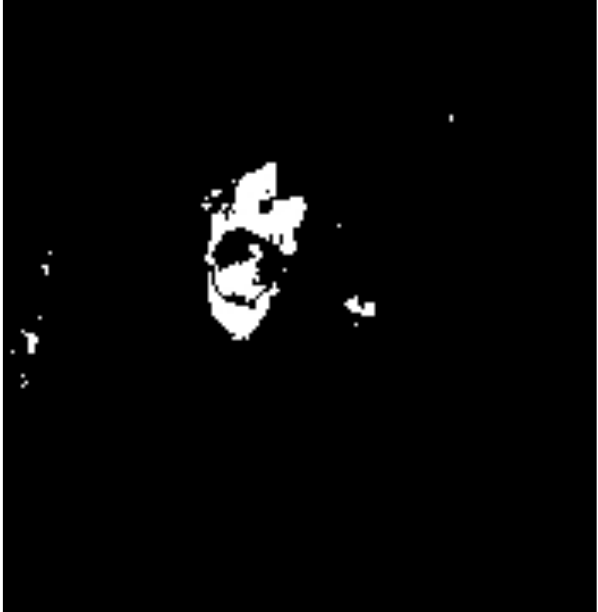}
			\caption{$\hat{\mathbf{d}}_{\mathrm{WC}}$}
			\label{fig:wcCD_sen9_1}
	\end{subfigure}
	\begin{subfigure}{\subfwidth}
			\centering
			\includegraphics[width=\figsize]{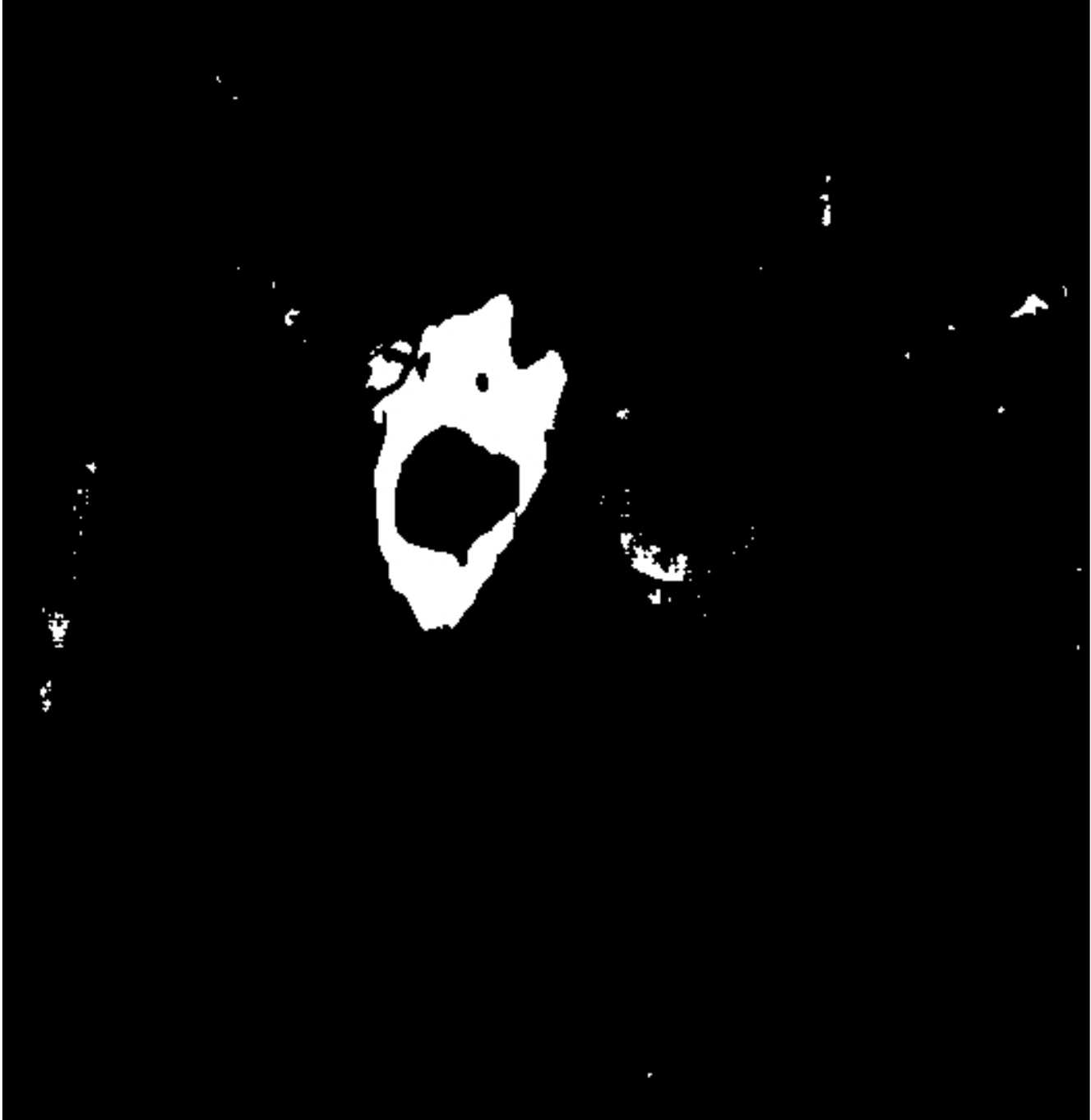}
			\caption{$\hat{\mathbf{d}}_{\mathrm{RF}}$}
			\label{fig:reCD_sen9_1}
	\end{subfigure}
	\caption{Scenario $\mathcal{S}_{9}$: \protect\subref{fig:imT1_sen9_1} Landsat-8 $30$m MS-$5$ observed image $\mathbf{Y}_{1}$ acquired on 09/22/2015, \protect\subref{fig:imT2_sen9_1} Sentinel-2 $10$m MS-$4$ observed image $\mathbf{Y}_{2}$ acquired on 04/12/2016, \protect\subref{fig:wcCD_sen9_1} change mask $\hat{\mathbf{d}}_{\mathrm{WC}}$ estimated by the WC approach from a pair of $30$m MS-$3$ degraded images and \protect\subref{fig:reCD_sen9_1} change mask $\hat{\mathbf{d}}_{\mathrm{RF}}$ estimated by the proposed approach from a $10$m MS-$6$ change image $\Delta\hat{\mathbf{X}}$.}%
	\label{fig:sen9_1}%
\end{figure}

This scenario correspond to a modified instance of scenario $\mathcal{S}_{4}$ (images of different spatial and spectral resolutions) with some non-overlapping bands (as for the previous scenario). The results obtained for one configuration are depicted in Figure \ref{fig:sen9_1}. In this case, the change image $\Delta\hat{\mathbf{X}}$ is characterized by a spatial resolution defined by the highest one among the observed image with a spectral resolution higher than both observed images.  Once again, the results show the accuracy of the proposed method in terms of detection and spatial resolution of the estimated change map.

\subsubsection{Scenario $\mathcal{S}_{10}$}

\begin{figure}[h!]
	\centering
	\begin{subfigure}{\subfwidth}
			\centering
			\includegraphics[width=\figsize]{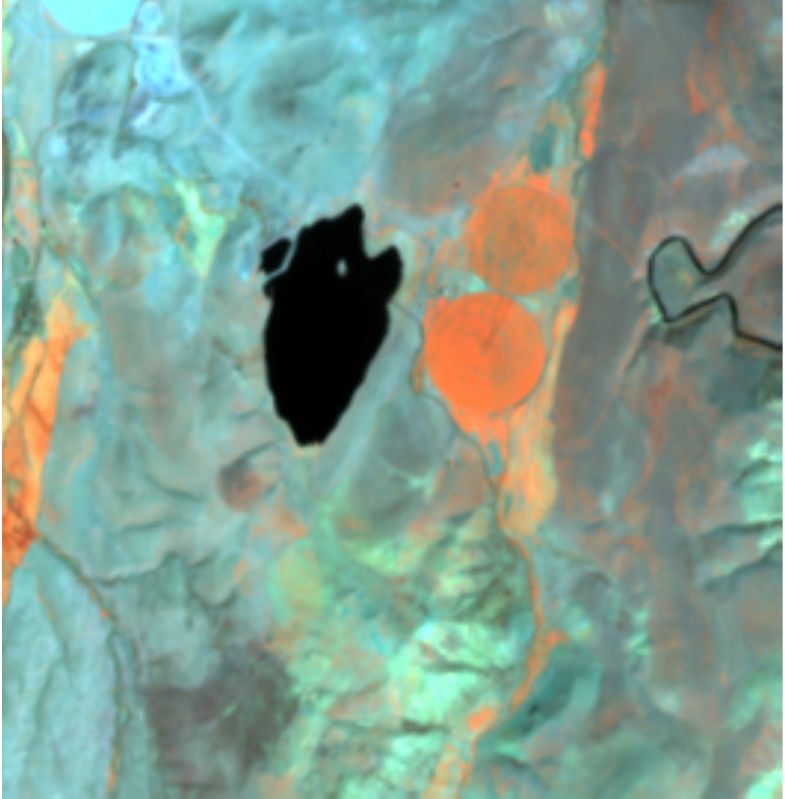}
			\caption{$\mathbf{Y}_{1}$}
			\label{fig:imT1_sen10_1}
	\end{subfigure}
	\begin{subfigure}{\subfwidth}
			\centering
			\includegraphics[width=\figsize]{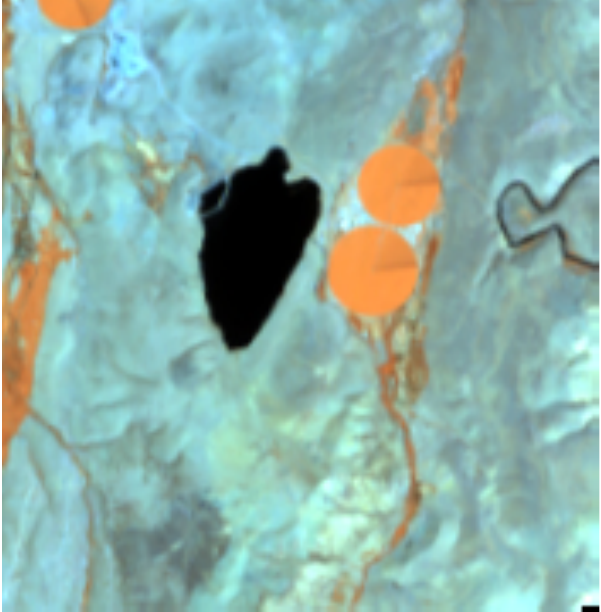}
			\caption{$\mathbf{Y}_{2}$}
			\label{fig:imT2_sen10_1}
	\end{subfigure}
		\begin{subfigure}{\subfwidth}
			\centering
			\includegraphics[width=\figsize]{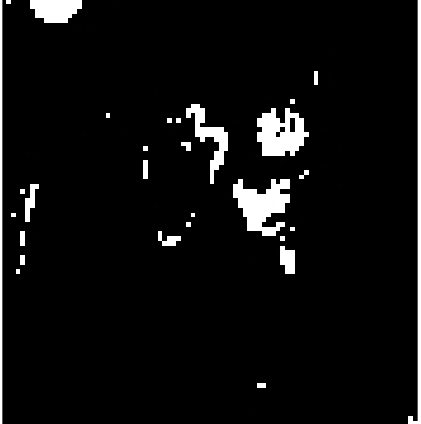}
			\caption{$\hat{\mathbf{d}}_{\mathrm{WC}}$}
			\label{fig:wcCD_sen10_1}
	\end{subfigure}
	\begin{subfigure}{\subfwidth}
			\centering
			\includegraphics[width=\figsize]{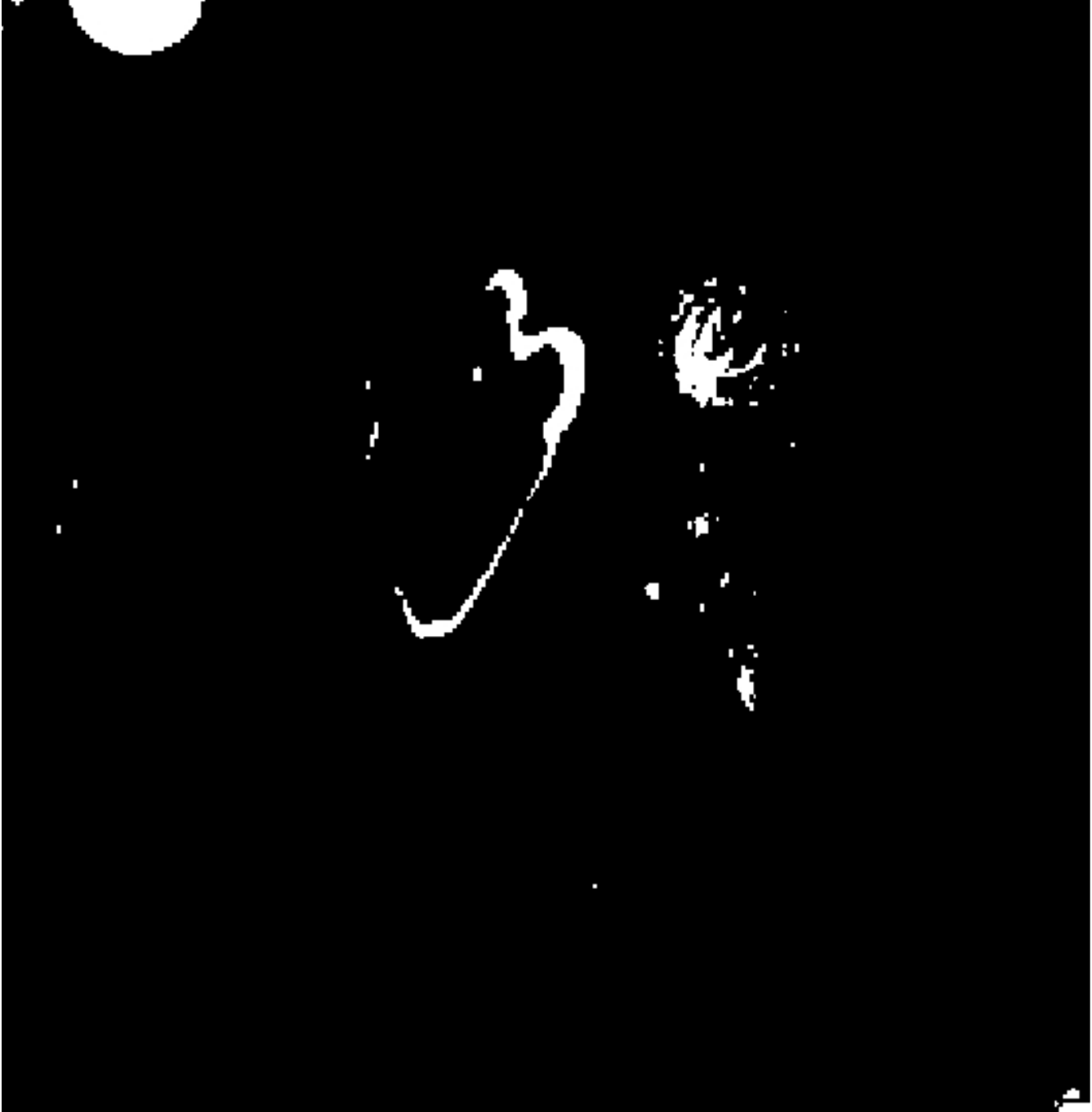}
			\caption{$\hat{\mathbf{d}}_{\mathrm{RF}}$}
			\label{fig:reCD_sen10_1}
	\end{subfigure}
	\caption{Scenario $\mathcal{S}_{10}$: \protect\subref{fig:imT1_sen10_1} Sentinel-2 $20$m MS-$6$ observed image $\mathbf{Y}_{1}$ acquired on 04/12/2016, \protect\subref{fig:imT2_sen9_1} EO-1 ALI $30$m MS-$9$ observed image $\mathbf{Y}_{2}$ acquired on 06/08/2011,\protect\subref{fig:wcCD_sen10_1} change mask $\hat{\mathbf{d}}_{\mathrm{WC}}$ estimated by the WC approach from a pair of $60$m MS-$4$ degraded images and \protect\subref{fig:reCD_sen10_1} change mask $\hat{\mathbf{d}}_{\mathrm{RF}}$ estimated by the proposed approach from a $10$m MS-$11$ change image $\Delta\hat{\mathbf{X}}$.}%
	\label{fig:sen10_1}%
\end{figure}

The last scenario combines of the difficulties previously encountered: images of different spatial and spectral resolution, characterized by a non-integer relative downsampling factor and non-overlapping spectral bands. As for scenarios $\mathcal{S}_{6}$ and $\mathcal{S}_{7}$, the change image $\Delta\hat{\mathbf{X}}$ and change mask $\hat{\mathbf{d}}_{\mathrm{RF}}$ recovered by the proposed RF-based CD method is of higher spatial resolution than the two observed images. In addition, as for scenario $\mathcal{S}_{8}$ and $\mathcal{S}_{9}$, the change image is also defined at a higher spectral resolution. Conversely, the WC approach derives a change image of lower spatial and spectral resolutions before conducting CVA. Figure \ref{fig:sen10_1} depicts the results obtained by both methods. On this particularly challenging scenario, the proposed approach demonstrates its superiority in recovering relevant changes and in localizing them accurately.

% in a variation of \emph{Scenarios} 7. It stands for two images with different spectral resolutions with some non-overlapping bands and spatial resolutions attached by a non-integer downsampling factor. The same stategy as for \emph{Scenario} 6 is employed here in order to overcome the non-integer downsampling factor problem. Additionaly, a spectral superresolution step is required in the proposed method in order to bring all estimated variables to the same spectral resolution, which increase the difficulty to solve the problem.
% Figure \ref{fig:sen10_1} represent on example of this scenario together with CD map results of the proposed and WC method. The proposed method present the highest spatial and spectral resolution CD estimation $\Delta\mathbf{X}$ , in this case, a 10m 11-Bands MS image, againist the 60m 4-band MS of the WC method estimation. This important dual resolution gap contributes a lot in the difference on the false alarm/detection rate between both results proving, once more, the efficiency and accuracy of the proposed method.

\section{Conclusion}
\label{sec:conclusion}
This paper derived a robust fusion framework to perform change detection between optical images of different spatial and spectral resolutions. The versatility of the proposed approach allowed all possible real scenarios to be handled efficiently. The technique was based on the definition of two high spatial and spectral resolution latent images related to the observed images via a double physically-inspired forward model. The difference between these two latent images was assumed to be spatially sparse, implicitly locating the changes at a high resolution scale. Inferring these two latent images was formulated as an inverse problem which was solved within a $2$-step iterative scheme. Depending on the considered scenario, these $2$ steps can be interpreted as ubiquitous signal \& image processing problems (namely spatial super-resolution, spectral deblurring, denoising or multi-band image fusion) for which closed-form solutions or efficient algorithms had been recently proposed in the literature. Real images acquired by four different sensors were used to illustrate the accuracy and the flexibility of the proposed method, as well as its superiority with respect to the state-of-the-art change detection methods. Future works will assess the robustness of the proposed technique w.r.t. nonlinear effects (e.g., due to atmospheric effects, geometric and radiometric distortions). Detecting changes between optical and non-optical data is also under investigation.

\appendix
\section{Matrix normal distribution}
\label{app:matrix_normal_distribution}
The probability density function $p(\mathbf{X}|\mathbf{M},\mathbf{\Sigma}_{r},\mathbf{\Sigma}_{r})$ of a matrix normal distribution $\mathcal{M}\mathcal{N}_{r,c}(\mathbf{M},\mathbf{\Sigma}_{r},\mathbf{\Sigma}_{c})$ is given in \cite{guptamatrix1999}
\begin{center} $ p\left(\mathbf{X}|\mathbf{M},\mathbf{\Sigma}_{r},\mathbf{\Sigma}_{r}\right) = \frac{ \exp \left(-\frac{1}{2}tr \left[
\mathbf{\Sigma}_{c}^{-1} \left(\mathbf{X}-\mathbf{M}\right)^{T} \mathbf{\Sigma}_{r}^{-1} \left(\mathbf{X}-\mathbf{M}\right) \right]\right)}{\left(2\pi\right)^{rc/2}\left|\mathbf{\Sigma}_{c}\right|^{r/2}\left|\mathbf{\Sigma}_{r}\right|^{c/2}}$
\end{center}
where $\mathbf{M} \in \mathbb{R}^{r\times c}$ is the mean matrix, $\mathbf{\Sigma}_{r} \in \mathbb{R}^{r\times r}$  is the row covariance matrix and $\mathbf{\Sigma}_{c} \in \mathbb{R}^{c\times c}$ is the column covariance matrix.

\section{Acknowledgments}
Part of this work has been supported by Coordena\c{c}\~{a}o de Aperfei\c{c}oamento de Ensino Superior (CAPES), Brazil, and EU FP7 through the ERANETMED JC-WATER program [MapInvPlnt Project ANR-15-NMED- 0002-02].

%\section*{References}
\bibliographystyle{elsarticle-harv}
\bibliography{strings_all_ref,HSbib_cleaned}

\end{document}